\documentclass[aps,floats,pre,superscriptaddress,showpacs,twocolumn]{revtex4-1}

\usepackage{amsfonts,amsmath,amssymb} \usepackage{bm} \usepackage{dcolumn}
\usepackage{amsthm}
\usepackage{epsfig} \usepackage{latexsym}
\usepackage{mathrsfs}
\usepackage[utf8]{inputenc}
\usepackage[dvipsnames]{xcolor}

\begin{document}

\title{Breakdown of quantum-classical correspondence and dynamical generation of entanglement
}
\author{Chushun Tian}
\email{ct@itp.ac.cn}
\affiliation{CAS Key Laboratory of Theoretical Physics and Institute of Theoretical Physics, Chinese Academy of Sciences, Beijing 100190, China}
\author{Kun Yang}
\email{kunyang@magnet.fsu.edu}
\affiliation{National High Magnetic Field Laboratory and Department of Physics, Florida State University, Tallahassee, FL 32306, USA}

\date{\today}

\begin{abstract}

The {\it exchange} interaction arising from the particle indistinguishability is of central importance to {physics of many-particle quantum systems}. Here we study analytically the {dynamical generation of quantum entanglement induced by this interaction in an isolated system, namely, an ideal Fermi gas confined in a chaotic cavity,} which evolves unitarily from a non-Gaussian pure state. We find that the breakdown of the quantum-classical correspondence of particle motion{, via dramatically changing the spatial structure of many-body wavefunction,} leads to profound changes of the entanglement structure. Furthermore, for a class of initial states, such change leads {to the approach to thermal equilibrium everywhere in the cavity, with the well-known Ehrenfest time in quantum chaos as the thermalization time. Specifically, the quantum expectation values of various correlation functions at different spatial scales are all determined by the Fermi-Dirac distribution. In addition, by using the reduced density matrix (RDM) and the entanglement entropy (EE) as local probes, we find that the gas inside a subsystem is at equilibrium with that outside, and its thermal entropy is the EE, even though the whole system is in a pure state. As a by-product of this work, we provide an analytical solution supporting an important conjecture on thermalization, made and numerically studied by Garrison and Grover in: Phys. Rev. X \textbf{8}, 021026 (2018), and strengthen its statement}.

\end{abstract}

\maketitle

\section{Introduction}
\label{sec:introduction}

The quantum entanglement is a fundamental and counterintuitive property of quantum many-body systems, and is finding applications in an increasingly broad range of branches of science and technology \cite{Horodecki09,Fradkin13,Chuang00,Yang19,Degen17}. Remarkably, it has been found \cite{Gemmer04,Popescu06,Lebowitz06} to give rise to the emergence of thermal equilibrium phenomena in a system coupled to an environment from the overwhelming majority of pure states describing the isolated composite, namely, the system $+$ the environment. This finding, called ``canonical typicality'', sheds new light on the {long-debated} foundational issue of statistical physics \cite{von Neumann29}, namely, whether and how an isolated system undergoing unitary pure-state evolution can exhibit thermal phenomena, commonly conceived to be the long-time behaviors of a virtual ensemble of isolated systems prepared under the same macroscopic conditions. There have been increasing interests in searching the relations between the quantum entanglement and the fundamentals of statistical physics and applications of such relations in various modern topics \cite{Gogolin16,Nandkishore15}. Notwithstanding this, many key aspects remain largely unexplored.

First, there are diverse sources that can generate the quantum entanglement. The canonical typicality crucially relies on that a system and an environment are entangled via a direct interaction, which accounts for an interaction term in the total Hamiltonian. Yet, even though the direct interaction is absent, the quantum entanglement can still arise, provided the constituting particles are indistinguishable, i.e., identical. This {(particle) indistinguishability-induced entanglement} cannot be attributed to a Hamiltonian, rather, is attributed to the (anti)symmetry of many-particle wavefunctions upon exchanging two particles, namely, the { exchange interaction} \cite{Landau37} of indistinguishable bosons (fermions). {Although there have been many theoretical and experimental investigations of this type of quantum entanglement (see, e.g., Refs.~\cite{Adesso20,Preiss20,Marzolino20}) and its potential applications have even been proposed \cite{Hu10,Jacob20}}, not until recently have the studies of its roles in pure-state equilibrium {\cite{Singh14,Lai15,Tian18,Magan16,DasSarma16,Rigol17,Mueller18}} and nonequilibrium {\cite{Sen17,Zaanen17}} statistical physics been initiated. In particular, kinematic studies based on both numerical experiments and analytical theories {\cite{Lai15,Tian18,Mueller18}} have shown that this entanglement leads the overwhelming majority of Fock states to behave like a statistical ensemble at thermal equilibrium. This pure-state statistical phenomenon, called ``eigenstate typicality'' (see Sec.~\ref{sec:eigenstate_typicality} for detailed introduction), has deep connections to the so-called {{\it limit shape}} of random geometric objects discovered by mathematicians \cite{Vershik94,Vershik96,Vershik04,Okounkov16}, and is conceptually different from the canonical typicality. Thus advancing the fundamental principle of standard, ensemble-based, statistical physics, namely, that many-particle system's statistical behaviors depend strongly on the exchange interaction, to pure-state statistical physics potentially {opens} up a highly interdisciplinary research area.

\begin{figure}[t]
\includegraphics[width=8.6cm]{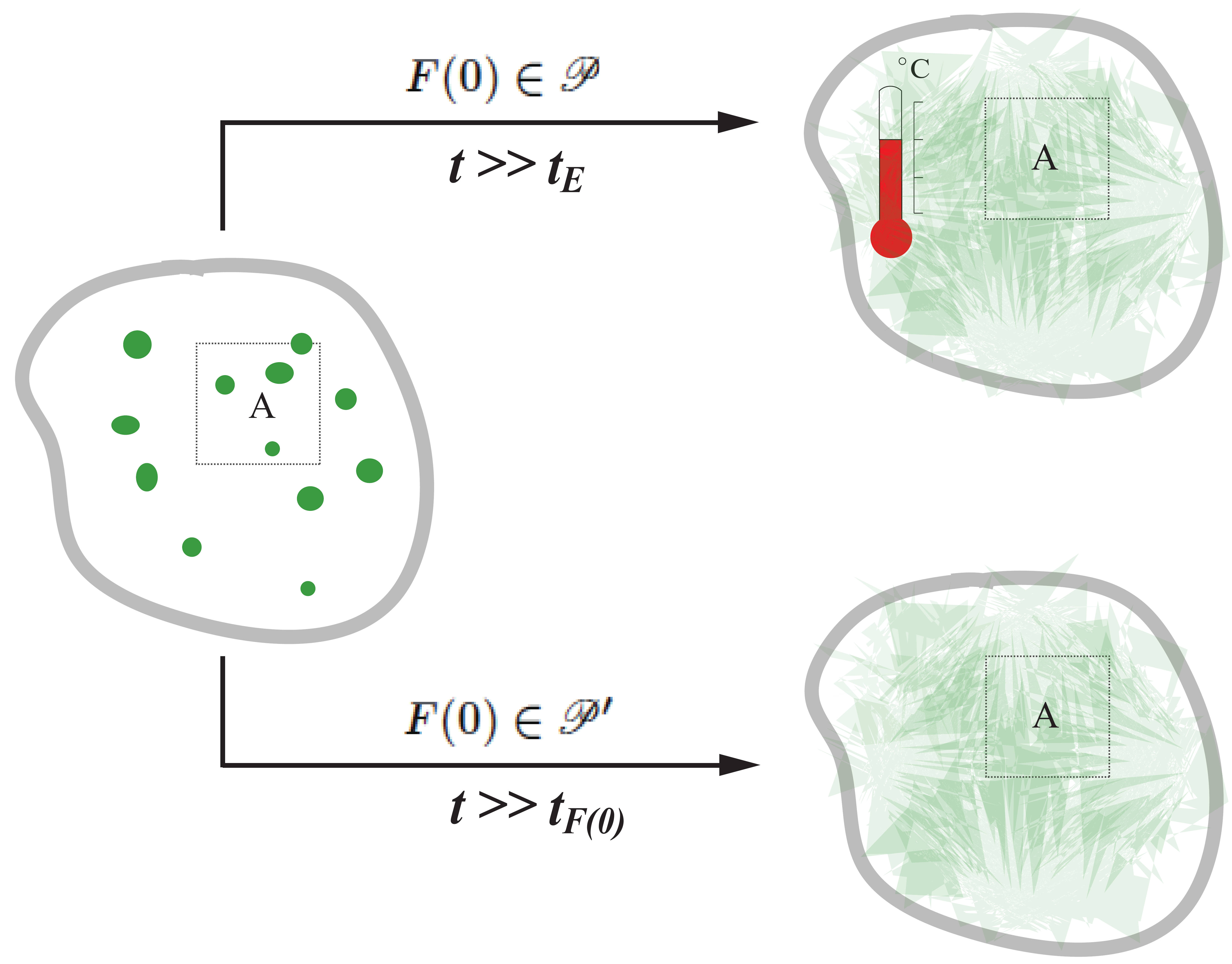}
\caption{Evolution of the entanglement structure of an ideal Fermi gas confined in a chaotic cavity. Left: Initially, the overlap of different particle wavepackets (green spots) and that between a wavepacket and the boundary of the subsystem A (dashed line) are small, resulting in a low-level entanglement and a small EE. Right: As the quantum-classical correspondence of particle motion breaks down, particle waves spread over the entire cavity and strongly overlap with each other as well as with the boundary of A, resulting in a high-level entanglement and a large EE. Undergoing this change in the entanglement structure, the entire Fermi gas is thermalized, if the initial state $F(0)$ is in $\mathscr{P}$, while equilibrates if $F(0)$ is in $\mathscr{P}'$. See the text for details.}
\label{fig:1}
\end{figure}

Second, the canonical typicality addresses the kinematic aspect of the roles of quantum entanglement. It makes no references to system's dynamical properties, notably, integrable or chaotic. This situation is similar to that in standard equilibrium statistical physics \cite{Landau37}, but is in sharp contrast to that in other proposals \cite{Deutsch91,Srednicki94,Rigol08,Rigol16,Borhonovi16} for pure-state equilibrium statistical physics, all of which have many-body chaos arising from the direct interaction as the starting point. Thus investigations of the dynamic aspect, more precisely, the roles of the dynamics of quantum entanglement, especially the indistinguishability-{induced} entanglement, in the emergence of pure-state equilibrium statistical physics, are of urgent and fundamental importance.

In fact, on the numerical side, there have been considerable studies of the {EE dynamics} in a variety of chaotic or integrable quantum systems \cite{Sarkar99,Tanaka02,Cardy05,Balasubramanian11,Huse13}. However, all those systems are subjected to a direct interaction. On the analytical side, to address the entanglement dynamics in generic systems remains an intellectual challenge. It is often assumed that a quantum system has some strong properties, e.g., (quantum) integrable, conformal, holographic, and one-dimensional ($1$D) \cite{Cardy05,Balasubramanian11,Maldacena13,Liu14,Mueller16}, and (or) that a quantum state is Gaussian (see Refs.~\cite{Peschel09,Hackl18} and references therein). However, in realistic systems these properties are often (partially) absent.

{Third, kinematic studies based on various typicality considerations \cite{Gemmer04,Popescu06,Lebowitz06,Lai15,Mueller18,Liu18} have established thermalization so far only for a subsystem. This differs from the (original version of) eigenstate thermalization \cite{Deutsch91,Srednicki94,Rigol08} in a fundamental aspect. That is, the latter is for the whole system. An exception \cite{Tian18} is that, for chaotic systems, both short- and long-ranged one-particle correlation at a typical eigenstate are thermal, in contrast to that only the short-ranged correlation is thermal for integrable systems \cite{Lai15,Mueller18}. This difference may be regarded as a precursor that in a chaotic (an integrable) system, thermalization can be established for whole system (a subsystem). Most importantly, it implies dramatic impacts of the dynamic aspects of entanglement on thermalization.}

In this work, we consider an ideal quantum gas confined in a two-dimensional ($2$D) chaotic cavity of volume $V$ (Fig.~\ref{fig:1}) to explore the dynamics of the indistinguishability-{induced} entanglement and its relations to the emergence of pure-state equilibrium statistical physics. This gas is composed of $N(\gg 1)$ indistinguishable free fermions. They are subjected to the exchange interaction only and thus is truly ideal. As such, the chaoticity of this isolated many-particle system arises solely from the collision between a particle and the cavity boundary.
So this chaoticity is of one-body nature, in sharp contrast to many-body chaos {--- arising from direct interaction between particles ---} that is widely adopted as a starting point in the studies of the foundations of statistical physics \cite{Deutsch91,Srednicki94,Rigol08,Rigol16,Borhonovi16,Krylov79}. Note that the relevance of one-body chaos to the studies of the foundations of statistical physics has been emphasized by many researchers; see Ref.~\cite{Dorfman99} and references therein. However, those studies are strictly based on classical mechanics, and thus cannot address the exchange interaction and the ensuing quantum entanglement. {In fact, the combined effects of one-body chaos and exchange interaction remain a largely unexplored realm \cite{Tian16,Galitski20}.}

We develop an analytical theory of the dynamics of the indistinguishability-{induced} entanglement accompanying the pure-state evolution. The initial state, $F(0)$, considered in this work is the antisymmetrization of $N$ spatially localized wavepackets, with the overlap of two wavepackets being very small in general (Fig.~\ref{fig:1}, left). This gives rise to a {low-level} entanglement. Such initial states are non-Gaussian. They are superposed by Fock states in a microcanonical energy shell and divided into two classes. In one class, denoted as $\mathscr{P}$, the majority of weight goes to Fock states bearing the aforementioned eigenstate typicality, while in the other class, denoted as $\mathscr{P}'$, Fock states bearing no eigenstate typicality have a significant weight. We refer to Sec.~\ref{sec:entanglement_entropy_typical_states} for the mathematical description of $\mathscr{P}$ and $\mathscr{P}'$ as well as of the eigenstate typicality.

{We employ various particle correlations, the RDM --- of arbitrary location, size and geometry --- and the EE to probe the dynamical generation of the entanglement. These quantities have different mathematical structures and describe different aspects of the entanglement. The correlation function can be expressed as the quantum expectation of some operator at the evolving pure state $F(t)$ at time $t$. Various particle correlation functions at different spatial scales altogether provide rich information of entanglement structure in space. Moreover, in pure-state statistical physics \cite{von Neumann29,Deutsch91,Srednicki94,Rigol08,Rigol16}, comparing such quantum expectation with corresponding thermal values is nowadays a canonical method of diagnosing thermalization. In particular, nonlocal observables such as long-ranged correlation functions can be used to diagnose whether thermalization is established for the whole system. The RDM cannot be expressed directly as the quantum expectation of some operator. However, when we expand it in some operator basis, the expansion coefficients are multi-particle correlation functions (see Sec.~\ref{sec:formulation} for details). Equivalently, one may think about the RDM as a peculiar linear combination of quantum expectations of multi-particle correlations, with operator-value coefficients. In this sense, the mathematical structure of EE is extraordinary, because it cannot be expressed as the linear combination of quantum expectations of any operators \cite{Raju20}. The RDM and the EE characterize the entanglement between a subsystem and its complement, and together with short-ranged correlation functions can serve as local probes of thermal properties of the gas.}

{We find (cf.~Fig.~\ref{fig:1}) that the entanglement is generated in the course of pure-state evolution: $F(0)\rightarrow F(t)$, and all probes of entanglement} equilibrate at a time scale signalling the breakdown of the quantum-classical correspondence of the particle motion, with profound changes of the {spatial structure of many-body wavefunction and the ensuing} entanglement structure. Moreover, for initial states in $\mathscr{P}$ we find that {the whole gas is thermalized and, strikingly, the thermalization time is} the Ehrenfest time $t_E$ \cite{Larkin69,Zaslavsky81,Izrailev90}. Thus it is suggested that the combination of one-body chaos and the exchange interaction suffices to give rise to quantum thermalization. This is conceptually different from other scenarios for thermalization of an isolated quantum system \cite{Borhonovi16,Deutsch91,Srednicki94,Rigol08,Rigol16}, for which many-body chaos arising from the direct interaction is indispensable. Moreover, our findings suggest that the approach to thermal equilibrium in an isolated quantum system is accompanied by the {dynamical generation} of the indistinguishability-{induced} entanglement, resembling the second law in thermodynamics.

The rest of the paper is organized as follows. In Sec.~\ref{sec:summary_main_results} we summarize the main analytical results{, present a physical picture for the evolution of the entanglement structure, and discuss the implications of our results on a conjecture on thermalization imposed and studied numerically in Ref.~\cite{Grover18}}. In Sec.~\ref{sec:entanglement_entropy_typical_states} we first review the notion of the eigenstate typicality. Then we provide the mathematical description of the two classes of initial states, $\mathscr{P}$ and $\mathscr{P}'$, and discuss their experimental preparations. Finally we formulate three problems that are closely related and to be studied in this work: they concern respectively the dynamics of the spatial correlation of particles, the RDM and the EE. In Sec.~\ref{sec:evolution_correlation_function} we solve the first problem for the initial state in $\mathscr{P}$. This section is a substantial extension of our earlier preprint \cite{Tian16}. A preliminary result reported in that work is strengthened significantly by improving the original derivations. Armed with the results obtained in Sec.~\ref{sec:evolution_correlation_function}, we solve the second and the third problems in Sec.~\ref{sec:entanglement_entropy_evolution} for the initial state in $\mathscr{P}$. In Sec.~\ref{sec:some_generalizations} we use the scheme developed in Secs.~\ref{sec:evolution_correlation_function} and \ref{sec:entanglement_entropy_evolution} to study the three problems for the initial state in $\mathscr{P}'$. In Sec.~\ref{sec:comparison_standard_statistical_physics} we compare the thermalization scenario implied by the results obtained in Secs.~\ref{sec:evolution_correlation_function} and \ref{sec:entanglement_entropy_evolution} with the paradigm of standard statistical physics. We make concluding remarks in Sec.~\ref{sec:conclusions_discussions}. Some additional technical details {and further discussions} are given in Appendices \ref{sec:quantum_recurrence_time}-\ref{sec:proof}.

\section{Summary of main analytical results and their implications}
\label{sec:summary_main_results}

\subsection{Results for $\boldsymbol{F(0)\in \mathscr{P}}$}
\label{sec:results_F0_P}

We first summarize the main analytical results for the initial state $F(0)\in \mathscr{P}$. We find that, as the pure state follows the Schr\"odinger equation to {evolve: $F(0)\rightarrow F(t)$}, various probes of quantum entanglement relax at the time scale of the Ehrenfest time, well known to signal the breakdown of the quantum-classical correspondence of wavepacket dynamics in quantum chaos \cite{Larkin69,Zaslavsky81,Izrailev90},
\begin{equation}\label{eq:19}
    t_E=\frac{1}{\lambda_L}\ln \frac{A}{\hbar}.
\end{equation}
Here $\lambda_L$ is the Lyapunov exponent characterizing the exponential instability of the classical single-particle motion, and $A$ is the characteristic classical action {which is} much larger than $\hbar$. Both $\lambda_L$ and $A$ are determined by the average single-particle energy, the particle mass $m$, and the cavity size $L\sim\sqrt{V}$, and $\lambda_L$ is order of the inverse free flight time at that energy. Very recently, experimental \cite{Tian18a} and theoretical \cite{Stanford14,Galitski17} investigations on physics occurring at the time scale of $t_E$ in a variety of quantum systems have been boosted. However, we are not aware of any reports on the roles of $t_E$ in the entanglement dynamics. Moreover, very little \cite{Tian16} has been known about its roles in pure-state statistical physics, despite that in as early as 1940s N. S. Krylov realized the fundamental importance of $t_E$ in the foundations of standard statistical physics \cite{Krylov79}.

We analytically study the quantum expectation value of the spatial correlation of $j$ {($1\leq j\leq N$)} particles, that are annihilated at spatial points $\{{\bf r}\}\equiv \{{\bf r}_1,{\bf r}_2,\cdots, {\bf r}_j\}$ and created at $\{{\bf r}'\}\equiv\{{\bf r}'_1,{\bf r}'_2,\cdots, {\bf r}'_j\}$ {with all ${\bf r},{\bf r}'$ being arbitrary in the cavity},
at the evolving state $F(t)$, which is defined as
\begin{eqnarray}\label{eq:76}
    M^{(j)}_{\{{\bf r}\}\{{\bf r}'\}}
    (t)\equiv\langle F(t)| a^\dagger_{{\bf r}'_1}\cdots a^\dagger_{{\bf r}'_j}a_{{\bf r}_1}\cdots a_{{\bf r}_j}|F(t)\rangle,
\end{eqnarray}
where $a_{{\bf r}_i}$ ($a^\dagger_{{\bf r}_i}$) is the annihilation (creation) operator at the spatial point ${\bf r}_i$. We show below that
\begin{eqnarray}
\label{eq:87}
  &&M^{(j)}_{\{{\bf r}\}\{{\bf r}'\}}(t)\stackrel{t\gg t_E}{\longrightarrow}\sum_P\sigma(P)\nonumber\\
  &&\quad\quad\times\prod_{k=1}^j\frac{1}{V}\int dm(\nu)J_0\left(\frac{|{\bf r}_{k}-{\bf r}'_{P(k)}|}{\lambda_{\varepsilon_\nu}}\right)n_{FD}(\varepsilon_\nu).\,\,\,\,\,
\end{eqnarray}
Here the sum is over all permutations $P$, with $\sigma(P)$ being the signature of $P$. $J_{\bar\nu}(x)$ denotes the Bessel function of order ${\bar{\nu}}$. $\lambda_{\varepsilon}$ is {particle's de Broglie wavelength at energy $\varepsilon$. $\nu$ labels} the single-particle eigenstate {with corresponding eigenenergy $\varepsilon_\nu$}. Importantly, the relaxed value (namely, the right-hand side) depends on only few parameters: the temperature $T$ and the chemical potential $\mu$ in the Fermi-Dirac distribution: $n_{FD}(\varepsilon)\equiv 1/(e^{\frac{\varepsilon-\mu}{T}}+1)$ \cite{note1}, which are determined by the {central energy of the microcanonical shell $E$, the particle number $N$}, and the spectral density $dm(\nu)$ and the volume $V$ of the cavity, {irrespective} of the detailed constructions of $F(0)$ and the system. {Therefore, the whole confined gas is thermalized.}

Then, we study the RDM for the subsystem A of arbitrary location, size and geometry, defined as
\begin{equation}\label{eq:187}
  \hat\rho_A(t)\equiv{\rm Tr}_{\bar{A}}(\hat\rho(t)), \quad \hat\rho(t)\equiv|F(t)\rangle\langle F(t)|,
\end{equation}
where the trace ${\rm Tr}_{\bar{A}}$ is restricted on the complement {of the subsystem, $\bar{A}$}. Using Eq.~(\ref{eq:87}), we find that the RDM relaxes to a Gaussian state, although before the relaxation the RDM is non-Gaussian. Moreover, the relaxed (operator) value is determined completely by the relaxed one-particle correlation function given by Eq.~(\ref{eq:87}) (for $j=1$). Formally,
\begin{eqnarray}\label{eq:155}
    \hat\rho_A(t)\stackrel{t\gg t_E}{\longrightarrow}Gaussian\,\,{RDM\,\,[T,\mu,dm(\nu),V]}
\end{eqnarray}
{with the bracket standing for the parameters on which the relaxed RDM depends,} and the explicit form of the right-hand side is given by Eq.~(\ref{eq:31}). We emphasize that, unlike other works {\cite{Peschel09,Klich06a}}, here $T,\mu$ in the relaxed RDM are {\it not} the effective temperature and chemical potential which are determined by the eigenvalues of the one-particle correlation --- when viewed as an operator --- restricted on A {and depend on subsystem's size and geometry in general \cite{Peschel09}. As pointed out in Ref.~\cite{Peschel09}, a Gaussian RDM with an effective temperature ``{\it is not a true Boltzmann operator}''. Instead, $T,\mu$ in Eq.~(\ref{eq:155})} are genuine thermodynamic parameters characterizing {thermal properties of the {\it entire} gas at equilibrium, and are determined completely by $E,N$. We also emphasize that the relaxed RDM in Eq.~(\ref{eq:155}), though being Gaussian and governed by thermal properties of the entire gas, is not necessarily a thermal (canonical or grand canonical) ensemble, because its covariance matrix has a very complicated dependence on subsystem's size and geometry and thermal parameters $T,\mu$ in general. As we will discuss in Sec.~\ref{sec:generalized_garrison_grover_problem}, the reduction of such Gaussian RDM to the thermal ensemble arises, when the subsystem A is deep inside the cavity.}

{The EE associated to $\hat\rho_A(t)$ is} defined as
\begin{equation}\label{eq:30}
    S_A(t)\equiv-{\rm Tr}_A\left(\hat\rho_A(t)\ln\hat\rho_A(t)\right),
\end{equation}
where the trace ${\rm Tr}_A$ is restricted on A. Using the results for the relaxed RDM and one-particle correlation function, {we find that
\begin{equation}\label{eq:80}
    S_A(t)\stackrel{t\gg t_E}{\longrightarrow}thermal\,\,value,
\end{equation}
with the relaxed value depending on $T,\mu,dm(\nu),V$, and its explicit expression depends on the location, the geometry and the size of A in general.}

{Furthermore, we find the closed form of the relaxed EE for a special but broad class of subsystems A. It requires that A is deep inside the bulk, so that its volume $V_A\ll V$ but is sufficiently large, and is either a polygon or convex (such as a disk). The condition regarding the geometry is likely technical but not physical. We perform the calculations for the discrete lattice space (with a lattice constant $a$ smaller than all particle wavelengths)}, and then pass to the continuum limit $a\rightarrow 0$, obtaining
\begin{eqnarray}
    S_A(t)\stackrel{t\gg t_E}{\longrightarrow}\left\{\begin{array}{c}
                                                       N_A{S_a}=(V_A/a^2){S_a},\quad a>0; \\
                                                       V_AS_0,\quad a=0,\quad\quad\quad\quad\quad\quad\quad
                                                     \end{array}
    \right.\label{eq:110}
\end{eqnarray}
where {$N_A
$ is the subsystem volume in the lattice}. Equation (\ref{eq:110}) implies that the relaxed EE obeys the volume law, with ${S_a}$ and $S_0$ being the relaxed EE density corresponding to the lattice and continuous space, respectively. For the lattice space, we find that the relaxed EE density
\begin{eqnarray}
\label{eq:111}
  {S_a}&=&-\int\!\!\!\!\int_{-\pi}^{\pi}\frac{d\theta_1d\theta_2}{(2\pi)^2}\big({\cal C}(\theta_1,\theta_2)\ln{\cal C}(\theta_1,\theta_2)\nonumber\\
  &+&(1-{\cal C}(\theta_1,\theta_2))\ln(1-{\cal C}(\theta_1,\theta_2))\big),
\end{eqnarray}
with
\begin{eqnarray}\label{eq:113}
    {\cal C}(\theta_1,\theta_2)=\frac{a^2}{V}\sum_{n_1,n_2\in \mathbb{Z}}e^{i(n_1\theta_1+n_2\theta_2)}\quad\quad\nonumber\\
    \times\int dm(\nu)J_0\left(\frac{a}{\lambda_{\varepsilon_\nu}}\sqrt{n_1^2+n_2^2}\right)n_{FD}(\varepsilon_\nu),
\end{eqnarray}
which is thermal. In the continuum limit: $a\rightarrow 0$, we find that, strikingly, ${S_a}/a^2$ is identical to
\begin{eqnarray}
\label{eq:72}
    S_0=-\int\frac{d{\bf p}}{(2\pi\hbar)^2}\Big(n_{FD}(\frac{{\bf p}^2}{2m})\ln n_{FD}(\frac{{\bf p}^2}{2m})\nonumber\\
  +(1-n_{FD}(\frac{{\bf p}^2}{2m}))\ln(1-n_{FD}(\frac{{\bf p}^2}{2m}))\Big),
\end{eqnarray}
which is the {thermal entropy density of an unconfined ideal Fermi gas in standard statistical physics \cite{Landau37}. The Fermi-Dirac distribution has the familiar form, $n_{FD}(\frac{{\bf p}^2}{2m})$, corresponding to the unconfined ideal gas. This special Fermi-Dirac distribution was analytically derived for a many-body eigenstate before \cite{Srednicki94}, as a key characteristic of eigenstate thermalization. Contrary to the present work, that result holds only for systems where the constituting fermions have direct interaction. In Ref.~\cite{Lai15}, the distribution $n_{FD}(\frac{{\bf p}^2}{2m})$ was derived for a many-body eigenstate of free Fermi gas on a torus, which, however, has a fundamental difference from the present system, i.e., exhibits translation invariance with ${\bf p}$ as the corresponding good quantum number. We are not aware of any reports on this result for chaotic systems, and will discuss its physical implications in Sec.~\ref{sec:generalized_garrison_grover_problem}. We shall also see in Sec.~\ref{sec:continuum_limit} that this result is connected to Widom's theorem for the Fredholm determinant of the high-dimensional integral equation with translational kernels \cite{Widom60}.}

{The results summarized above} suggest that, despite that the pure-state evolution is unitary and that the fermions have no direct interaction, the system exhibits quantum thermalization, and the thermalization time is $t_E$. {Because the entire gas is at thermal equilibrium, this is different from subsystem thermalization at a pure state for systems with \cite{Gemmer04,Popescu06,Lebowitz06} or without \cite{Lai15,Liu18} a direct interaction. Also, here the Fermi-Dirac distribution emerges in a way different from earlier works \cite{Srednicki94,Borhonovi16,Gribakin99,Benenti01}, where a direct interaction is required}.

\subsection{Results for $\boldsymbol{F(0)\in \mathscr{P}'}$}
\label{sec:results_F0_PP}

For the initial state $F(0)\in \mathscr{P}'$, we find that the particle correlation, the RDM and the EE all relax also, but at a different time scale, $t_{F(0)}$. This time scale also signals the quantum-classical correspondence breakdown in the presence of one-body chaos, and, similar to $t_E$, has a logarithmic dependence on $\hbar$. However, for $t_{F(0)}$ the pre-logarithm factor and the action rescaling $\hbar$ depend on the details of the constructions of $F(0)$. The relaxed values of various probes of quantum entanglement depend on the details of the constructions of $F(0)$ also; see Eqs.~(\ref{eq:161}) and (\ref{eq:162}) for their explicit expressions. These results imply that the system equilibrates, but is not thermalized.

\subsection{Evolution of entanglement structure}
\label{sec:physical_picture_evolution_entanglement_structure}

The results summarized above indicate that the breakdown of the quantum-classical correspondence of {particle motion} gives rise to a profound change in the entanglement structure{, due to the profound change in the spatial structure of many-body wavefunction (Fig.~\ref{fig:1}). For simplicity we focus on $F(0)\in\mathscr{P}$ in this subsection.}

{Let us start from the EE. At} short time the particles are localized wavepackets, which do not overlap with the boundary of the subsystem A in general. So we may ignore this overlap and obtain a product state $F(t)=\Psi_A(t)\Psi_{\bar A}(t)$, where the factors: $\Psi_A(t)$, $\Psi_{\bar A}(t)$ is the many-body wavefunction in $A$ and $\bar A$, respectively. So the EE vanishes. Owing to the quantum-classical correspondence, i.e., that each quantum particle behaves essentially as a classical one, the unitary pure-state evolution merely results in the change of the configuration of the wavepacket centers, namely, the detailed form of $\Psi_A(t)$ and $\Psi_{\bar A}(t)$, but does not destroy the product structure of $F(t)$. Thus the quantum-classical correspondence leads to a low-level entanglement of fermions, although they are indistinguishable. As the time increases, more and more wavepackets spread out of A or {\it vice versa}. So the product structure is destroyed and the EE increases. Eventually, when the quantum-classical correspondence breaks down, all wavepackets spread, owing to the chaoticity, to the entire cavity. This gives rise to a high-level entanglement, with the EE saturation as a manifestation.

{Furthermore, from Fig.~\ref{fig:1} we see that initially not only the EE vanishes, but also many parts of the gas are disentangled. The situations are changed completely when the quantum-classical correspondence breaks down. Indeed, because the observation and source points in the correlation function are arbitrary, the relaxed correlation function given by Eq.~(\ref{eq:87}) implies that when the thermal equilibrium is established, any two parts of the gas are entangled, otherwise letting the observation point be in one part and the source point be in the other the correlation function would depend on the detailed constructions of wavefunction instead of having a universal expression.}

{So} we may consider that the pure-state dynamics corresponds to the evolution from a ``semiclassical state'' whose entanglement level is low to a ``quantum state'' whose entanglement level is high.

{\subsection{Implications for a conjecture of Garrison and Grover}
\label{sec:generalized_garrison_grover_problem}}

{Our results for $t\gg t_E$ provide an analytical support to a conjecture on thermalization, imposed and studied numerically by Garrison and Grover in Ref.~\cite{Grover18}, and strengthen the statement of that conjecture. First of all, we observe that Eq.~(\ref{eq:87}) shows that all operators: $\hat{\mathfrak{O}}\equiv a^\dagger_{{\bf r}'_1}\cdots a^\dagger_{{\bf r}'_j}a_{{\bf r}_1}\cdots a_{{\bf r}_j}$ satisfy the following general formula,
\begin{equation}\label{eq:190}
  \langle F(t\gg t_E)|\hat{\mathfrak{O}}|F(t\gg t_E)\rangle=\frac{{\rm Tr}\left(\hat{\mathfrak{O}} e^{-\beta(\hat{\mathfrak{H}}-\mu \hat{\mathfrak{N}})}\right)}{{\rm Tr}\left(e^{-\beta(\hat{\mathfrak{H}}-\mu \hat{\mathfrak{N}})}\right)}.
\end{equation}
Here the trace is on the entire cavity space denoted as $\mathfrak{C}$, the many-particle Hamiltonian $\hat{\mathfrak{H}}\equiv -\frac{\hbar^2}{2m}\int_{\mathfrak{C}} d{\bf r}a^\dagger_{{\bf r}}\partial_{\bf r}^2a_{{\bf r}}$ with $\mathfrak{H}_{{\bf r'}{\bf r}}$ being its matrix element, and the particle number operator $\hat{\mathfrak{N}}\equiv \int_{\mathfrak{C}} d{\bf r}a^\dagger_{{\bf r}}a_{{\bf r}}$. The right-hand side of Eq.~(\ref{eq:190}) may be {\it formally} interpreted as the average of $\hat{\mathfrak{O}}$ with respect to the grand canonical ensemble. Letting $\hat{\mathfrak{O}}$ be $\hat{\mathfrak{H}}$ and $\hat{\mathfrak{N}}$, the left-hand side is $E$ and $N$, correspondingly. In this way we can determine the thermal parameters $\beta$ and $\mu$ as functions of $E,N$ as well as $V$.}

{Then, we project ($\hat{\mathfrak{H}}-\mu \hat{\mathfrak{N}}$) onto a subsystem A deep inside the cavity but sufficiently large. The ensuing operator
\begin{eqnarray}\label{eq:191}
  \hat{\mathfrak{H}}_A-\mu\hat{\mathfrak{N}}_A&\equiv&\int_{A} d{\bf r}a^\dagger_{{\bf r}}\left(-\frac{\hbar^2\partial_{\bf r}^2}{2m}-\mu\right)a_{{\bf r}}\nonumber\\
  &=&\int \frac{d{\bf p}}{(2\pi\hbar)^2}\left(\frac{\hbar^2{\bf p}^2}{2m}-\mu\right)a^\dagger_{{\bf p}}a_{{\bf p}},
\end{eqnarray}
where in the second line we have taken the advantage of large subsystem size to pass to the continuous Fourier representation. On the other hand, according to Eq.~(\ref{eq:155}) $\hat{\rho}_A(t\gg t_E)$ is Gaussian, and Eqs.~(\ref{eq:72}) and (\ref{eq:191}) further enforces its explicit expression to be
\begin{equation}\label{eq:192}
  \hat{\rho}_A(t\gg t_E)=\frac{e^{-\beta(\hat{\mathfrak{H}}_A-\mu \hat{\mathfrak{N}}_A)}}{{\rm Tr}_A\left(e^{-\beta(\hat{\mathfrak{H}}_A-\mu \hat{\mathfrak{N}}_A)}\right)}.
\end{equation}
}

{Now we note that $F(t\gg t_E)$ is a superposition of many eigenstates (which are Fock states). Let us consider a special case, where $F(t\gg t_E)$ has only single eigenstate component, and set $\mu=0$ (namely, the canonical ensemble) in Eqs.~(\ref{eq:190}) and (\ref{eq:192}). The validity of such simplified Eqs.~(\ref{eq:190}) and (\ref{eq:192}) for general local and nonlocal operators $\hat{\mathfrak{O}}$ in non-integrable systems is what Garrison and Grover conjectured \cite{Grover18}. Thus our findings provide a concrete example to support their conjecture analytically, and suggest a strengthened conjecture, for which the eigenstate in the original statement, namely, Eq.~(2b) in Ref.~\cite{Grover18}, is replaced by a pure state evolving for sufficiently long time, and the canonical ensemble by the grand canonical ensemble.}

Closing this section, it is worth mentioning that, strictly speaking, these results hold only for $t$ much smaller than the quantum recurrence time. However, the latter is extremely large (see Appendix \ref{sec:quantum_recurrence_time}) and therefore we ignore the quantum recurrence throughout.

\begin{figure}[h]
\includegraphics[width=8.6cm]{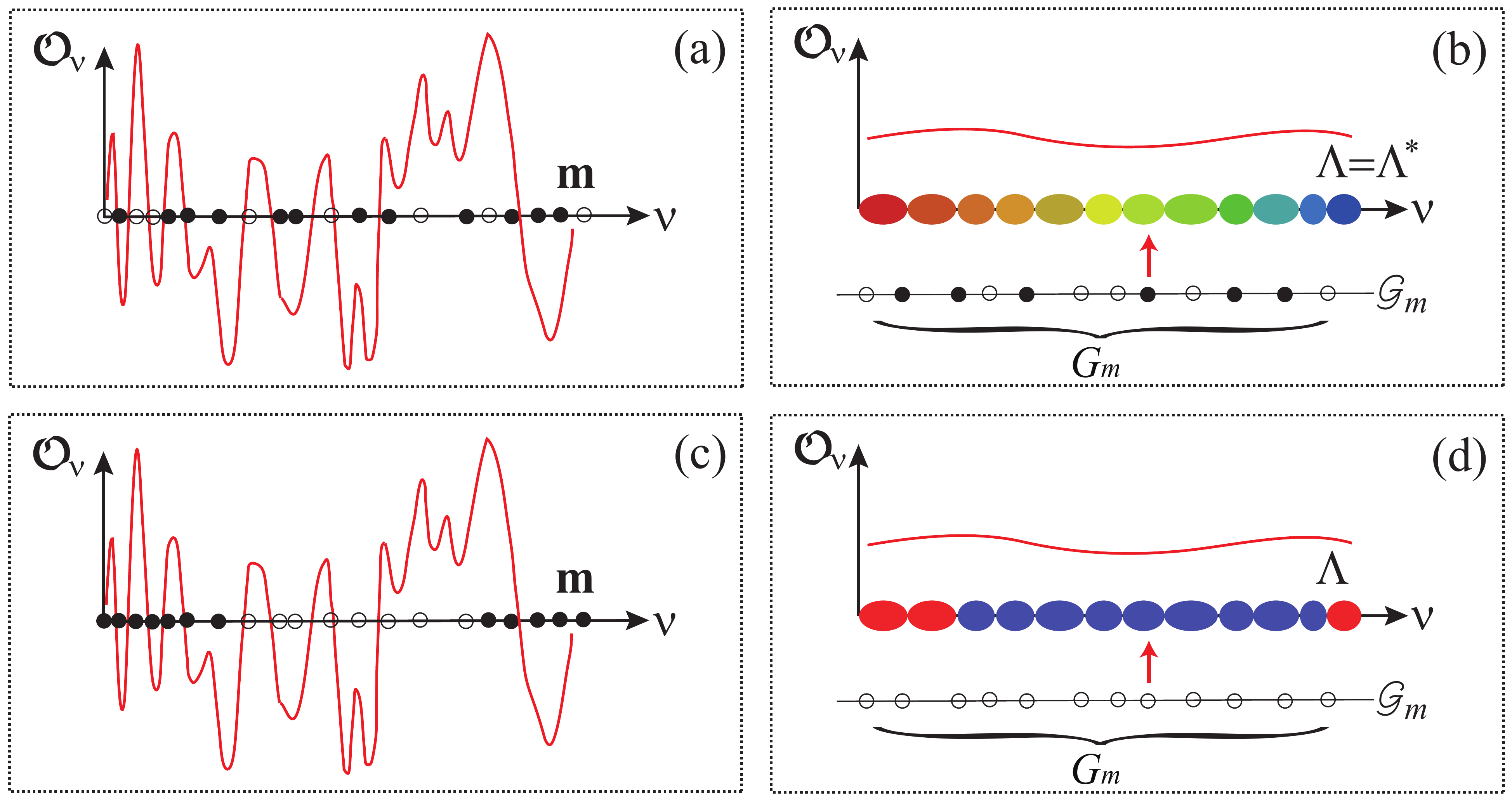}
\caption{(a,c) The occupation number pattern of an individual Fock state has a fine structure, ${\bf m}$, which is resolved by an observable violating the condition Eq.~(\ref{eq:115}). (b,d) Because a generic observable satisfies that condition, a less fine structure, $\Lambda({\bf m})$, is resolved instead. (b) For typical ${\bf m}$, $\Lambda({\bf m})=\Lambda^*$ and is thermal (schematically shown by gradual change in the cluster color representing the value of $N^*_m/G_m$ of the subspace ${\cal G}_m$). (d) For atypical ${\bf m}$, $\Lambda({\bf m})$ is athermal (schematically shown by nongradual change in the cluster color). Solid (open) circles denote
(un)occupied eigenstates $\nu$.}
\label{fig:2}
\end{figure}

\section{Notions and settings}
\label{sec:entanglement_entropy_typical_states}

{The Fock space is a basic tool for the studies of quantum many-particles systems. Yet, not until recently, has it been found to carry a hidden thermal structure, the eigenstate typicality \cite{Lai15}, for a simple many-particle integrable system, namely, free fermions on a torus. Originally, it states that for a generic local Hamiltonian an overwhelming majority of (thereby typical) highly excited many-body eigenstates (namely, those whose excitation energy scales with system size, or with a finite excitation energy density) {\it a.k.a.} Fock states, the RDM of a sufficiently small subsystem is thermal, with a temperature corresponding to that of the excitation energy density, and other parameters (like chemical potential) determined by the density of all conserved quantities in general (if any).
The eigenstate typicality is closely related to the eigenstate thermalization \cite{Deutsch91,Srednicki94,Rigol08}, but there are also some important differences. In fact, it was called the subsystem eigenstate thermalization \cite{Liu18} or the weak version of eigenstate thermalization \cite{Mueller18} later on. The notion of eigenstate typicality has been extended to more general translation-invariant and non-translational-invariant systems, even in the presence of direct interaction, and other quantities, notably, nonlocal observables \cite{Tian18,Mueller18}. In particular, the extension to nonlocal observables opens a door to investigate thermalization of the whole system instead of a subsystem from typicality perspectives.}

{In this section, we first briefly review the kinematic notion of eigenstate typicality of free fermion systems. It} allows us to provide a mathematical description of the initial state and a proposal of its experimental realization. With these preparations, we formulate three closely related problems to be studied in this work, that address different {dynamic aspects of} entanglement.

{\subsection{Eigenstate typicality}}
\label{sec:eigenstate_typicality}

{Here we restrict ourselves to the ideal Fermi gas confined in a chaotic cavity.} For this system, because of chaoticity there is only one good quantum number associated to the single-particle quantum motion, namely, the single-particle eigenenergy $\varepsilon_\nu$, with $\nu$ labelling the corresponding single-particle eigenstate. Then, each Fock state, ${\bf m}$, is represented by a specific occupation number pattern $\{n_\nu\}$, exemplified by Fig.~\ref{fig:2}(a) and (c), (thus we shall not distinguish $\{n_\nu\}$ and ${\bf m}$ hereafter.) where $n_\nu=0,1$ is the occupation number at the eigenstate $\nu$.

At first glance, it {seems} impossible to associate an individual Fock state ${\bf m}$ or occupation number pattern $\{n_\nu\}$ to any thermodynamic notions, since the pattern is seemingly quite arbitrary. Yet, as shown analytically and confirmed by numerical experiments in Refs.~\cite{Lai15,Tian18}, this is not true. A key observation is that there are two structures associated to each pattern: one is the fine structure, the occupation number pattern ${\bf m}$ [Fig.~\ref{fig:2}(a) and (c)], and the other, denoted as $\Lambda({\bf m})$, is less fine and can be resolved only through observables properly chosen but forming a broad class, thus dubbed ``the observable-resolved structure'' [Fig.~\ref{fig:2}(b) and (d)]. For illustrations let us take a general one-body observable $\hat{\mathfrak{O}}$. Its quantum expectation value at the Fock state ${\bf m}$ has the general form
\begin{equation}\label{eq:116}
  \langle {\bf m}|\hat{\mathfrak{O}}|{\bf m}\rangle=\sum_\nu n_\nu \mathfrak{O}_\nu,
\end{equation}
where $\mathfrak{O}_\nu$ depends on $\nu$. Provided $\mathfrak{O}_\nu$ varies rapidly with $\nu$ [Fig.~\ref{fig:2}(a) and (c)], a fine tuning in the pattern $\{n_\nu\}$ alters significantly the right-hand side of Eq.~(\ref{eq:116}). Thus the expectation value $\langle {\bf m}|\hat{\mathfrak{O}}|{\bf m}\rangle$ can detect the fine tuning in $\{n_\nu\}$. However, it turns out that for chaotic cavity generic $\hat{\mathfrak{O}}$ behaves in the opposite way \cite{Tian18}, i.e., $\mathfrak{O}_\nu$ varies smoothly with $\nu$ so that
\begin{equation}\label{eq:115}
  \mathfrak{O}_\nu\approx \mathfrak{O}_{\nu'},\quad for\, nearest\, \nu,\nu'.
\end{equation}
Because of this the space of $\nu$ can be ``naturally'' decomposed into many subspaces, denoted as ${\cal G}_m$ and represented by the clusters in Fig.~\ref{fig:2}(b) and (d): in each subspace (labelled by $m$), both $\mathfrak{O}_\nu$ and $\varepsilon_\nu$ are approximately constant, denoted as $\mathfrak{O}_m$ and $\varepsilon_m$, respectively, and there is a large number of single-particle eigenstates. With this decomposition Eq.~(\ref{eq:116}) reduces to
\begin{equation}\label{eq:118}
  \langle {\bf m}|\hat{\mathfrak{O}}|{\bf m}\rangle=\sum_m N_m \mathfrak{O}_m,\quad N_m=\sum_{\nu\in m} n_\nu.
\end{equation}
From this expression we see that a fine tuning in the pattern $\{n_\nu\}$ does not lead to essential changes in the quantum expectation value. This implies that as long as the condition Eq.~(\ref{eq:115}) is satisfied, the expectation value $\langle {\bf m}|\hat{\mathfrak{O}}|{\bf m}\rangle$ cannot resolve the fine structure $\{n_\nu\}$, but a less fine structure $\{N_m\}\equiv \Lambda({\bf m})$ represented by the set of clusters in Fig.~\ref{fig:2}(b) and (d). $\{N_m\}$ is constrained by
{\begin{eqnarray}\label{eq:117}
    \sum_m N_m&=&N=\sum_\nu n_\nu,\nonumber\\
    \sum_m N_m \varepsilon_m&=&E\approx\sum_\nu n_\nu\varepsilon_\nu\equiv E_{\bf m},
\end{eqnarray}
where $E_{\bf m}$ is the exact many-body eigenenergy corresponding to the eigenstate ${\bf m}$, and $E$ is its approximation.} Note that although there are no degenerate many-body eigenstates, i.e., $\sum_{\nu}n_\nu\varepsilon_\nu\neq \sum_{\nu}n'_\nu\varepsilon_\nu$ for $\{n_\nu\}\neq \{n'_\nu\}$, different many-body eigenstates can have the same value of $\sum_mN_m \varepsilon_m$. It should be emphasized that $\Lambda({\bf m})$ does not depend on the explicit form of $\hat{\mathfrak{O}}$, except that it has to satisfy the condition Eq.~(\ref{eq:115}).

{We remark that the above decomposition of the single-particle eigenstate ($\nu$) space into subspaces resembles von Neumann's grouping of eigenstates in the presence of ``macroscopic observations'' \cite{note2}. The essence of such grouping is that in each group of states ``{\it every macroscopic operator possesses the same eigenvalue, for otherwise carrying out all macroscopically possible observations would allow us to distinguish completely between all of the $\omega_1,\omega_2,\cdots$ (i.e., an absolutely precise determination of the state, which in general is not the case)}'' ($\omega_1,\omega_2,\cdots$ stand for system's eigenstates in the original paper.) \cite{note2}. Von Neumann's state, ``macroscopic operator'', and group of states may respectively regarded as the single-particle eigenstate, the $\hat{\mathfrak{O}}$ satisfying the criterion Eq.~(\ref{eq:115}), and the subspace ${\cal G}_m$ in the present work. The only difference is that in von Neumann's work, the eigenstates are those of the Hamiltonian describing the whole system, which thus are ${\bf m}$ in the present context. However, because of Eq.~(\ref{eq:116}) this difference is inessential here. It should be emphasized that the decomposition and the criterion Eq.~(\ref{eq:115}) for the operator $\hat{\mathfrak{O}}$ to resolve $\Lambda({\bf m})$ are regardless of chaoticity. Of course, for integrable systems the index $\nu$ should be replaced by the set of conserved quantities or good quantum numbers; see Appendix \ref{sec:observable_resolved_Structure_fermionic_oscillator} for example, where results about $\Lambda({\bf m})$ obtained by numerical experiments and their connections to the limit shape in number theory are reviewed. In fact, the original decomposition, criterion and $\hat{\mathfrak{O}}$ introduced in Ref.~\cite{Lai15} have been found to play important roles in the approach to steady states in a driven integrable system \cite{Sen17}, which is very different from systems studied in either Ref.~\cite{Lai15} or the present work. This decomposition, or the observable resolved structure $\Lambda({\bf m})$, is essentially a coarse graining in the good quantum number space, but at the level of an individual Fock state or many-body eigenstate.}

It is easy to see that the number of ${\bf m}$ carrying the same $\Lambda$ is $\prod_m\frac{G_m!}{N_m!(G_m-N_m)!}\equiv W[\Lambda]$, where $G_m$ is the number of single-particle eigenstates in the subspace ${\cal G}_m$. This number, when viewed as a functional of $\Lambda$, exhibits a sharp peak at some $\Lambda^*=\{N_m^*\}$. That is, an overwhelming number of Fock states carry the same observable-resolved structure $\Lambda^*$. To find $\Lambda^*$ explicitly we define $S=\ln W[\Lambda]$. By definition of $\Lambda^*$ we then have
\begin{equation}\label{eq:119}
  \frac{\partial S}{\partial N_m}\Big|_{\Lambda=\Lambda^*}=\alpha+\beta \varepsilon_m,
\end{equation}
where $\alpha,\beta$ are the Lagrange multipliers introduced by the two constraints in Eq.~(\ref{eq:117}), and
\begin{equation}\label{eq:20}
  \frac{\partial S}{\partial E}\Big|_{\Lambda=\Lambda^*}=\beta,\quad \frac{\partial S}{\partial N}\Big|_{\Lambda=\Lambda^*}=\alpha.
\end{equation}
With the substitution of $W[\Lambda]$ into Eq.~(\ref{eq:119}), we find that
\begin{equation}\label{eq:120}
  N_m^*/G_m=\frac{1}{e^{\beta\varepsilon_m-\alpha}+1}.
\end{equation}
Equations (\ref{eq:20}) and (\ref{eq:120}) give the thermodynamic relation and the Fermi-Dirac distribution $n_{FD}$, with $\beta=1/T$, $\alpha=-\mu/T$ and $S=\ln W[\Lambda^*]$ being the thermal entropy \cite{note6}. It is crucial that the thermodynamic relations and the Fermi-Dirac distribution can be probed only if observables satisfy the condition Eq.~(\ref{eq:115}), whereas standard statistical physics makes no reference to observables. This is a very reflection of the fundamental differences between pure-state statistical physics and standard statistical physics.

Let the Fock space constrained by Eq.~(\ref{eq:117}) be equipped with a uniform probability measure, and ${\bf m}$ be drawn randomly from this
measure. Equation (\ref{eq:118}) and the analysis above suggest that $\langle {\bf m}|\hat{\mathfrak{O}}|{\bf m}\rangle$ has a {\it typical} value with respect to this measure, because an overwhelming number of ${\bf m}$ satisfy $\Lambda({\bf m})=\Lambda^*$. This {is the mathematical basis of} eigenstate (of the ideal quantum gas) typicality. A Fock state ${\bf m}$ is said to bear this typicality and thus be typical, if it satisfies $\Lambda({\bf m})=\Lambda^*$ [Fig.~\ref{fig:2}(b)], otherwise is said to be {\it atypical} [Fig.~\ref{fig:2}(d)]. It should be emphasized that this typicality merely refers to the individual Fock state, namely, the many-body eigenstate of the ideal Fermi gas, and thus is of kinematic nature. Nevertheless, the eigenstate typicality is expected to have fundamental dynamical consequences, which is a main topic of this work. Let us also mention that the eigenstate typicality is completely different from the canonical typicality \cite{Gemmer04,Popescu06,Lebowitz06}, which has nothing to do with the particle indistinguishability. {Finally, we note that very recently there have been increasing interests in searching atypical states in many-particle systems and investigating their consequences on quantum thermalization. However, most attentions have been paid to the effects of quantum scar (e.g., Refs.~\cite{Vafek17,Turner18,Turner18a,Moudgalya18,LinMotrunich18,Ok19}). The existence of atypical Fock states and their superposition suggests that there are diverse mechanisms giving rise to atypical states in chaotic many-particle systems.}

\subsection{Description and experimental preparation of initial states}
\label{sec:classification_initial_states}

We now introduce a space of pure states using a subset of Fock states as the bases, defined as
\begin{eqnarray}\label{eq:114}
  \mathscr{H}_S\equiv \{\tilde F|\tilde F\,\, spanned\,\, by\,\, {\bf m}\in {\mathscr{F}_S}\},
\end{eqnarray}
where ${\mathscr{F}_S}$ is a microcanonical energy shell with center energy $E$ and the particle number for each ${\bf m}$ is $N$. This shell is narrow but includes many Fock states ${\bf m}$. Let $|\tilde F\rangle=\sum_{{\bf m}\in {\mathscr{F}_S}}\tilde C_{\bf m}|{\bf m}\rangle$, where the coefficients $\tilde C_{\bf m}$ are complex numbers satisfying $\sum_{{\bf m}\in {\mathscr{F}_S}}|\tilde C_{\bf m}|^2=1$. With the eigenstate typicality introduced in Sec.~\ref{sec:eigenstate_typicality}, we can divide $\mathscr{H}_S$ into two disconnected sets $\mathscr{H}_{S1}$ and $\mathscr{H}_{S2}$. In $\mathscr{H}_{S1}$, the majority of weight $|\tilde C_{\bf m}|^2$ goes to typical ${\bf m}$; whereas in $\mathscr{H}_{S2}$, the weight of atypical ${\bf m}$ is significant. Therefore,
\begin{eqnarray}\label{eq:131}
  \mathscr{H}_S=\mathscr{H}_{S1} \cup \mathscr{H}_{S2},\quad \mathscr{H}_{S1} \cap \mathscr{H}_{S2}=\emptyset.
\end{eqnarray}
Within each subset, $\tilde F$ can have very different spatial structure. Notably, it can be the superposition of $N$ localized wavepackets (of different widths and localization centers in general), and for most of wavepacket pairs the constituting wavepackets do not overlap. A representative of the amplitude profile of such $\tilde F$ is given in Fig.~\ref{fig:1}. For the convenience below, we define this special spatial structure of many-body wavefunction as the $*$-structure. Then the initial state $F(0)$ considered in this work belongs to either the set
\begin{eqnarray}\label{eq:132}
  \mathscr{P}\equiv \{F(0)|F(0)\in \mathscr{H}_{S1}\,and\,has \, *-structure.\}
\end{eqnarray}
or
\begin{eqnarray}\label{eq:159}
  \mathscr{P}'\equiv \{F(0)|F(0)\in \mathscr{H}_{S2}\,and\,has \, *-structure.\}.
\end{eqnarray}
Due to Eq.~(\ref{eq:131}) these two sets are disconnected,
\begin{eqnarray}\label{eq:167}
  \mathscr{P} \cap \mathscr{P}'=\emptyset.
\end{eqnarray}
Provided that we equip $\mathscr{P}\cup\mathscr{P}'$ with a uniform measure, then most $F(0)$ belongs to $\mathscr{P}$, because most Fock states in $\mathscr{H}_S$ are typical. The $*$-structure makes $F(0)$ possess certain semiclassical features discussed in the last section. Moreover, it is obvious that $F(0)$ is non-Gaussian. We also note that due to the Pauli principle for large $N$ the center energy of ${\mathscr{F}_S}$ has to be sufficiently large.

The preparation of the initial state $F(0)\in \mathscr{P}$ or $\mathscr{P}'$ may be well within the experimental reach of state-of-the-art ultracold-atom techniques, and a protocol is as follows. At the first step, the experimental technique \cite{Hadzibabic12} based on holographic, phase only method allows one to produce $2$D optical traps of arbitrary shape. This is particularly useful for creating a stadium-like cavity; the motion of a particle trapped by this cavity is chaotic. At the second step, we load a number of cesium atoms in the chaotic cavity, with desired total kinetic energy.
The atoms are subjected to a short-ranged interaction, and the interaction strength is tuned by the magnetic field that controls the distance from the Feshbach resonance \cite{Chin10}. Let this interacting Fermi gas equilibrate. Since this interacting system is chaotic, it thermalizes following the scenario of the eigenstate thermalization \cite{Deutsch91,Srednicki94,Rigol08}.
Moreover, it can be shown (see Appendix~\ref{sec:discussions_experiment_initial_state}) that the particle number distribution over the single-particle eigenstates of the cavity, namely, the eigenenergy spectrum $\{\varepsilon_\nu\}$, is Fermi-Dirac. At the third step, we slowly turn off the interaction by tuning the magnetic field and a state results, which is
a superposition of some typical ${\bf m}$ with close total energy belonging to the energy shell ${\mathscr{F}_S}$. As such, we achieve a state in $\mathscr{H}_{S1}$. At the last step, we use the light scattering method to measure atom's positions with a spatial resolution, which is controlled by the wavelength and intensity of the light. The resolution is required to be much larger than the de Broglie wavelength of the atom. 
This measurement projects each atom onto a localized
wavepacket, but does {\em not} affect the energy distribution of the gas because the resolution is much larger than the de Broglie wavelength. So the $*$-structure is achieved and we realize a $F(0)\in \mathscr{P}$. Provided that at the second step we do not let the interacting Fermi gas equilibrate, we realize a $F(0)\in \mathscr{P}'$ instead (see Appendix~\ref{sec:discussions_experiment_initial_state} for further discussions).

\subsection{Formulation of the problem set}
\label{sec:formulation}

For an initial state $F(0)$ with expansion coefficients $C_{\bf m}$, it evolves unitarily, $F(0)\rightarrow F(t)$, following
\begin{equation}\label{eq:S15}
    |F(t)\rangle=\sum_{{\bf m}\in {\mathscr{F}_S}} e^{-iE_{{\bf m}} t/\hbar}C_{{\bf m}} |{{\bf m}}\rangle.
\end{equation}
{From this an elementary but important fact follows.} Namely, the unitary evolution neither annihilates any Fock state component of $F(0)$, nor creates any new Fock state component. However, this provides no knowledge about the spatial structure of $F(t)$. The evolution of this structure has fundamental consequences on the entanglement, as we will see below.

Owing to its $*$-structure, it is obvious that $F(0)$ is far from equilibrium and, as at such state a generic particle is away from and thereby does not overlap with most of the others, the initial entanglement {induced} by the indistinguishability must be low. In this work we will study how such entanglement evolves with $F(t)$. The first problem has the particle correlation as the probe of that entanglement and is formulated as follows:\\

{\it Problem 1. Picking up a $F(0)$ from $\mathscr{P}$ or $\mathscr{P}'$ and letting it evolve unitarily, how do the particle correlation functions defined by Eq.~(\ref{eq:76}) behave in the course of time?}\\

Furthermore, we wish to use the RDM and the corresponding EE as the probes of that entanglement. To this end {let the $2$D space} be divided into a number of small plaques, each of which has a size $a$. Thus a discrete lattice results, and in this lattice space the space occupied by the cavity is $\mathbb{Z}^2\cap\mathfrak{C}$. Then we divide the cavity into two parts, a subsystem A, which is far away from the cavity boundary, and its complement. The volume, namely, the total number of the lattice points of A is $N_A$. (We are not able to formulate the two problems below in the continuous space. But we shall show that the continuum limit of their solutions is well-defined.)

Following Ref.~\cite{Korepin04} we can expand the RDM of A, $\hat\rho_A(t)$, at (the lattice version of) the evolving state $F(t)$ in terms of bases of the form: $|\Psi_I\rangle\langle\Psi_J|$, where $\Psi_I$ is a state describing an occupation number pattern of fermions at some single-particle state, e.g., the eigenstate of the position operator $\hat {\bf r}_i$ at a site ${\bf r}_i\in A$. So Eq.~(\ref{eq:187}) is cast into
\begin{equation}\label{eq:24}
    \hat\rho_A(t)=\sum_{I,J}{\rm Tr}\left(\hat\rho(t)|\Psi_J\rangle\langle\Psi_I|\right)|\Psi_I\rangle\langle\Psi_J|.
\end{equation}
By definition of $\Psi_I$ we can rewrite $|\Psi_I\rangle\langle\Psi_J|$ as
\begin{equation}\label{eq:27}
    |\Psi_I\rangle\langle\Psi_J|=\prod_i|n^I_{i}\rangle\langle n^J_{i}|,
\end{equation}
where $n^I_{i}$ ($n^J_{i}$) is the occupation number at the aforementioned single-particle state $i$ at $\Psi_I$ ($\Psi_J$). It is easy to check that
\begin{eqnarray}
\label{eq:28}
  && a^\dagger_{i}=|1_{i}\rangle\langle 0_{i}|,\quad a_{i}=|0_{i}\rangle\langle 1_{i}|,\nonumber\\
  && a^\dagger_{i}a_{i}=|1_{i}\rangle\langle 1_{i}|,\quad a_{i}a^\dagger_{i}=|0_{i}\rangle\langle 0_{i}|,
\end{eqnarray}
where $a_{i}$ ($a^\dagger_{i}$) is the annihilation (creation) operator at single-particle state $i$ and $0_i$ ($1_i$) stands for that the state $i$ is not occupied (occupied by single particle). By these expressions we can trade the expansion in terms of $|\Psi_I\rangle\langle\Psi_J|$ in Eq.~(\ref{eq:24}) to an expansion in terms of $\prod_{i} O_{i}$, with $O_{i}$ taking an operator value from the set: $\{a^\dagger_{i},a_{i},a^\dagger_{i}a_{i},a_{i}a^\dagger_{i}\}$. This gives
\begin{eqnarray}\label{eq:29}
    \hat\rho_A(t)&=&\sum_{\{O_i\}}{\rm Tr}\left(\hat\rho(t)\prod_{i} O^\dagger_{i}\right)\prod_{i} O_{i}\nonumber\\
    &=&\sum_{\{O_i\}}\langle F(t)|\prod_{i} O^\dagger_{i}|F(t)\rangle\prod_{i} O_{i},
\end{eqnarray}
where the sum is over all allowed operator values of $O_{i}$. So the second problem is:\\

{\it Problem 2. Letting $F(0)$ be the same as that in Problem 1, how does $\hat\rho_A(t)$ given by Eq.~(\ref{eq:29}) behave in the course of time?}\\

Equations (\ref{eq:S15}) and (\ref{eq:29}) provide a framework for the study of the dynamics of all macroscopic observables defined on the subsystem A. In particular, it allows us to study the EE defined by Eq.~(\ref{eq:30}). Initially, because of the $*$-structure most particles do not overlap with the boundary of A (for generic geometry). As a result, the initial EE, $S_A(0)$, is low, for which a (sub)area law might be expected. So the third problem arises naturally:\\

{\it Problem 3. Letting $F(0)$ be the same as that in Problems 1 and 2, how does $S_A(t)$ defined by Eq.~(\ref{eq:30}) behave in the course of time?}\\

Let us make several remarks on the three problems above:

First, although the formulation of Problem 1 to some extent resembles von Neumann's ideology for a statistical description of isolated quantum systems \cite{von Neumann29}, which is also built upon the dynamics of the quantum expectation value of observables, none of his results can be used here for two reasons. (i) Because those results deal with long-time behaviors, they are mute on short-time dynamics and are of kinematic nature \cite{Lebowitz10}. (ii) For the present system, owing to the absence of a direct interaction, there exist many-body eigenstate quadruples: (${\bf m}_1,{\bf m}_2,{\bf m}'_1,{\bf m}'_2$), such that any two of them are different and they satisfy: $E_{{\bf m}_1}-E_{{\bf m}_2}=E_{{\bf m}'_1}-E_{{\bf m}'_2}$. This spoils a key condition \cite{note_von_Neumann} for establishing von Neumann's results. The formulation of Problem 1 also resembles the setup of the celebrated numerical experiment on quantum thermalization \cite{Rigol08}. However, there is a key difference, namely, the absence of a direct interaction in the present system. Thus one may expect the mechanism for thermalization in the present system, if it does happen, to have many conceptual differences from previous scenario \cite{Deutsch91,Srednicki94,Rigol08} for thermalization in isolated quantum systems.

Second, Problem 2 pushes the studies of the statistical behaviors of macroscopic observables in a subsystem forward to the studies of the more complete statistical object, namely, the reduced density of matrix. In this sense Problem 2 is in spirit parallel to Boltzmann's kinetic theory, which addresses the evolution of the statistical distribution. The fundamental difference is that the statistical distribution here concerns only the subsystem, while the entire system is described by a pure state, rather than a statistical distribution.

Third, for $F(0)\in \mathscr{P}$ some thermal properties are already hidden in $F(0)$, because the majority of the weight $|C_{\bf m}|^2$ goes to typical ${\bf m}$. As we will see, it is to make those properties visible that appropriate macroscopic observables $\hat{\mathfrak{O}}$ and its time evolution are required. In other words, the dynamics of appropriate $\hat{\mathfrak{O}}$ might convert those properties hidden in microoscopic $F(0)$ into genuine thermal equilibrium phenomena occurring at the macroscopic level. In fact, one may regard $\hat\rho_A(t)$ as a macroscopic observable as well, since according to Eq.~(\ref{eq:29}) it is {a linear combination of quantum expectation values: $\langle F(t)|\prod_{i} O^\dagger_{i}|F(t)\rangle$ (with operator-valued coefficients)}. So appropriate $\hat{\mathfrak{O}}$ and its dynamics are indispensable ingredients for the formulation of a statistical description of an isolated quantum system, consistent with von Neumann's ideology \cite{von Neumann29}. Because of
\begin{eqnarray}\label{eq:134}
    &&\langle F(t)|\hat{\mathfrak{O}}|F(t)\rangle\nonumber\\
    &=&\sum_{{\bf m},{\bf m}'\in {\mathscr{F}_S}} e^{i(E_{{\bf m}'}-E_{{\bf m}}) t/\hbar}C^*_{{\bf m}'}C_{{\bf m}} \langle{{\bf m}'}|\hat{\mathfrak{O}}|{{\bf m}}\rangle,
\end{eqnarray}
the dynamics of a macroscopic observable is closely related to {dephasing as the phase $(E_{{\bf m}'}-E_{{\bf m}}) t/\hbar$ (${\bf m}\neq {\bf m}'$) increases with $t$}. This raises the fundamental issue {of whether quantum thermalization or equalibration arises from dephasing in the present context, as what was proposed for the eigenstate thermalization of interacting systems \cite{Rigol08,Rigol16}. To address this issue it is important to understand whether and to what extent the thermalization or equilibration time depends on the initial state and the observable, because the characteristic time of dephasing is the Heisenberg time, namely, the time to resolve individual many-body eigenstate, which has no such dependence. In fact, the Heisenberg time provides an upper bound for the thermalization or equilibration time. Whether this is a sharp bound is currently under investigations for interacting many-particle systems (see, e.g., Ref.~\cite{Santos21} and the references therein). In addition, the time scale for a subsystem to thermalize or equilibrate is of fundamental importance in pure-state statistical physics \cite{Gogolin16}. The solutions of the three problems provide insights into these issues for many-particle systems without direct interaction.}

\section{Dynamics of correlation functions: $\boldsymbol{F(0) \in \mathscr{P}}$}
\label{sec:evolution_correlation_function}

In this and the next sections we consider the initial state $F(0)\in\mathscr{P}$. In this section we study the spatial correlation functions of {$j$ particles}. We focus on the spatial correlation with the observation and source points
far away from the cavity boundary.

\subsection{One-particle correlation}
\label{sec:one_particle}

\subsubsection{General formalism}
\label{sec:formalism_one_particle_correlation}

{According to Eq.~(\ref{eq:76}), the} one-particle correlation function between two spatial points ${\bf r},{\bf r}'\in A$ at the state $F(t)$ {is}
\begin{equation}\label{eq:15}
    {M_{{\bf r}{\bf r}'}^{(1)}(t)\equiv}M_{{\bf r}{\bf r}'}(t)\equiv\langle F(t)| a^\dagger_{{\bf r}'}a_{{\bf r}}|F(t)\rangle,
\end{equation}
{where the superscript: $(1)$ is omitted henceforth. We} also define $\hat M(t)\equiv \{M_{{\bf r}{\bf r}'}(t)\}$. In Appendix~\ref{sec:derivations_von_Neumann} we show that the latter follows the von Neumann equation,
\begin{equation}\label{eq:2}
    \partial_t\hat M(t)=-\frac{i}{\hbar}[H(\hat{\bf q},\hat{\bf p}),\hat M(t)],
\end{equation}
where the single-particle Hamiltonian $H(\hat{\bf q},\hat{\bf p})=\frac{\hat{\bf p}^2}{2m}+V(\hat{\bf q})$, with $V(\hat{\bf q})$ being the potential that effects a $2$D cavity to confine particles and $\hat{\bf q}$ ($\hat{\bf p}$) being the position (momentum) operator. Equation (\ref{eq:2}) is implemented by the initial condition
\begin{eqnarray}
\label{eq:121}
  M_{{\bf r}{\bf r}'}(0) = \sum_{\nu'\nu}C_{\nu'\nu} \psi_{\nu'}({\bf r})\psi_\nu^*({\bf r}').
\end{eqnarray}
Here $\psi_\nu({\bf r})$ is the wavefunction of the single-particle eigenstate $\nu$, and the coefficients $C_{\nu'\nu}$ depend on the initial state $F(0)$ [for its explicit form, see Eq.~(\ref{eq:S19})]. Passing to the Wigner representation,
\begin{equation}\label{eq:3}
    M_{{\bf r}{\bf r}'}(t)\equiv\int d{\bf p} e^{-\frac{i}{\hbar}({\bf r}-{\bf r}')\cdot {\bf p}}
\mathfrak{M}({\bf q},{\bf p};t)
\end{equation}
with ${\bf q}\equiv \frac{{\bf r}+{\bf r}'}{2}$, we can rewrite Eq.~(\ref{eq:2}) as
\begin{equation}\label{eq:14}
    \left(\partial_t-\{H({\bf q},{\bf p}),\,\cdot\,\}_{{\rm M}}\right)\mathfrak{M}({\bf q},{\bf p};t)=0.
\end{equation}
Here $\{H,\,\cdot\,\}_{{\rm M}}$ stands for the Moyal bracket \cite{Moyal49}
\begin{equation}\label{eq:137}
  \{H,\,\cdot\,\}_{{\rm M}}\equiv \frac{2}{\hbar}H\sin\left(\frac{\hbar}{2}(\overleftarrow{\partial}_{{\bf q}}\cdot\overrightarrow{\partial}_{{\bf p}}-\overleftarrow{\partial}_{{\bf p}}\cdot\overrightarrow{\partial}_{{\bf q}})\right)(\cdot),
\end{equation}
with the derivatives $\overleftarrow{\partial}_{{\bf q},{\bf p}}$ acting on $H({\bf q}, {\bf p})$ and $\overrightarrow{\partial}_{{\bf q},{\bf p}}$ on a function on the phase space. We assume $V$ and all phase-space functions involved to be real analytic (i.e., $C^\infty$) so that the Moyal bracket is well defined; this assumption is technical and inessential to physical results presented in this paper. Note that this quantum evolution is of single-particle nature, and the many-body properties of $F$ enter into the initial condition $\mathfrak{M}({\bf q},{\bf p};0)$. It is important that, unlike the previous investigations of the second law of thermodynamics \cite{Zurek03}, here no decoherence terms, which arise from, e.g., the interaction between a quantum chaotic system and a reservoir, are added to the Moyal bracket: the quantum dynamics is strictly unitary.

Upon being expanded in terms of $\hbar$, the Moyal bracket reads
\begin{eqnarray}
    \{H,\,\cdot\,\}_{{\rm M}}=\{H,\,\cdot\,\}_{\rm P}+\delta \hat H,
\label{eq:4}
\end{eqnarray}
where $\{\cdot,\,\cdot\}_{\rm P}$ stands for the Poisson bracket and
\begin{eqnarray}
    \delta \hat H=\sum_{n=1}^\infty \frac{(-\hbar^2/4)^n}{(2n+1)!} H\left(\overleftarrow{\partial}_{{\bf q}}\cdot\overrightarrow{\partial}_{{\bf p}}-\overleftarrow{\partial}_{{\bf p}}\cdot\overrightarrow{\partial}_{{\bf q}}\right)^{2n+1}.\quad
\label{eq:10}
\end{eqnarray}
In the classical limit $\hbar\rightarrow 0$, we keep only the leading expansion, namely, the Poisson bracket. Consequently, Eq.~(\ref{eq:14}) reduces to the Liouville equation.
With the help of Eqs.~(\ref{eq:4}) and (\ref{eq:10}) it can be readily shown that for any real analytic function $k(H)$,
\begin{eqnarray}
    \{H,\,k(H)\}_{{\rm M}}=0,
\label{eq:139}
\end{eqnarray}
irrespective of the detailed form of $H$, implying that the single-particle energy is conserved during the quantum evolution, as a consequence of the absence of a direct interaction. Therefore, the evolutions at different phase-space energy shell, no matter classical ($\hbar=0$) or quantum ($\hbar>0$), are independent. So we can decompose the $4$-dimensional phase space into infinite number of $3$-dimensional phase-space energy shells (to be distinguished from $\mathscr{F}_S$), each of which is labelled by the single-particle energy $\varepsilon$. Thus we have the following change in the coordinate systems
\begin{equation}\label{eq:140}
  ({\bf q},{\bf p})\rightarrow (\varepsilon,
  {\bf x}_\parallel),
\end{equation}
where
${\bf x}_\parallel$ is the coordinate of the phase point in the energy shell. Correspondingly, we rewrite $\mathfrak{M}({\bf q},{\bf p};t)$ in the new coordinate system as
\begin{equation}\label{eq:142}
\mathfrak{M}({\bf q},{\bf p};t)=\mathfrak{M}(\varepsilon,{\bf x}_\parallel;t)\equiv\mathfrak{M}_\varepsilon({\bf x}_\parallel;t),
\end{equation}
where $\varepsilon$ is put in the subscript as a bookkeeping of its invariance during the evolution.

To proceed we introduce two Green's functions, $G_\varepsilon({\bf x}_\parallel,{\bf x}'_\parallel;t)$ and $g_\varepsilon({\bf x}_\parallel,{\bf x}'_\parallel;t)$, for motion in the energy shell $\varepsilon$, defined as
\begin{eqnarray}\label{eq:S25}
\left({\begin{array}{c}
      \partial_t-\{H,\,\cdot\,\}_{{\rm M}} \\
      \partial_t-\{H,\,\cdot\,\}_{{\rm P}}
                            \end{array}
    }\right)\left(\begin{array}{c}
                                                               G_\varepsilon \\
                                                               g_\varepsilon
                                                             \end{array}\right)
=\delta({\bf x}_\parallel-{\bf x}'_\parallel)\delta(t).
\end{eqnarray}
The second equation is the Liouville equation, which can be rewritten as
\begin{eqnarray}\label{eq:141}
\left(\partial_t+\frac{d{\bf x}_\parallel}{dt}\cdot \partial_{{\bf x}_\parallel}\right)g_\varepsilon
=\delta({\bf x}_\parallel-{\bf x}'_\parallel)\delta(t),
\end{eqnarray}
where $\frac{d{\bf x}_\parallel}{dt}$ is the phase-space velocity. By further introducing the $\circ$-product: $(A_\varepsilon\circ B_\varepsilon)({\bf x}_\parallel,{\bf x}''_\parallel)\equiv\int d{\bf x}'_\parallel A_\varepsilon({\bf x}_\parallel,{\bf x}'_\parallel)B_\varepsilon({\bf x}'_\parallel,{\bf x}''_\parallel)$, which is essentially the convolution and can be readily shown to be associative, i.e., $(A_\varepsilon\circ B_\varepsilon)\circ C_\varepsilon=A_\varepsilon\circ (B_\varepsilon\circ C_\varepsilon)$, we obtain
\begin{eqnarray}\label{eq:S27}
    G_\varepsilon(t)=g_\varepsilon(t)+\int_0^t dt'g_\varepsilon(t-t')\circ\delta\hat H G_\varepsilon(t')
\end{eqnarray}
from Eq.~(\ref{eq:S25}). It carries the same structure as the Dyson equation, with $\delta \hat H$ playing the role of the interaction. We suppress the phase-space coordinates of the energy shell to make formulae compact. Iterating Eq.~(\ref{eq:S27}), we can formally expand $G_\varepsilon(t)$ in $\delta \hat H$,
\begin{eqnarray}\label{eq:S62}
    G_\varepsilon(t)&=&g_\varepsilon(t)+\int
    dt_1g_\varepsilon(t-t_1)\circ\delta \hat H g_\varepsilon(t_1)\nonumber\\
    &+&\int\!\!\!\!\int
    dt_1dt_2g_\varepsilon(t-t_1)\circ\delta \hat H g_\varepsilon(t_1-t_2)\circ\delta \hat H g_\varepsilon(t_2)\nonumber\\
    &+&\cdots,
\end{eqnarray}
where the first term corresponds to the classical Liouville evolution and the other terms are quantum. Equation (\ref{eq:S62}) gives an expansion of $\mathfrak{M}_\varepsilon(t)$ for $t>0$,
\begin{eqnarray}
    \mathfrak{M}_\varepsilon(t)=\sum_{K=0}^\infty \mathfrak{M}_{\varepsilon,K+1}(t),\quad\quad\quad\quad\quad\quad\quad\label{eq:S28}\\
    K=0: \mathfrak{M}_{\varepsilon,1}(t)=g_\varepsilon(t)\circ \mathfrak{M}_\varepsilon(0),\quad\quad\quad\quad\quad\nonumber\\
    K\in \mathbb{N}: \mathfrak{M}_{\varepsilon,K+1}(t)=\int_0^tdt_1\cdots dt_K g_\varepsilon(t-t_1)\quad\quad\quad\quad\nonumber\\
    \circ\delta \hat H g_\varepsilon(t_1-t_2)\circ\delta \hat H g_\varepsilon(t_2-t_3)\circ\cdots \circ\delta \hat H g_\varepsilon(t_K)\circ \mathfrak{M}_\varepsilon(0).\nonumber
\end{eqnarray}
Equations (\ref{eq:3}) and (\ref{eq:S28}) provide a general formalism for calculating the one-particle correlation function.

For the $2$D motion it is convenient to choose
\begin{equation}\label{eq:146}
  {\bf x}_\parallel\equiv({\bf q},\vartheta),
\end{equation}
where the angle $\vartheta$ denotes the direction of ${\bf p}$. With this choice Eq.~(\ref{eq:3}) reduces to
\begin{equation}\label{eq:143}
    M_{{\bf r}{\bf r}'}(t)=m\int d\varepsilon d\vartheta e^{-\frac{i}{\hbar}({\bf r}-{\bf r}')\cdot {\bf p}({\bf q},\vartheta)}
\mathfrak{M}_\varepsilon({\bf q},\vartheta;t),
\end{equation}
where ${\bf p}$ is a function of ${\bf q}$ and $\vartheta$.  For $N\gg 1$, thanks to the Pauli principle most particles have very large $\varepsilon$. Consequently, the contributions from small $\varepsilon$ to the energy integral are small and will be ignored hereafter.

\subsubsection{Quantum-classical correspondence breakdown and the Ehrenfest time}
\label{sec:Ehrenfest_time}

Equation (\ref{eq:S28}) expresses the quantum evolution $\mathfrak{M}_\varepsilon(t)$ in terms of the classical Liouville evolution, i.e., Green's function $g_\varepsilon$, and its ``interaction'' with the quantum operator $\delta \hat H$. Because of the dynamical instability associated to classical (single-particle) trajectories, a volume element expands exponentially in the unstable direction with a rate $\lambda_L(\varepsilon)$, namely, the Lyapunov exponent. Thanks to the Liouville theorem this volume element shrinks exponentially in the stable direction with the same rate. The shrinking process makes $\mathfrak{M}_\varepsilon$ display finer and finer structures during the Liouville evolution, and thus varies more and more rapidly along the stable direction, over a scale decaying exponentially $\sim e^{-\lambda_L(\varepsilon)t}$. We now show that this has an important consequence.

For the present system, Eq.~(\ref{eq:10}) reduces to
\begin{eqnarray}
    \delta \hat H=\sum_{n=1}^\infty \frac{(-\hbar^2/4)^n}{(2n+1)!} V\left(\overleftarrow{\partial}_{{\bf q}}\cdot\overrightarrow{\partial}_{{\bf p}}\right)^{2n+1}.
\label{eq:144}
\end{eqnarray}
The characteristic value of $|\partial_{\bf q}V/\partial_{\bf q}^{2n+1}V|^{\frac{1}{2n}}$ may be estimated as $L$. Thanks to the $*$-structure, the momentum scale over which $\mathfrak{M}_\varepsilon$ varies may be estimated as $\sqrt{2m\varepsilon}$ for $t=0$, while as $\sqrt{2m\varepsilon}e^{-\lambda_L(\varepsilon)(t)}$ for later $t$. Taking these into account, we have \cite{Dittrich98}
\begin{eqnarray}
\delta \hat H \mathfrak{M}_\varepsilon(t)=\sum_{n=1}^\infty \frac{(-\hbar^2/4)^n}{(2n+1)!} V\left(\overleftarrow{\partial}_{{\bf q}}\cdot\overrightarrow{\partial}_{{\bf p}}\right)^{2n+1}\mathfrak{M}_\varepsilon(t)\quad\nonumber\\
\sim\sum_{n=1}^\infty \frac{(-1/4)^n}{(2n+1)!}\left(\frac{\hbar e^{\lambda_L(\varepsilon) t}}{A(\varepsilon)}\right)^{2n}{\cal O}(\partial_{{\bf q}}V\cdot\partial_{{\bf p}}\mathfrak{M}_\varepsilon(t)),\quad\quad
\label{eq:12}
\end{eqnarray}
where $A(\varepsilon)\equiv L\sqrt{2m\varepsilon}=\hbar L/\lambda_\varepsilon$. In principle, $A(\varepsilon)$ depends on $n$. However, it turns out that this dependence does not change the physical results below. (In fact, the analysis here can be generalized to treat this dependence.). So we shall not discuss it further and assume the $n$-independence of $A(\varepsilon)$ throughout this work. Most importantly, [for $\varepsilon$ dominating the integral in Eq.~(\ref{eq:143})] $A(\varepsilon)\gg\hbar$ because $L\gg\lambda_\varepsilon$.

In Eq.~(\ref{eq:12}), the coefficient $\frac{\hbar e^{\lambda_L(\varepsilon) t}}{A(\varepsilon)}$ is small for sufficiently short time. However, it grows exponentially in time. Provided that
\begin{equation}\label{eq:13}
    \hbar e^{\lambda_L(\varepsilon) t}/A(\varepsilon)={\cal O}(1)\Rightarrow t=\frac{1}{\lambda_L(\varepsilon)}\ln\frac{A(\varepsilon)}{\hbar}\equiv t_\varepsilon,
\end{equation}
Eq.~(\ref{eq:12}) becomes comparable to $\partial_{{\bf q}}V\cdot\partial_{{\bf p}}\mathfrak{M}_\varepsilon$. This signals that the quantum terms in Eq.~(\ref{eq:S28}) start to dominate over the classical term, i.e., the quantum-classical correspondence breaks down. Because $F(0)$ is superposed mainly by typical Fock states, the distribution of most particles over different phase-space energy shells leads to a weak variation of $\lambda_L(\varepsilon)\sim\sqrt{\varepsilon/m}/L$ with $\varepsilon$. Thus we may let $\varepsilon$ at $\lambda_L(\varepsilon)$ be the average single-particle energy,
and ignore the variation of $\lambda_L$ with $\varepsilon$ hereafter. The logarithm in Eq.~(\ref{eq:13}) has even weaker dependence on $\varepsilon$, and thus its variation with $\varepsilon$ is ignored also. As such, $t_\varepsilon$ reduces to the Ehrenfest time $t_E$ given by Eq.~(\ref{eq:19}).

Owing to the equivalence between the Liouville and the Hamiltonian equation, the above picture of the quantum-classical correspondence breakdown is equivalent to the canonical picture based on classical trajectories. Indeed, consider a classical trajectory in a chaotic cavity. Because of the Heisenberg uncertainty the direction of the initial momentum has a small angular resolution $\delta\vartheta_0$, which may be estimated as $\hbar/A$ on general grounds. When the trajectory is reflected by the cavity wall this resolution is magnified due to the dynamical instability, giving $\delta\vartheta_t=\delta\vartheta_0 e^{\lambda_L t}$. When $\delta\vartheta_t$ is comparable to $2\pi$, the direction of the momentum cannot be resolved and the concept of a classical trajectory ceases to work. This time is given by $\lambda_L^{-1}\ln (2\pi/\delta\vartheta_0)$, which is just $t_E$.

\subsubsection{The correlation function for short time}
\label{sec:short_time_correlation_function}

Using Eqs.~(\ref{eq:19}) and (\ref{eq:12}), we find that
\begin{eqnarray}
\delta \hat H \mathfrak{M}_\varepsilon(t)=\sum_{n=1}^\infty e^{2n\lambda_L(t-t_E)} {\cal O}\left(\{V,\mathfrak{M}_\varepsilon(t)\}_{\rm P}\right).\,\,
\label{eq:S39}
\end{eqnarray}
The exponents on the right-hand side render $\delta \hat H\mathfrak{M}_\varepsilon(t)$ negligibly small for $t\ll t_E$. In this regime we can ignore all quantum terms in Eq.~(\ref{eq:S28}), obtaining $\mathfrak{M}_\varepsilon(t)=\mathfrak{M}_{\varepsilon,1}(t)$. Substituting it into Eq.~(\ref{eq:143}) gives for $t\ll t_E$,
\begin{eqnarray}
    M_{{\bf r}{\bf r}'}(t)=m\int d\varepsilon d\vartheta e^{-\frac{i}{\hbar}({\bf r}-{\bf r}')\cdot {\bf p}({\bf q},\vartheta)} \mathfrak{M}_{\varepsilon,1}({\bf q},\vartheta;t).
    \label{eq:9}
\end{eqnarray}
From the Liouville equation it is equivalent to
\begin{eqnarray}
    M_{{\bf r}{\bf r}'}(t)=m\int d\varepsilon d\vartheta e^{-\frac{i}{\hbar}({\bf r}-{\bf r}')\cdot {\bf p}({\bf q},\vartheta)} \mathfrak{M}_{\varepsilon}({\bf q}_{-t},\vartheta_{-t};0),\quad\,\,
    \label{eq:11}
\end{eqnarray}
where $({\bf q}_{-t},\vartheta_{-t})$ stands for the energy-shell coordinates, such that a trajectory initiating from them {evolves to} $({\bf q},\vartheta)$ at time $t$.

\begin{figure}[b]
\includegraphics[width=8.6cm]{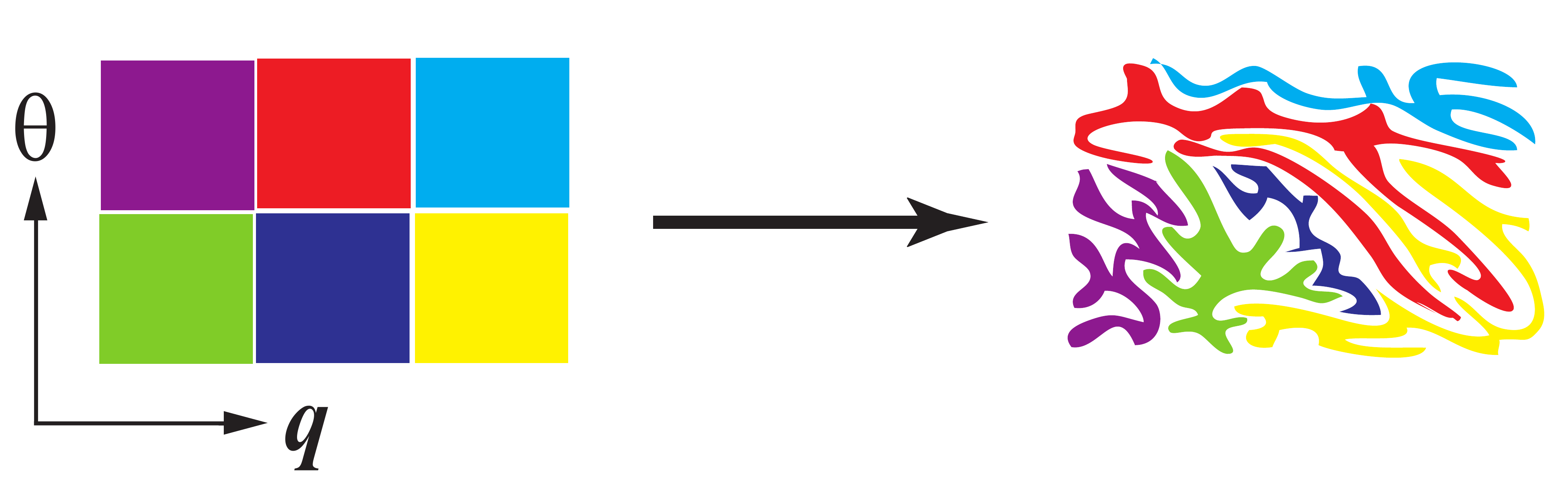}
\caption{{We partition the energy shell into small grids (left), each of which is deformed in the course of time (right).}}
\label{fig:3}
\end{figure}

To calculate Eq.~(\ref{eq:11}) we partition {the energy shell $\varepsilon$ into small grids (Fig.~\ref{fig:3}, left)}, within each of which $\mathfrak{M}_\varepsilon(0)$ is {approximately} a constant. {The number of grids is denoted as ${\cal N}$, and scales as some (positive) power of $A/\hbar$. Because the scale} over which $\mathfrak{M}_{\varepsilon}({\bf q}_{-t},\vartheta_{-t};0)$ varies in $\vartheta$ shrinks exponentially as $e^{-\lambda_Lt}${, we divide the angular ($\vartheta$) interval: $[0,2\pi]$ into subintervals}, each of which has a size of $e^{-\lambda_Lt}$. Within each {subinterval, $\mathfrak{M}_{\varepsilon}({\bf q}_{-t},\vartheta_{-t};0)$ is a constant.} So to perform the {$\vartheta$ integral in a subinterval}, we can pull $\mathfrak{M}_{\varepsilon}({\bf q}_{-t},\vartheta_{-t};0)$ out of {the integral} and then use Lagrange's mean-value theorem. Consequently, Eq.~(\ref{eq:11}) reduces to
\begin{eqnarray}\label{eq:S34}
M_{{\bf r}{\bf r}'}(t)=e^{-\lambda_L t}h({\bf r},{\bf r}';t)
\end{eqnarray}
{(for $\lambda_L^{-1}\lesssim t\ll t_E$)} with
\begin{eqnarray}\label{eq:S30}
    h({\bf r},{\bf r}';t)=\int d\varepsilon h_\varepsilon({\bf r},{\bf r}';t)\quad\quad\quad\quad\nonumber\\
    h_\varepsilon=\sum_k e^{-\frac{i}{\hbar}({\bf r}-{\bf r}')\cdot {\bf p}({\bf q},\vartheta_k)}\mathfrak{M}_{\varepsilon}({\bf Q}_{k,-t},\vartheta_{k,-t};0).\quad
\end{eqnarray}
Here $k$ labels {the subinterval}. The angular value $\vartheta_k$ results from the application of Lagrange's mean-value theorem in the {subinterval} $k$. $({\bf Q}_{k,-t},\vartheta_{k,-t})$ stands for the coordinates of the center of the grid, a trajectory initiating from which {evolves to} $({\bf q},\vartheta_k)$ at $t$. $\mathfrak{M}_\varepsilon({\bf Q}_{k,-t},\vartheta_{k,-t};0)$ is the constant value in the {subinterval} $k$.
Finally, due to chaoticity the grids are randomly sampled from ${\cal N}$ grids. Thus $h_\varepsilon$ must vary randomly with $t$, whose explicit form depends on ${\bf r},{\bf r}'$ and $F(0)$ and thereby is nonuniversal.

\subsubsection{The correlation function for long time}
\label{sec:relaxation_correlation_function}

\begin{center}
{\it 4.1. Impacts of quantum terms in Eq.~(\ref{eq:S28})}
\end{center}

For $t\gg t_E$, Eq.~(\ref{eq:S39}) shows that it is necessary to study the quantum terms, namely, the terms with $K\in \mathbb{N}$, in the expansion Eq.~(\ref{eq:S28}). {For the ($K+1$)th term with $K\gg t/t_E$}, which is an integral of $K$ time variables $t_s(s=1,\cdots,K)${, because of $t\geq t_1\geq \cdots\geq t_K\geq t_{K+1}\equiv 0$} there must be at least one $s$ such that $t_{s}-t_{s+1}\ll t_E$. For such $s$, thanks to Eq.~(\ref{eq:S39}) when $\delta \hat H {g_\varepsilon}(t_s-t_{s+1})$ acts on a phase-space function, a negligibly small quantum correction results. So the expansion is truncated, and we need to study only the impacts of the truncated expansion on $M_{{\bf r}{\bf r}'}(t)$.

Consider the $(K+1)$th term in the truncated expansion, with every $t_s-t_{s+1}\gtrsim t_E$. By the energy-shell partition above and that $\delta \hat H$ is {a differential operator} with respect to ${\bf p}$ [cf.~Eq.~(\ref{eq:144})], $\delta \hat H {g_\varepsilon}(t_K)\circ \mathfrak{M}_\varepsilon(0)$ does not vanish, at most, only at the boundary regime of two grids and oscillates around zero in $\vartheta$. {Because the classical motion of phase points gives rise to the deformation of a grid (Fig.~\ref{fig:3}, right), at} $t_K\gtrsim t_E$ the boundary regime is deformed, with a width $\sim e^{-\lambda_L t_K}={\cal O}(\frac{\hbar}{A})$ in $\vartheta$ axis. When the product of the sequence $\delta \hat H {g_\varepsilon}(t_s-t_{s+1})$ acts on $\delta \hat H {g_\varepsilon}(t_K)\circ \mathfrak{M}_\varepsilon(0)$, the oscillations in the boundary regime remain. However, the width is smaller by a factor of $e^{-\lambda_L (t_1-t_K)}={\cal O}((\frac{\hbar}{A})^{(K-1)})$. Therefore, at given ${\bf q}$ the considered quantum term includes many oscillations, each of which takes place in an extremely narrow regime. Because {$e^{-\frac{i}{\hbar}({\bf r}-{\bf r}')\cdot {\bf p}({\bf q},\vartheta)}$} varies with $\vartheta$ over a scale $\gtrsim\frac{\hbar}{A}$, when the quantum term is multiplied by this factor and the integral over $\vartheta$ is performed, the oscillations are averaged out.
As a result,
\begin{eqnarray}\label{eq:73}
    m\int d\varepsilon d\vartheta e^{-\frac{i}{\hbar}({\bf r}-{\bf r}')\cdot {\bf p}({\bf q},\vartheta)}\quad\quad\quad\quad\quad\quad\nonumber\\
    \times\mathfrak{M}_{\varepsilon, K+1}({\bf q},\vartheta;t)=0,\quad for\, t\gg t_E.
\end{eqnarray}
So the quantum evolution of $M_{{\bf r}{\bf r}'}(t)$ --- but not $\mathfrak{M}_\varepsilon({\bf q},\vartheta;t)$ --- at $t\gg t_E$ is still determined by the first term in the expansion Eq.~(\ref{eq:S28}).\\

\begin{center}
{\it 4.2. Relaxation of the correlation function}
\end{center}

Therefore, to find the behaviors of $M_{{\bf r}{\bf r}'}(t)$ at $t\gg t_E$ we just need to extend the studies of Eq.~(\ref{eq:11}) to the regime $t\gg t_E$. {Recall that a grid in the phase-space energy shell is deformed in the course of time (Fig.~\ref{fig:3}, right). For $t\gg t_E$ all grids intersect with the $\vartheta$ line passing the phase point $({\bf q},0)$ (namely, the line with $\vartheta$ varying from $0$ to $2\pi$ while ${\bf q}$ fixed), and the angular measure of the intersection between a deformed grid and that line} equilibrates, which is $\sim {\cal N}^{-1}$. Because of ${\cal N}^{-1}/e^{-\lambda_L t}\equiv{\cal N}_1(t)\gg 1$ such a measure must be contributed by ${\cal N}_1(t)$ disconnected {sets in the $\vartheta$ line. In other words, a grid is deformed so highly that it has ${\cal N}_1(t)$ intersections with that line. Owing to the mixing property, these intersections are uniformly distributed. As a result, when we partition the interval $[0,2\pi]$ (or the $\vartheta$ line) into subintervals} in the same way as the short-time case, we find that the {subinterval} size $e^{-\lambda_L t}$ is smaller than the angular scale over which {$e^{-\frac{i}{\hbar}({\bf r}-{\bf r}')\cdot {\bf p}({\bf q},\vartheta)}$} varies. Thus {for ${\cal N}_2(t)\sim e^{\lambda_L t}$ (with the prefactor being time independent and thus omitted) nearest $\vartheta_k$, $e^{-\frac{i}{\hbar}({\bf r}-{\bf r}')\cdot {\bf p}}$ takes the same value}. So, we can organize the {subinterval} indices $\{k\}$ into $\frac{{\cal N}{\cal N}_1(t)}{{\cal N}_2(t)}$ groups; each group, labelled by $G$, includes ${\cal N}_2(t)$ nearest $\vartheta_k$ {corresponding to the same value of $e^{-\frac{i}{\hbar}({\bf r}-{\bf r}')\cdot {\bf p}}$, denoted as} $e^{-\frac{i}{\hbar}({\bf r}-{\bf r}')\cdot {\bf p}({\bf q},\vartheta_G)}$. As such, $h_\varepsilon$ in Eq.~(\ref{eq:S30}) is replaced by
\begin{eqnarray}\label{eq:49}
    h_\varepsilon=\sum_G e^{-\frac{i}{\hbar}({\bf r}-{\bf r}')\cdot {\bf p}({\bf q},\vartheta_G)}
    \sum_{k\in G}\mathfrak{M}_\varepsilon({\bf Q}_{k,-t},\vartheta_{k,-t};0).\quad\,\,
\end{eqnarray}
As the mixing property renders the grids uniformly sampled by the ${\cal N}_2(t)\gg 1$ coordinates: $({\bf Q}_{k,-t},\vartheta_{k,-t})$, the second factor can be simplified as
\begin{eqnarray}
    &&\frac{1}{{\cal N}_2(t)}\sum_{k\in G}\mathfrak{M}_\varepsilon({\bf Q}_{k,-t},\vartheta_{k,-t};0)\nonumber\\
    &\stackrel{t\gg t_E}{\sim}&
    \int\!\!\!\!\int d{\bf q}'d\vartheta'\mathfrak{M}_\varepsilon({\bf q}',\vartheta';0),\,\,\,\quad
    \label{eq:S42}
\end{eqnarray}
where in the second line we omit an irrelevant normalization factor. With its substitution, Eq.~(\ref{eq:49}) reduces to
\begin{eqnarray}\label{eq:58}
    h_\varepsilon\sim e^{\lambda_L t}\sum_G e^{-\frac{i}{\hbar}({\bf r}-{\bf r}')\cdot {\bf p}({\bf q}, \vartheta_G)}\int\!\!\!\!\int d{\bf q}'d\vartheta'\mathfrak{M}_\varepsilon({\bf q}',\vartheta';0),\nonumber\\
    for\, t\gg t_E. \qquad\qquad\qquad
\end{eqnarray}
Combining it with the expressions of $M_{{\bf r}{\bf r}'}(t)$ given in Eq.~(\ref{eq:S34}), we find that the {exponentially growing factor balances the exponentially decaying one,} giving
\begin{equation}\label{eq:61}
    M_{{\bf r}{\bf r}'}(t)\stackrel{t\gg t_E}{\longrightarrow}const..
\end{equation}
This implies the relaxation of $M_{{\bf r}{\bf r}'}(t)$.

It should be emphasized that, unlike Eq.~(\ref{eq:S34}) {which involves only the leading classical term in Eq.~(\ref{eq:S62}), Eq.~(\ref{eq:61}) involves the entire expansion in $\delta \hat H$ in Eq.~(\ref{eq:S62}),} and results from that all quantum corrections to the right-hand side of Eq.~(\ref{eq:9}) vanish. This phenomenon is of quantum origin and resembles a phenomenon in level statistics at the frequency scale much smaller than the inverse Ehrenfest time \cite{Tian04}, where all quantum terms of the level-level correlator {vanish} and only classical terms remain. Moreover, the explicit expressions of the right-hand sides of Eqs.~(\ref{eq:S34}) and (\ref{eq:61}) differ because the former (latter) is for short (long) time.

Summarizing, we have found that, for $t\gg t_E$, the contributions of all quantum terms of $\mathfrak{M}_\varepsilon(t)$ to $M_{{\bf r}{\bf r}'}(t)$ vanish, i.e., $M_{{\bf r}{\bf r}'}(t)$ is governed by the Liouville evolution of $\mathfrak{M}_\varepsilon(t)$, and as a result $M_{{\bf r}{\bf r}'}(t)$ relaxes. It should be emphasized that this relaxation process refers to the macroscopic observable $M_{{\bf r}{\bf r}'}(t)$, but not to the phase-space function $\mathfrak{M}_\varepsilon(t)$. Indeed, like in standard statistical physics, to pass from the latter to the former integrating out (some) phase-space coordinates is inevitable, and we have seen above that the $\vartheta$ integral in Eq.~(\ref{eq:3}) is essential to justify the relaxation of $M_{{\bf r}{\bf r}'}(t)$. {This relaxation phenomenon resembles a well-known result in the studies of the foundations of classical statistical physics \cite{Dorfman99}. There, when a probability distribution in phase space evolves, a macroscopic observable obtained by averaging with respect to that distribution can relax to an equilibrium value which is the average with respect to some smooth probability distribution, but the distribution does not relax to that smooth distribution and is even not smooth at long time. In mathematical literatures this is called {\it weak limit}.}

\subsubsection{Relaxed value of the correlation function}
\label{sec:relaxation_value_correlation_function}

We proceed to find the relaxed value of $M_{{\bf r}{\bf r}'}(t)$. In principle, it is possible to use Eqs.~(\ref{eq:S34}), (\ref{eq:S30}) and (\ref{eq:58}) to find this value. Below we adopt a simpler method.

Let us substitute Eq.~(\ref{eq:S15}) into Eq.~(\ref{eq:15}) to obtain
\begin{eqnarray}\label{eq:16}
    M_{{\bf r}{\bf r}'}(t)=\sum_{{\bf m},{\bf m}'}C_{{\bf m}'}^*C_{{\bf m}}e^{i(E_{{\bf m}'}-E_{{\bf m}})t/\hbar}\langle {\bf m}'| a^\dagger_{{\bf r}'}a_{{\bf r}}|{\bf m}\rangle,\quad
\end{eqnarray}
and perform the time average: $\lim_{T_0\rightarrow\infty} \int_0^{T_0} \frac{dt}{T_0} M_{{\bf r}{\bf r}'}(t)$. This gives the relaxed value, i.e.,
\begin{equation}\label{eq:75}
  M_{{\bf r}{\bf r}'}(t)\stackrel{t\gg t_E}{\longrightarrow}\sum_{{\bf m}\in {\mathscr{F}_S}}|C_{{\bf m}}|^2\langle {\bf m}| a^\dagger_{{\bf r}'}a_{{\bf r}}|{\bf m}\rangle.
\end{equation}
The remaining task is to calculate the right-hand side.

It is easy to show that
\begin{equation}\label{eq:74}
  \langle {\bf m}| a^\dagger_{{\bf r}'}a_{{\bf r}}|{\bf m}\rangle=\sum_\nu n_\nu \psi_\nu({\bf r})\psi^*_\nu({\bf r}').
\end{equation}
With its substitution Eq.~(\ref{eq:75}) reduces to
\begin{eqnarray}\label{eq:17}
    M_{{\bf r}{\bf r}'}(t)\stackrel{t\gg t_E}{\longrightarrow}\sum_{{\bf m}\in {\mathscr{F}_S}}|C_{{\bf m}}|^2\sum_\nu n_\nu C_\nu({\bf r},{\bf r}'),
\end{eqnarray}
which relates the relaxed value to the occupation number pattern $\{n_\nu\}$ corresponding to the Fock state ${\bf m}$ and the autocorrelation of a single-particle eigenfunction defined as $C_\nu({\bf r},{\bf r}')\equiv\psi_\nu({\bf r})\psi^*_\nu({\bf r}')$.

For $F(0)\in \mathscr{P}$ most probability weight of $|C_{{\bf m}}|^2$ goes to typical ${\bf m}$. It has been shown in Ref.~\cite{Tian18} that $C_\nu({\bf r},{\bf r}')$ satisfies the relation Eq.~(\ref{eq:115}), with $\hat{\mathfrak{O}}=a^\dagger_{{\bf r}'}a_{{\bf r}}$ and $\mathfrak{O}_\nu=C_\nu({\bf r},{\bf r}')$. Using this result it has been shown \cite{Tian18} that for typical ${\bf m}$,
\begin{eqnarray}\label{eq:25}
    \sum_\nu n_\nu C_\nu({\bf r},{\bf r}')=\int dm(\nu)C_\nu({\bf r},{\bf r}')n_{FD}(\varepsilon_\nu)
\end{eqnarray}
follows, where $dm(\nu)$ gives the number of single-particle eigenstates in the interval: $(\nu,\nu+d\nu)$ of the $\nu$ space. In Appendix \ref{sec:S3} we show that for a $2$D chaotic cavity,
\begin{equation}\label{eq:26}
    C_\nu({\bf r},{\bf r}')=\frac{1}{V}J_0\left(\frac{|{\bf r}-{\bf r}'|}{\lambda_{\varepsilon_\nu}}\right).
\end{equation}
Combining Eqs.~(\ref{eq:17}), (\ref{eq:25}) and (\ref{eq:26}), we obtain
\begin{equation}\label{eq:1}
    M_{{\bf r}{\bf r}'}(t)\stackrel{t\gg t_E}{\longrightarrow}\frac{1}{V}\int dm(\nu)J_0\left(\frac{|{\bf r}-{\bf r}'|}{\lambda_{\varepsilon_\nu}}\right)n_{FD}(\varepsilon_\nu),
\end{equation}
which is the special case of Eq.~(\ref{eq:87}) at $j=1$. We see that, as long as $F(0)\in \mathscr{P}$, the relaxed value is independent of $F(0)$. Instead, it depends only on the thermodynamic quantities $T,\mu$ and the spectral structure described by $dm(\nu)$. This is a hallmark of quantum thermalization. Correspondingly, the relaxation time $t_E$ is the thermalization time. It is important that this thermalization time is much smaller than the time to resolve an individual many-body eigenenergy, {namely, the Heisenberg time} $\hbar/\Delta_{\mathscr{F}_S}$, with $\Delta_{\mathscr{F}_S}$ being the level spacing of the many-body eigenstates in the microcanonical energy shell ${\mathscr{F}_S}$. Therefore, thermalization occurs long before the occurrence of {dephasing}.

\subsection{Multi-particle correlation}
\label{sec:j_particle_correlation}

The multi-particle correlation function involving $j\geq 2$ particles that are annihlated at spatial points $\{{\bf r}\}\equiv \{{\bf r}_1,{\bf r}_2,\cdots, {\bf r}_j\}$ and created at $\{{\bf r}'\}\equiv\{{\bf r}'_1,{\bf r}'_2,\cdots, {\bf r}'_j\}$ are defined by Eq.~(\ref{eq:76}). We further define the set of all such elements $\hat M^{(j)}(t)\equiv \{M^{(j)}_{\{{\bf r}\}\{{\bf r}'\}}
(t)\}$.

In Appendix~\ref{sec:derivations_von_Neumann} we show that $\hat M^{(j)}(t)$ satisfies the following von Neumann equation
\begin{equation}\label{eq:79}
    \partial_t\hat M^{(j)}(t)=-\frac{i}{\hbar}\sum_{k=1}^j \left[H(\hat{{\bf q}}_k,\hat{{\bf p}}_k),\hat M^{(j)}(t)\right].
\end{equation}
Passing to the Wigner representation,
\begin{eqnarray}\label{eq:83}
    M^{(j)}_{\{{\bf r}\}\{{\bf r}'\}}(t)&\equiv&\int d{\bf p}_1\cdots d{\bf p}_j e^{-\frac{i}{\hbar}\sum_{k=1}^j({\bf r}_k-{\bf r}'_k)\cdot {\bf p}_k}\nonumber\\
&\times&\mathfrak{M}^{(j)}(\{{\bf q}\},\{{\bf p}\};t),
\end{eqnarray}
where $\{{\bf q}\}\equiv \{{\bf q}_1,\cdots,{\bf q}_j\}$, with ${\bf q}_k\equiv\frac{{\bf r}_k+{\bf r}'_k}{2}$, and $\{{\bf p}\}\equiv \{{\bf p}_1,\cdots,{\bf p}_j\}$, we can rewrite Eq.~(\ref{eq:79}) as
\begin{equation}\label{eq:84}
    \left(\partial_t-\sum_{k=1}^j\{H({\bf q}_k,{\bf p}_k),\,\cdot\,\}_{{\rm M}}\right)\mathfrak{M}^{(j)}(\{{\bf q}\},\{{\bf p}\};t)=0.
\end{equation}
Owing to the similarity of these two equations to Eqs.~(\ref{eq:3}) and (\ref{eq:14}), we can generalize the method developed in Sec.~\ref{sec:one_particle} to the multi-particle case. Because the analysis is parallel, we shall give only the final results. In particular, as a generalization of Eq.~(\ref{eq:S34}), we find that
\begin{eqnarray}\label{eq:81}
M^{(j)}_{\{{\bf r}\}\{{\bf r}'\}}(t)=e^{-j\lambda_L t}h^{(j)}(\{{\bf r}\},\{{\bf r}'\};t),\nonumber\\
for\, \lambda_L^{-1}\lesssim t\ll t_E,\quad\quad\quad
\end{eqnarray}
with
\begin{eqnarray}\label{eq:145}
    h^{(j)}(\{{\bf r}\},\{{\bf r}'\};t)=\int d\varepsilon_1\cdots d\varepsilon_j h_{\varepsilon_1\cdots \varepsilon_j}^{(j)}(\{{\bf r}\},\{{\bf r}'\};t).\quad
\end{eqnarray}
Here $h_{\varepsilon_1\cdots \varepsilon_j}^{(j)}$ varies randomly with $t$, whose explicit form depends on ${\bf r},{\bf r}'$ and $F(0)$ and thereby is nonuniversal.
For $t\gg t_E$, this multi-particle correlation function relaxes, i.e.,
\begin{equation}\label{eq:82}
    M^{(j)}_{\{{\bf r}\}\{{\bf r}'\}}(t)\stackrel{t\gg t_E}{\longrightarrow}const..
\end{equation}

Similar to the one-particle case, to find the relaxed value we perform the time average of Eq.~(\ref{eq:76}). As a result,
\begin{equation}\label{eq:85}
  M^{(j)}_{\{{\bf r}\}\{{\bf r}'\}}(t)\stackrel{t\gg t_E}{\longrightarrow}\sum_{{\bf m}\in {\mathscr{F}_S}}|C_{{\bf m}}|^2\langle {\bf m}| a^\dagger_{{\bf r}'_1}\cdots a^\dagger_{{\bf r}'_j}a_{{\bf r}_1}\cdots a_{{\bf r}_j}|{\bf m}\rangle,
\end{equation}
which generalizes Eq.~(\ref{eq:75}). Because ${\bf m}$ is a Gaussian state, we can use Wick's theorem to factorize $\langle {\bf m}|a^\dagger_{{\bf r}'_1}\cdots a^\dagger_{{\bf r}'_j}a_{{\bf r}_1}\cdots a_{{\bf r}_j}|{\bf m}\rangle$ into the product of one-particle correlation functions, obtaining
\begin{eqnarray}
\label{eq:86}
  &&\langle {\bf m}|a^\dagger_{{\bf r}'_1}\cdots a^\dagger_{{\bf r}'_j}a_{{\bf r}_1}\cdots a_{{\bf r}_j}|{\bf m}\rangle\nonumber\\
  &=&\sum_P\sigma(P)\prod_{k=1}^j \langle {\bf m}|a^\dagger_{{\bf r}'_{P(k)}}a_{{\bf r}_k}|{\bf m}\rangle.
\end{eqnarray}
Upon substituting it into Eq.~(\ref{eq:85}), we find that
\begin{eqnarray}\label{eq:124}
  M^{(j)}_{\{{\bf r}\}\{{\bf r}'\}}(t)&\stackrel{t\gg t_E}{\longrightarrow}&\sum_{{\bf m}\in {\mathscr{F}_S}}|C_{{\bf m}}|^2 \nonumber\\
  &\times&\sum_P\sigma(P)\prod_{k=1}^j
  \sum_\nu n_\nu C_\nu({\bf r}_{{\bf r}_k},{\bf r}'_{P(k)}).\quad
\end{eqnarray}
Then we substitute Eqs.~(\ref{eq:25}) and (\ref{eq:26}) into Eq.~(\ref{eq:124}). As a result, the second line of Eq.~(\ref{eq:124}) is independent of ${\bf m}$, and Eq.~(\ref{eq:87}), which generalizes Eq.~(\ref{eq:1}), follows. That this relaxed value or the right-hand side of Eq.~(\ref{eq:87}) is independent of $F(0)$ reflects again that the ideal Fermi gas is thermalized at $t_E$, as long as $F(0)\in \mathscr{P}$.

\section{Dynamics of RDM and EE: $\boldsymbol{F(0) \in \mathscr{P}}$}
\label{sec:entanglement_entropy_evolution}

Armed with the results for the correlation functions obtained in Sec.~\ref{sec:evolution_correlation_function}, we proceed to study the dynamics of the RDM, $\hat{\rho}_A(t)$, and the EE, $S_A(t)$. We keep in mind that for the studies of these two quantities we {first} work in the lattice space $\mathbb{Z}^2\cap\mathfrak{C}$ {and then pass to the continuum limit in the final results}.

\subsection{Relaxation of RDM and EE}
\label{sec:entanglement_entropy_short_time}

We rewrite Eq.~(\ref{eq:29}) as
\begin{eqnarray}\label{eq:59}
  \hat\rho_A(t)&=&{\sum_{\{O_i\}}}^+\langle F(t)|{\prod_{i}} O^\dagger_{i}|F(t)\rangle{\prod_{i}} O_{i}\nonumber\\
    &+&{\sum_{\{O_i\}}}^-\langle F(t)|{\prod_{i}} O^\dagger_{i}|F(t)\rangle{\prod_{i}} O_{i}.
\end{eqnarray}
Here the superscript $+(-)$ in the sum stands for that the number of creation operators appearing in the operator configuration $\{O_i\}$ is (not) equal to that of annihilation operators. Because $F(t)$ is superposed by Fock states with fixed particle number $N$, the second term vanishes, give
\begin{eqnarray}\label{eq:186}
  \hat\rho_A(t)={\sum_{\{O_i\}}}^+\langle F(t)|{\prod_{i}} O^\dagger_{i}|F(t)\rangle{\prod_{i}} O_{i}.
\end{eqnarray}
For every expansion coefficient $\langle F(t)|{\prod_{i}} O^\dagger_{i}|F(t)\rangle$, we can use the anticommuntative relations to organize it as the linear superposition of multi-particle correlation functions, which are the lattice version of those defined by Eq.~(\ref{eq:76}). With the help of the results obtained in Sec.~\ref{sec:evolution_correlation_function}, we find that $\langle F(t)|{\prod_{i}} O^\dagger_{i}|F(t)\rangle$ relaxes at the time scale of $t_E$. As a result, the RDM relaxes also, i.e.,
\begin{eqnarray}
\hat\rho_A(t)\stackrel{t\gg t_E}{\longrightarrow}{\sum_{\{O_i\}}}^+M[\{O^\dagger_{i}\}]{\prod_{i}} O_{i}.
\label{eq:60}
\end{eqnarray}
Here $M[\{O^\dagger_{i}\}]$ is the relaxed expansion coefficient of $\langle F(t)|{\prod_{i}} O^\dagger_{i}|F(t)\rangle$, which can be calculated in the same way as before and is found to be
\begin{equation}\label{eq:63}
  M[\{O^\dagger_{i}\}]=\sum_{{\bf m}\in\mathscr{F}_S}|C_{\bf m}|^2\langle {\bf m}|{\prod_{i}} O^\dagger_{i}|{\bf m}\rangle.
\end{equation}
Because $\langle {\bf m}|{\prod_{i}} O^\dagger_{i}|{\bf m}\rangle$ is the same for all typical ${\bf m}$ in $\mathscr{F}_S$, and the majority of the weight $|C_{\bf m}|^2$ goes to typical ${\bf m}$, Eq.~(\ref{eq:63}) is simplified as
\begin{eqnarray}
\label{eq:62}
M[\{O^\dagger_{i}\}]=\langle {\bf m}_{\mathscr{F}_S}|{\prod_{i}} O^\dagger_{i}|{\bf m}_{\mathscr{F}_S}\rangle,
\end{eqnarray}
where ${\bf m}_{\mathscr{F}_S}$ can be any typical ${\bf m}$ in ${\mathscr{F}_S}$. With its substitution Eq.~(\ref{eq:60}) reduces to
\begin{eqnarray}\label{eq:31}
    \hat\rho_A(t)\stackrel{t\gg t_E}{\longrightarrow}
    \sum_{\{O_i\}}\langle {\bf m}_{{\mathscr{F}_S}}|\prod_{i} O^\dagger_{i}|{\bf m}_{{\mathscr{F}_S}}\rangle\prod_{i} O_{i}.
\end{eqnarray}
We emphasize that the relaxed RDM on the right-hand side does not depend on the choice of ${\bf m}_{\mathscr{F}_S}$.

Furthermore, one can always use Wick's theorem to cast $\langle {\bf m}_{{\mathscr{F}_S}}|\prod_{i} O^\dagger_{i}|{\bf m}_{{\mathscr{F}_S}}\rangle$ into the sum of all possible products of one-particle correlation functions at the Fock state ${\bf m}_{\mathscr{F}_S}$, each of which has the form
\begin{equation}\label{eq:148}
  ({\cal M}_{N_A})_{ij}\equiv\langle {\bf m}_{\mathscr{F}_S}|a^\dagger_{i}a_{j}|{\bf m}_{\mathscr{F}_S}\rangle=a^2\sum_\nu n_\nu C({\bf r}_i,{\bf r}_j),
\end{equation}
with ${\bf r}_i$ being the coordinate of plaque $i$. [For the convenience below, we also define a $N_A\times N_A$ matrix,
\begin{equation}\label{eq:149}
  \hat {\cal M}_{N_A}\equiv\{({\cal M}_{N_A})_{ij},{\bf r}_i,{\bf r}_j\in A,i,j=1,2,\cdots,N_A\}
\end{equation}
at given ${\bf m}_{\mathscr{F}_S}$.] So, according to Eq.~(\ref{eq:1}), the relaxed RDM, i.e., $\hat{\rho}_A(t\gg t_E)$, depends only on the macroscopic parameters of the Fermi gas in the cavity (not in the subsystem), namely, {$T,\mu,dm(\nu)$ and $V$}. This justifies the statement of Eq.~(\ref{eq:155}). Thus for $t\gg t_E$ any macroscopic quantity defined on the subsystem is completely determined by these quantities. {Because this result holds for subsystems of arbitrary location, geometry and size, the whole gas is at thermal equilibrium everywhere, and the thermal properties of the gas inside and outside a subsystem are the same}.

We remark that the dynamics of the RDM here is fundamentally different from that in some quenched systems \cite{Peschel09}. There the RDM is always Gaussian and thus its dynamics is completely determined by that of the one-particle correlation function. In contrast, here, because $F(t)$ is non-Gaussian at any $t$, the RDM is non-Gaussian before it relaxes, and this relaxation is determined by the dynamics of all multi-particle correlation functions. Moreover, the Gaussian nature of the relaxed RDM here arises from $F(0)\in \mathscr{H}_{S1}$. As we shall see in Sec.~\ref{sec:some_generalizations}, this nature is lost for $F(0)\in\mathscr{H}_{S2}$, even though the RDM still relaxes.

As a straightforward application of Eq.~(\ref{eq:31}), we have {Eq.~(\ref{eq:80})}. So the EE relaxes at the time scale of $t_E$, and the relaxed value is thermal. {This result holds for subsystems of arbitrary location, geometry and size.} In the next two subsections, we will find an explicit analytic expression of this relaxed value {for a special class of subsystems, which are deep inside the bulk but sufficiently large and have a specific geometry}.

\subsection{A warmup: relaxed value of EE for $\boldsymbol{1}$D subsystem geometry}
\label{sec:entanglement_entropy_long_time_special}

For finite $a$ so that a discrete lattice results, we have a special subsystem geometry, which is $1$D and consists of $N_A\gg 1$ contiguous plaques. This corresponds to a quasi $1$D subsystem with a finite width in the continuous space. For this special subsystem, nontrivial correlation behaviors occur only in the longitudinal direction, and the relaxed EE does not vanish, as long as the subsystem does not shrink {to} a genuine $1$D line, i.e., $a=0$. This subsection {serves as} a preparation for the studies of the relaxed EE in $2$D subsystem, which will be pursued in the next subsection.

By using Eqs.~(\ref{eq:25}) and (\ref{eq:26}) we obtain
\begin{equation}\label{eq:36}
    ({\cal M}_{N_A})_{ij}=\frac{a^2}{V}\int dm(\nu)J_0\left(\frac{a}{\lambda_{\varepsilon_\nu}}(i-j)\right)n_{FD}(\varepsilon_\nu).
\end{equation}
Therefore, $\hat {\cal M}_{N_A}$ defined by Eq.~(\ref{eq:149}) is a Toeplitz matrix. That is, it satisfies $({\cal M}_{N_A})_{ij}=c_{i-j}$. Here
\begin{equation}\label{eq:37}
    c_{n}=\int_{-\pi}^\pi \frac{d\theta}{2\pi} e^{-in\theta}{\cal C}(\theta),\quad n\in\mathbb{Z},
\end{equation}
and ${\cal C}(\theta)$ is called the generating function{, given by
\begin{equation}\label{eq:39}
    {\cal C}(\theta)=\frac{a^2}{V}\sum_{n\in \mathbb{Z}}e^{in\theta}\int dm(\nu)J_0\left(\frac{an}{\lambda_{\varepsilon_\nu}}\right)n_{FD}(\varepsilon_\nu).
\end{equation}
Note that in deriving Eq.~(\ref{eq:39}) we have used Eq.~(\ref{eq:36}) and extended it to the regime where $i-j$ is order of or larger than the cavity size. {However, provided that the chain is deep inside the cavity, such extension plays no essential roles and thus is legitimate}}.
For a review of the Toeplitz matrix and the Toeplitz determinant we refer to Refs.~\cite{Grenander53,Boetchner90}. In Appendix \ref{sec:generating_function} we use Eq.~(\ref{eq:36}) to find the explicit expression of ${\cal C}(\theta)$, which is
\begin{eqnarray}
\label{eq:38}
  {\cal C}(\theta) &=& \frac{a^2}{V}\int dm(\nu)n_{FD}(\varepsilon_\nu)\frac{\lambda_{\varepsilon_\nu}}{a}\nonumber\\
  &\times&{\sum_{k\in \mathbb{Z}}}'\frac{1}{\sqrt{1-\left(\frac{\lambda_{\varepsilon_\nu}}{a}(\theta-2\pi k)\right)^2}}.
\end{eqnarray}
Here the prime stands for that the sum runs over $k$ with $\lambda_{\varepsilon_\nu}^2(\theta-2\pi k)^2\leq a^2$.

Because $\hat {\cal M}_{N_A}$ is real symmetric, we can diagonalize it by an orthogonal matrix $\hat V\equiv\{V_{ij}\},i,j=1,2,\cdots,N_A$, i.e.,
\begin{eqnarray}\label{eq:33}
    \hat {\cal M}_{N_A}=\hat V^T \left(
                                   \begin{array}{cccc}
                                     \frac{1+v_1}{2} & 0 & \cdots & 0 \\
                                     0 & \frac{1+v_2}{2} & \cdots & 0 \\
                                     \vdots & \vdots & \ddots & \vdots \\
                                     0 & 0 & \cdots & \frac{1+v_{N_A}}{2} \\
                                   \end{array}
                                 \right)\hat V,
\end{eqnarray}
where $\frac{1+v_i}{2}\in [0,1]$ are the eigenvalues and $T$ stands for the transpose. Thus if we choose the single-particle state $i$ so that
\begin{equation}\label{eq:34}
    a_i=\sum_{j=1}^{N_A}V_{ij}a_{{\bf r}_j},
\end{equation}
which will be considered in the remainder of this subsection, then
\begin{eqnarray}
\label{eq:35}
  &&\langle {\bf m}_{\mathscr{F}_S}|a_{i}|{\bf m}_{\mathscr{F}_S}\rangle=\langle {\bf m}_{\mathscr{F}_S}|a_{i}a_{j}|{\bf m}_{\mathscr{F}_S}\rangle=0,\nonumber\\
  &&\langle {\bf m}_{\mathscr{F}_S}|a^\dagger_{i}a_{j}|{\bf m}_{\mathscr{F}_S}\rangle=\delta_{ij}\frac{1+v_i}{2}.
\end{eqnarray}
Taking this into account, we reduce Eq.~(\ref{eq:31}) to
\begin{eqnarray}\label{eq:32}
    \hat\rho_A&\stackrel{t\gg t_E}{\longrightarrow}&\prod_{i=1}^{N_A}\left(\frac{1+v_i}{2} a^\dagger_{i}a_{i}+\frac{1-v_i}{2}
    a_{i}a^\dagger_{i}\right).
\end{eqnarray}
Equations (\ref{eq:36}), (\ref{eq:37}), (\ref{eq:38}) and (\ref{eq:32}) allow us to use the scheme of Ref.~\cite{Korepin04} to calculate the relaxed value of the EE, i.e., $S_A(t\gg t_E)$.

Since $a^\dagger_{i}a_{i}$ has the eigenvalue of $0,1$, the eigenvalues of $\hat\rho_A$ are
\begin{equation}\label{eq:42}
    \lambda_{x_1\cdots x_{N_A}}=\prod_{i=1}^{N_A}\frac{1+(-1)^{x_i}v_i}{2},\quad x_i=0,1,
\end{equation}
from which we obtain
\begin{eqnarray}\label{eq:43}
    &&S_A(t\gg t_E)=-\sum_{x_1\cdots x_{N_A}}\lambda_{x_1\cdots x_{N_A}}\ln \lambda_{x_1\cdots x_{N_A}}\nonumber\\
    &=&-\sum_{i=1}^{N_A}\left(\frac{1+v_i}{2}\ln\frac{1+v_i}{2}+\frac{1-v_i}{2}\ln\frac{1-v_i}{2}\right)\nonumber\\
    &=&\sum_{i=1}^{N_A}e(1,v_i),
\end{eqnarray}
with $e(x,v)\equiv-\frac{x+v}{2}\ln\frac{x+v}{2}-\frac{x-v}{2}\ln\frac{x-v}{2}$. Because of $\frac{1\pm v_i}{2}\in [0,1]$, one can use Cauchy's residue theorem to rewrite Eq.~(\ref{eq:43}) as
\begin{eqnarray}\label{eq:44}
    S_A(t\gg t_E)=\frac{1}{2\pi i}\oint_C e(1,\lambda)\frac{d}{d\lambda}\ln D_{N_A}(\lambda)d\lambda.
\end{eqnarray}
Here $C$ is a contour that encircles the line from $-(1+0^+)$ to $1+0^+$ and along which $e(1,\lambda)$ is analytic. All the zeros of the Toeplitz determinant
\begin{eqnarray}\label{eq:45}
    D_{N_A}(\lambda)={\rm det}\left[(\lambda+1)\mathbb{I}_{N_A}-2\hat {\cal M}_{N_A}\right]
\end{eqnarray}
reside along the line encircled by $C$, where $(\lambda+1)\mathbb{I}_{N_A}-2\hat {\cal M}_{N_A}$ is a Toeplitz matrix with
\begin{eqnarray}\label{eq:53}
    \tilde {\cal C}(\theta)=\lambda+1-2{\cal C}(\theta)
\end{eqnarray}
as its generating function.

Using the property: ${\cal C}(\theta)={\cal C}(-\theta)$ shown in Appendix \ref{sec:generating_function} and Eq.~(\ref{eq:38}), we see that $\tilde {\cal C}(\theta)$ either has no zeros or has zeros in pairs: $(\theta_r(\lambda),-\theta_r(\lambda))$ with $r$ labeling the pairs. In the former case, $\tilde {\cal C}(\theta)$ is regular, i.e., is nonzero everywhere and has zero index. With the help of Szeg$\ddot{\rm o}$'s theorem \cite{Grenander53,Boetchner90}, we can find the large $N_A$ asymptotic,
\begin{equation}\label{eq:47}
    \ln D_{N_A}(\lambda)\stackrel{N_A\gg 1}{\longrightarrow} N_A \int_{-\pi}^{\pi}\frac{d\theta}{2\pi}\ln \tilde {\cal C}(\theta).
\end{equation}
Thus we have
\begin{equation}\label{eq:48}
    \frac{d}{d\lambda}\ln D_{N_A}(\lambda)\stackrel{N_A\gg 1}{\longrightarrow} N_A\int_{-\pi}^{\pi}\frac{d\theta}{2\pi}\frac{1}{\lambda+1-2{\cal C}(\theta)}.
\end{equation}
In the latter case, $\tilde {\cal C}(\theta)$ is singular and can be factorized into the regular part $b(\theta)$ and the {\it zero factor} --- a special case of the so-called {\it Fisher-Hartwig symbol} \cite{Fisher68} --- as
\begin{eqnarray}
\label{eq:50}
  \tilde {\cal C}(\theta) &=& b(\theta)\prod_r\left((2-\cos(\theta-\theta_r))(2-\cos(\theta+\theta_r))\right)^{m_r},\nonumber\\
  b(\theta) &=& \frac{\lambda+1-2{\cal C}(\theta)}{\prod_r\left((2-\cos(\theta-\theta_r))(2-\cos(\theta+\theta_r))\right)^{m_r}},
\end{eqnarray}
where $m_r$ is the integral order of the zeros $\pm\theta_r(\lambda)$. One can use the Fisher-Hartwig conjecture \cite{Fisher68}, which has been proven \cite{Boetchner90} --- thus is a theorem --- for the present case with integral-order zeros as the only singularity, to calculate the asymptotic of $D_{N_A}$. As a result,
\begin{equation}\label{eq:51}
    \ln D_{N_A}(\lambda)\stackrel{N_A\gg 1}{\longrightarrow} N_A \int_{-\pi}^{\pi}\frac{d\theta}{2\pi} \ln b(\theta).
\end{equation}
Upon taking its derivative with respect to $\lambda$ we find that the denominator of $b(\theta)$ in Eq.~(\ref{eq:50}) does not contribute, because the contributions from the pair of zeros $\pm\theta_i(\lambda)$ cancel out. As a result, Eq.~(\ref{eq:48}) remains valid.

Let us substitute Eq.~(\ref{eq:48}) into Eq.~(\ref{eq:44}). Note that ${\cal C}(\theta)\geq 0$, which is obvious from Eq.~(\ref{eq:38}). Moreover, as shown in Appendix \ref{sec:generating_function}, ${\cal C}(\theta)\leq 1$ (the exceptions may exist, but at most constitute a set of zero Lebesgue measure and thus do not play any roles.). Thus when $\theta$ is fixed, the denominator on the right-hand side of Eq.~(\ref{eq:48}), as a function of $\lambda$, must have zeros encircled by $C$. Then, applying Cauchy's theorem we obtain
\begin{eqnarray}
\label{eq:52}
  S_A(t\gg t_E)&=&-N_A\int_{-\pi}^{\pi}\frac{d\theta}{2\pi}\big({\cal C}(\theta)\ln{\cal C}(\theta)\nonumber\\
  &+&(1-{\cal C}(\theta))\ln(1-{\cal C}(\theta))\big).
\end{eqnarray}
So the volume law: $S_A\propto N_A$ follows and the relaxed values of the EE corresponding to distinct microcanonical energy shell $\mathscr{F}_S$ differ only in the proportionality coefficient.


\subsection{Relaxed value of EE: $\boldsymbol{2}$D subsystem geometry}
\label{sec:entanglement_entropy_long_time_general}

Beyond  $1$D geometry the analytical method used in Sec.~\ref{sec:entanglement_entropy_long_time_special} encounter some fundamental difficulties, arising from that Szeg$\ddot{\rm o}$'s theorem and the Fisher-Hartwig conjecture do not apply, if $i$ (or $j$) in Eq.~(\ref{eq:36}) represents a coordinate of {a $2$D lattice}. Here we use two different methods to address the relaxed value of the EE in $2$D subsystems: one is mathematically rigorous, but only for square geometry, which essentially replaces Szeg\"o's theorem by its high-dimensional generalization --- Doktorsky's theorem \cite{Doktorsky84} --- in the method used in Sec.~\ref{sec:entanglement_entropy_long_time_special}, and the other is approximate, but for {more general $2$D geometry}.

\subsubsection{Square geometry}
\label{sec:method_I}

We consider the subsystem A which is a square with its vertices on $\mathbb{Z}^2\cap\mathfrak{C}$. The side length is $\sqrt{N_A}\in \mathbb{N}$. In $2$D the definitions of the Toeplitz matrix and the Toeplitz determinant are subjected to some modifications \cite{Boetchner90}. First, a Toeplitz matrix $T_A$ acts on a state $\phi\equiv\{\phi_i\}_{i\in A}$ in the Hilbert space $L^2(A)$ according to the rule: $(T_A\phi)_i\equiv \sum_{j\in A}c_{i-j}\phi_j$ for $i\in A$, where the coefficient $c_n$ [$n\equiv (n_1,n_2)\in\mathbb{Z}^2$] is given by the generating function ${\cal C}$ on $\mathbb{T}^2$ via
\begin{eqnarray}\label{eq:89}
    c_{n}=\int\!\!\!\!\int_{-\pi}^\pi \frac{d\theta_1d\theta_2}{(2\pi)^2} e^{-i(n_1\theta_1+n_2\theta_2)}{\cal C}(\theta_1,\theta_2).
\end{eqnarray}
Second, the determinant of $T_A$ is defined as the product of all eigenvalues of $T_A$.

Keeping these modifications in mind, one may readily check that Eqs.~(\ref{eq:44}) and (\ref{eq:45}) still hold, except that the generating function of the Toeplitz matrix: $(\lambda+1)\mathbb{I}_{N_A}-2\hat {\cal M}_{N_A}$ is now given by
\begin{eqnarray}\label{eq:90}
    \tilde {\cal C}(\theta_1,\theta_2)&=&\lambda+1-2{\cal C}(\theta_1,\theta_2),\nonumber\\
    {\cal C}(\theta_1,\theta_2)&=&\frac{a^2}{V}\sum_{n_{1,2}\in \mathbb{Z}}e^{i(n_1\theta_1+n_2\theta_2)}\nonumber\\
    &\times&\int dm(\nu)J_0\left(\frac{a}{\lambda_{\varepsilon_\nu}}\sqrt{n_1^2+n_2^2}\right)n_{FD}(\varepsilon_\nu),\quad
\end{eqnarray}
that replaces Eq.~(\ref{eq:53}), and the Toeplitz determinant $D_{N_A}(\lambda)$ in Eq.~(\ref{eq:45}) should be understood in the way as that described above. {Similar to the discussions on Eq.~(\ref{eq:39}), to derive the above expression for ${\cal C}(\theta_1,\theta_2)$ we have used (the $2$D version of) Eq.~(\ref{eq:36}) and extended it to the regime where $i-j$ is order of or larger than the cavity size. {However, provided that the subsystem is deep inside the cavity, implying that the ratio of subsystem-to-cavity volume is $\ll 1$, such extension plays no essential roles and thus is legitimate}.} In Appendix \ref{sec:generating_function_two_dimension}, we show that $\tilde {\cal C}(\theta_1,\theta_2)$ satisfies all conditions required by Doktorsky's theorem \cite{Doktorsky84}. The latter theorem gives the determinant of the Toeplitz matrix: $(\lambda+1)\mathbb{I}_{N_A}-2\hat {\cal M}_{N_A}$ for $N_A\gg 1$, which is
\begin{equation}\label{eq:91}
    \ln D_{N_A}(\lambda)\stackrel{N_A\gg 1}{\longrightarrow} N_A \int\!\!\!\!\int_{-\pi}^{\pi}\frac{d\theta_1d\theta_2}{(2\pi)^2}\ln \tilde {\cal C}(\theta_1,\theta_2).
\end{equation}
(In Appendix \ref{sec:generating_function_two_dimension} we explain how to choose the branch of the logarithm.) Let us substitute Eqs.~(\ref{eq:90}) and (\ref{eq:91}) into Eq.~(\ref{eq:44}), and perform the $\lambda$ integral first by Cauchy's residue theorem. As a result,
\begin{eqnarray}
\label{eq:92}
  S_A(t\gg t_E)&=&-N_A\int\!\!\!\!\int_{-\pi}^{\pi}\frac{d\theta_1d\theta_2}{(2\pi)^2}\big({\cal C}(\theta_1,\theta_2)\ln{\cal C}(\theta_1,\theta_2)\nonumber\\
  &+&(1-{\cal C}(\theta_1,\theta_2))\ln(1-{\cal C}(\theta_1,\theta_2))\big).
\end{eqnarray}
which is similar to Eq.~(\ref{eq:52}) and is the special case of Eqs.~(\ref{eq:110}) and (\ref{eq:111}) for $d=2$. In Appendix \ref{sec:generating_function_two_dimension} we will show that $0\leq{\cal C}(\theta_1,\theta_2)\leq 1$. (The upper bound is violated at most in a set of zero Lebesgue measure.) So the right-hand side is well defined. From Eq.~(\ref{eq:92}) again $S_A\propto N_A$ follows and the relaxed values of the EE corresponding to distinct $\mathscr{F}_S$ differ only in the proportionality coefficient. {It is important that this volume law is derived for the subsystem which is deep inside the cavity.
}

We make several remarks. First, for more general $2$D geometries, e.g., a polygon of general shape, as long as the condition: ${\cal C}(\theta_1,\theta_2)\leq 1$ is met, Eq.~(\ref{eq:92}) holds. Although on the physical ground such condition is very likely true, we are not able to prove this rigorously for a general polygon. However, below by using an approximate method we will derive Eq.~(\ref{eq:92}) for {more general} $2$D geometry. Second, {we will explain in Appendix \ref{sec:generating_function_two_dimension} that at $T=0$ Doktorsky's theorem does not apply. This opens up a door for scaling behaviors different from the volume law. Indeed, for non-chaotic systems a scaling law $\sim \sqrt{N_A}\ln N_A$ has been found previously for a square subsystem geometry \cite{Klich06}.}

\subsubsection{{More general} geometry}
\label{sec:method_II}

For any subsystem geometry, according to Eq.~(\ref{eq:31}) the relaxed RDM, $\hat{\rho}_A(t\gg t_E)$, is a Gaussian state. Therefore, it must take the following general form,
\begin{eqnarray}\label{eq:68}
  \hat{\rho}_A(t\gg t_E)=e^{-\hat{{\cal H}}_{\rm eff}}/{\rm Tr}_Ae^{-\hat{{\cal H}}_{\rm eff}},
\end{eqnarray}
and the effective Hamiltonian $\hat{{\cal H}}_{\rm eff}$ is a free particle Hamiltonian with the general form,
\begin{equation}\label{eq:169}
  \hat{{\cal H}}_{\rm eff}=\sum_{i,j=1}^{N_A}{\cal H}_{ij}a_i^\dagger a_j.
\end{equation}
From this one may readily obtain the one-particle correlation function on the subsystem A, which should be identical to $({\cal M}_{N_A})_{ij}$. Thus $\hat{{\cal H}}\equiv\{{\cal H}_{ij}\}$ can be determined, read
\begin{eqnarray}\label{eq:70}
\hat{{\cal H}}=\ln \left(\hat{{\cal M}}_{N_A}^{-1}-1\right).
\end{eqnarray}
{[Equations (\ref{eq:68}), (\ref{eq:169}) and (\ref{eq:70}) clearly show that although a RDM, like $\hat{\rho}_A(t\gg t_E)$, can be Gaussian and governed by $T,\mu$ via the thermal correlation matrix $\hat{{\cal M}}_{N_A}$, its inverse covariance matrix, i.e., $\hat{{\cal H}}$, in general has a very complicated dependence on subsystem's size and geometry and the thermal parameters $T,\mu$. Thus such RDM is not a genuine thermal ensemble, as mentioned in Sec.~\ref{sec:results_F0_P}.]}
With the substitution of Eqs.~(\ref{eq:68}) and (\ref{eq:70}) into Eq.~(\ref{eq:30}), we obtain
\begin{eqnarray}\label{eq:71}
  S_A(t\gg t_E)&=&-{\rm Tr}_{A}\Big(\hat{{\cal M}}_{N_A}\ln \hat{{\cal M}}_{N_A}\nonumber\\
  &+&(1-\hat{{\cal M}}_{N_A})\ln (1-\hat{{\cal M}}_{N_A})\Big).
\end{eqnarray}
To calculate it we use the replica trick. Specifically, we introduce an auxiliary quantity defined as
\begin{eqnarray}\label{eq:156}
  S_{A,R}&=&-\frac{1}{R}{\rm Tr}_{A}\Big(\hat{{\cal M}}_{N_A}(\hat{{\cal M}}_{N_A}^R-1)\nonumber\\
  &+&(1-\hat{{\cal M}}_{N_A})((1-\hat{{\cal M}}_{N_A})^R-1)\Big),\quad R\in \mathbb{N}.\quad
\end{eqnarray}
Thus we obtain
\begin{eqnarray}\label{eq:158}
  S_A(t\gg t_E)=\lim_{R\rightarrow 0}S_{A,R}.
\end{eqnarray}
Next, for $R\in \mathbb{N}$ we can organize $S_{A,R}$ in terms of the following expansion,
\begin{eqnarray}\label{eq:150}
  S_{A,R}=-\frac{1}{R}\sum_m d_m(R) {\rm Tr}_{A}\left(\hat{{\cal M}}_{N_A}^m\right),
\end{eqnarray}
where the expansion coefficients $d_m$ satisfy the following identity,
\begin{equation}\label{eq:157}
  \lim_{R\rightarrow 0}\frac{1}{R}\sum_m d_m(R) x^m=-e(1,1-2x).
\end{equation}
Written in terms of the matrix elements, Eq.~(\ref{eq:150}) is
\begin{eqnarray}\label{eq:151}
  S_{A,R}=-\frac{1}{R}\sum_m d_m(R)\sum_{i_1,\cdots,i_m=1}^{N_A}\prod_{j=1}^{m}\left({\cal M}_{N_A}\right)_{i_ji_{j+1}},\quad
\end{eqnarray}
with $i_{m+1}\equiv i_1$.

To proceed we expect that, provided that the Fermi gas is not at the ground state, the matrix element $({\cal M}_{N_A})_{ij}$ decays sufficiently fast as the distance between $i$ and $j$ increases. {If a subsystem is deep inside the bulk, and is either a polygon or convex (see Appendix \ref{sec:convexity_subsystem_geometry} for discussions on this constraint on the geometry),} we can extend the sum of $i_j$ ($j\geq 2$) over the subsystem, i.e., the set: $\{1,2,\cdots, N_A\}$, to the sum over the whole lattice $\mathbb{Z}^2$, i.e.,
\begin{eqnarray}\label{eq:152}
  S_{A,R}\approx-\frac{1}{R}\sum_m d_m(R) \sum_{i_1=1}^{N_A}\sum_{i_2,\cdots,i_m\in \mathbb{Z}^2}\prod_{j=1}^{m}\left({\cal M}_{N_A}\right)_{i_ji_{j+1}}.\nonumber\\
\end{eqnarray}
With the help of Eq.~(\ref{eq:90}), upon summing up all $i_j$ we obtain
\begin{eqnarray}\label{eq:153}
  S_{A,R}\approx -\frac{N_A}{R}\int\!\!\!\!\int_{-\pi}^\pi \frac{d\theta_1d\theta_2}{(2\pi)^2}\sum_m d_m(R){\cal C}(\theta_1,\theta_2)^m.\quad
\end{eqnarray}
Thanks to Eq.~(\ref{eq:157}), we have
\begin{eqnarray}\label{eq:154}
  \lim_{R\rightarrow 0}\frac{1}{R}\sum_m d_m(R){\cal C}(\theta_1,\theta_2)^m={\cal C}(\theta_1,\theta_2)\ln{\cal C}(\theta_1,\theta_2)\quad\nonumber\\
  +(1-{\cal C}(\theta_1,\theta_2))\ln (1-{\cal C}(\theta_1,\theta_2)).\quad\quad\quad
\end{eqnarray}
Combining it with Eqs.~(\ref{eq:158}) and (\ref{eq:153}) we obtain Eq.~(\ref{eq:92}). So Eqs.~(\ref{eq:110}) and (\ref{eq:111}) hold for general $2$D subsystem geometry.

\subsubsection{Continuum limit $a\rightarrow 0$ and relation to Widom's theorem}
\label{sec:continuum_limit}

So far we have considered the relaxed EE in the lattice space. Now we would like to pass to the continuum limit $a\rightarrow 0$. By introducing
\begin{equation}\label{eq:170}
  an_{1,2}\equiv x_{1,2},\quad \frac{\hbar\theta_{1,2}}{a}\equiv p_{1,2},
\end{equation}
we obtain
\begin{eqnarray}\label{eq:171}
    {\cal C}(\theta_1,\theta_2)\stackrel{a\rightarrow 0}{\longrightarrow}{\cal C}({\bf p})=\frac{1}{V}\int d{\bf x} e^{\frac{i}{\hbar}{\bf x}\cdot {\bf p}}\quad\nonumber\\
    \times\int dm(\nu)J_0\left(\frac{x}{\lambda_{\varepsilon_\nu}}\right)n_{FD}(\varepsilon_\nu)\quad\quad
\end{eqnarray}
from Eq.~(\ref{eq:113}), where ${\bf p}\equiv (p_1,p_2)$ and ${\bf x}\equiv (x_1,x_2)$. To perform the ${\bf x}$ integral, we switch to the polar coordinates. Upon integrating out the angle, we obtain
\begin{eqnarray}\label{eq:172}
    {\cal C}({\bf p})=\frac{2\pi}{V}\!\!\int\!\! dm(\nu)\!\!\int_0^\infty\!\! dx x J_0\left(\frac{x}{\lambda_{\varepsilon_\nu}}\right) J_0\left(\frac{px}{\hbar}\right)n_{FD}(\varepsilon_\nu).\nonumber\\
\end{eqnarray}
With the help of the following identity proved in Appendix \ref{sec:proof},
\begin{eqnarray}\label{eq:173}
    \int_0^\infty dx x J_0\left(ax\right) J_0\left(bx\right)=a^{-1}\delta(a-b),\, for\, a,b>0,\quad\quad
\end{eqnarray}
we simplify Eq.~(\ref{eq:172}) to
{\begin{eqnarray}\label{eq:174}
    {\cal C}({\bf p})=\frac{2\pi\hbar^2}{mV}n_{FD}(\frac{{\bf p}^2}{2m})\int dm(\nu)\delta\left(\varepsilon_\nu-\frac{{\bf p}^2}{2m}\right).
\end{eqnarray}
For a subsystem deep inside the cavity, the discreteness of the spectrum: $\{\varepsilon_\nu\}$ is inessential. Therefore, we may approximate $dm(\nu)$ by $\frac{Vm}{2\pi\hbar^2}d\varepsilon_\nu$, where $\frac{Vm}{2\pi\hbar^2}$ is the average spectral density. With this approximation Eq.~(\ref{eq:174}) reduces to
\begin{eqnarray}\label{eq:175}
    {\cal C}({\bf p})=n_{FD}(\frac{{\bf p}^2}{2m}).
\end{eqnarray}}
Combining Eqs.~(\ref{eq:111}), (\ref{eq:170}) and (\ref{eq:175}) gives
\begin{eqnarray}
\label{eq:176}
  \frac{{S_a}}{a^2}\stackrel{a\rightarrow 0}{\longrightarrow}-\int\frac{d{\bf p}}{(2\pi\hbar)^2}\Big(n_{FD}(\frac{{\bf p}^2}{2m})\ln n_{FD}(\frac{{\bf p}^2}{2m})\nonumber\\
  +(1-n_{FD}(\frac{{\bf p}^2}{2m}))\ln(1-n_{FD}(\frac{{\bf p}^2}{2m}))\Big),
\end{eqnarray}
from which the second line of Eq.~(\ref{eq:110}) and Eq.~(\ref{eq:72}) follow.

The continuum limit of the relaxed EE, described by the second line of Eq.~(\ref{eq:110}) and Eq.~(\ref{eq:72}), has a deep mathematical foundation. Consider the relaxed EE of a $2$D subsystem in the continuum limit without passing to the lattice space. In this limit Eq.~(\ref{eq:44}) remains valid, but the Toeplitz determinant, $D_{N_A}(\lambda)$, is replaced,
\begin{eqnarray}\label{eq:183}
    D_{N_A}(\lambda)\rightarrow D_{V_A}(\lambda)={\rm det}\left[(\lambda+1){\mathbb{I}}_{V_A}-2\hat M_{V_A}\right],\quad
\end{eqnarray}
where $(\hat M_{V_A})_{{\bf r}{\bf r}'}\equiv M_{{\bf r}{\bf r}'}(t\gg t_E)$ and $({\mathbb{I}}_{V_A})_{{\bf r}{\bf r}'}\equiv\delta({\bf r}-{\bf r}')$, with all ${\bf r},{\bf r}'$ restricted on the continuous space occupied by the subsystem A. The generating function of the operator: $(\lambda+1)\delta_{V_A}-2\hat M_{V_A}$ is given by its Fourier transformation with respect to (${\bf r}-{\bf r}'$), which can be readily found to be: $\lambda+1-2n_{FD}(\frac{{\bf p}^2}{2m})$, with ${\bf p}/\hbar$ as the Fourier wavenumber, by following the derivations of Eqs.~(\ref{eq:171})-(\ref{eq:175}). {Note that similar to previous discussions on the generating functions given by Eqs.~(\ref{eq:39}) and (\ref{eq:90}), respectively, for this result to hold A has to be deep inside the cavity.} Provided that the conditions of Widom's theorem \cite{Widom60} were satisfied, it gives
\begin{equation}\label{eq:184}
    \ln D_{V_A}(\lambda)\rightarrow V_A \int\frac{d{\bf p}}{(2\pi\hbar)^2}\ln \left(\lambda+1-2n_{FD}(\frac{{\bf p}^2}{2m})\right)
\end{equation}
for large subsystem volume $V_A$. {Note that for Widom's theorem to apply, A has to be either a polygon or convex \cite{Widom60}.} Substituting it into Eq.~(\ref{eq:44}), we recover the relaxed EE given by the second line of Eq.~(\ref{eq:110}) and Eq.~(\ref{eq:72}).

\subsection{Bounding the time profile of EE}
\label{sec:nonmonotonic_increase_EE}

We remark that the linear increase of the EE, $S_A(t)\propto t$, before equilibration has been observed in a number of quantum systems \cite{Sarkar99,Tanaka02,Cardy05,Balasubramanian11,Huse13}. However, unlike the present system, all those systems are subjected to the direct interaction. Moreover, analytical studies require a quantum system to carry some strong properties, such as the quantum integrability, the conformal symmetry and the holography \cite{Cardy05,Balasubramanian11,Maldacena13,Liu14,Mueller16}, and (or) the state to be Gaussian \cite{Hackl18}. As these properties are all absent in the present system, it is a natural question whether this linear increase still exists.

If this was true, $S_A(t)\propto t$ would hold up to the time scale of $t_E$. Since the increase finds its origin in one-body chaos, as discussed in Sec.~\ref{sec:physical_picture_evolution_entanglement_structure}, and the volume law holds for $t\gg t_E$, we expect the proportionality coefficient to be $N_A\lambda_L$ (up to an irrelevant universal numerical factor). So it follows that $S_A(t)=N_A\lambda_L t+const.$ for $t\lesssim t_E$. However, this expression then gives $S_A(t\sim t_E)$ that cannot match the thermal equilibrium value given by Eqs.~(\ref{eq:110}) and (\ref{eq:111}). This makes us to conjecture that the increase of the EE is nonlinear. While we cannot prove this conjecture, below we bound the time profile $S_A(t)$ from above. The ensuing upper bound has a time profile which is nonlinear.

To this end, we introduce an auxiliary evolving density of matrix for the subsystem A, $\hat\rho_{A,G}(t)$, such that it is Gaussian and gives a one-particle correlation function which is identical to $a^2M_{{\bf r_i}{\bf r}_j}(t)\equiv \left({\cal M}_{N_A}(t)\right)_{ij}$ at any $t$ and relaxes to $\left({\cal M}_{N_A}\right)_{ij}$ defined by Eq.~(\ref{eq:148}) at the time scale of $t_E$. For this density of matrix we can define the entropy in the same way as Eq.~(\ref{eq:30}),
\begin{equation}\label{eq:164}
    S_{A,G}(t)\equiv-{\rm Tr}_A\left(\hat\rho_{A,G}(t)\ln\hat\rho_{A,G}(t)\right),
\end{equation}
which can be readily found to be
\begin{eqnarray}\label{eq:166}
  S_{A,G}(t)&=&-{\rm Tr}_{A}\Big(\hat{{\cal M}}_{N_A}(t)\ln \hat{{\cal M}}_{N_A}(t)\nonumber\\
  &+&(1-\hat{{\cal M}}_{N_A}(t))\ln (1-\hat{{\cal M}}_{N_A})(t)\Big),
\end{eqnarray}
similar to Eq.~(\ref{eq:71}). For the two density of matrices, $\hat\rho_{A,G}(t)$ and $\hat\rho_{A}(t)$, one can define the {\it relative entropy} \cite{Vedral02}. By noting that the latter cannot be negative, one can readily show \cite{Hackl18} that
\begin{equation}\label{eq:165}
    S_{A}(t)\leq S_{A,G}(t).
\end{equation}
Combining Eqs.~(\ref{eq:166}) and (\ref{eq:165}), we establish the following bound for the evolving EE,
\begin{eqnarray}\label{eq:168}
  S_{A}(t)&\leq&-{\rm Tr}_{A}\Big(\hat{{\cal M}}_{N_A}(t)\ln \hat{{\cal M}}_{N_A}(t)\nonumber\\
  &+&(1-\hat{{\cal M}}_{N_A}(t))\ln (1-\hat{{\cal M}}_{N_A}(t))\Big).
\end{eqnarray}
Because of $\hat{{\cal M}}_{N_A}(t)\stackrel{t\gg t_E}{\longrightarrow} \hat{{\cal M}}_{N_A}$ this bound relaxes to the thermal EE given by Eq.~(\ref{eq:71}) at the time scale of $t_E$. So for long times this inequality is actually an equality. For short times it is easy to see that the bound has a nonlinear dependence on $t$.


\begin{table*}[t]
\centering
\caption{Comparison of different thermalization scenarios.}
\begin{tabular}{lllll}
\hline
\hline
&& Ensemble-based && Pure state-based \\
\hline
Target && Ensemble distribution && Quantum expectation of observable \\
Particle interaction && Direct && Exchange \\
Chaos && Many-body && One-body \\
Origin of {heat} && {Particle collision} && Eigenstate typicality \\
{Approach to thermal equilibrium} && {Ensemble distribution} && {Quantum expectation} \\
&& {$\rightarrow$ thermal distribution} && {$\rightarrow$ to thermal value} \\
Thermalization time && Particle interaction-dependent && Ehrenfest time \\
``The second law'' && Increase of entropy && {Dynamical generation of} entanglement \\
\hline\hline
\end{tabular}
\label{tab:1}
\end{table*}

\section{Generalizations to $\boldsymbol{F(0) \in \mathscr{P}'}$}
\label{sec:some_generalizations}

So far we have considered the initial state $F(0)\in\mathscr{P}$. In this section we study Problems 1, 2 and 3 in Sec.~\ref{sec:formulation} for $F(0)\in\mathscr{P}'$ and generalize some results in Secs.~\ref{sec:evolution_correlation_function} and \ref{sec:entanglement_entropy_evolution} to this case.

First of all, it is easy to check that the general formalism developed in Sec.~\ref{sec:formalism_one_particle_correlation} for calculating $M_{{\bf r}{\bf r}'}(t)$ still applies. The difference is the detailed form of the initial phase-space function $\mathfrak{M}_\varepsilon({\bf q},\vartheta;0)$. In particular, in the present case because of $F(0)\in \mathscr{H}_{S2}$ the majority of the weight $|C_{\bf m}|^2$ goes to atypical Fock state $\bf m$. As a result, $\mathfrak{M}_\varepsilon({\bf q},\vartheta;0)$ can have a quite arbitrary distribution over $\varepsilon$. Inheriting from this, the time scale for the quantum-classical correspondence associated to the motion at different (phase-space) energy shell, $t_\varepsilon$ defined in Eq.~(\ref{eq:13}), has a quite arbitrary distribution over $\varepsilon$ also. So the largest $t_\varepsilon$ depends on available $\varepsilon$ and is determined by the detailed constructions of $F(0)$, which is thereby denoted as $t_{F(0)}$. Similar to the Ehrenfest time, $t_{F(0)}$ depends on $\hbar$ logarithmically and {is also much smaller than the Heisenberg time} $\hbar/\Delta_{\mathscr{F}_S}$. But the pre-logarithm factor and the classical action rescaling $\hbar$ in $t_{F(0)}$ are very different from those in the Ehrenfest time. Reproducing the analysis in Secs.~\ref{sec:short_time_correlation_function} and \ref{sec:relaxation_correlation_function}, we see that at given energy $\varepsilon$, the quantity: $\int d\vartheta e^{-\frac{i}{\hbar}({\bf r}-{\bf r}')\cdot {\bf p}({\bf q},\vartheta)}
\mathfrak{M}_\varepsilon({\bf q},\vartheta;t)$ relaxes at the time scale of $t_\varepsilon$. Since the evolutions at different $\varepsilon$ are independent, combining this result with Eq.~(\ref{eq:143}) we find that $M_{{\bf r}{\bf r}'}(t)$ relaxes at the time scale of $t_{F(0)}$. The relaxed value can be found by using the method in Sec.~\ref{sec:relaxation_value_correlation_function}, which is still given by the right-hand side of Eq.~(\ref{eq:17}). Thus we have
\begin{eqnarray}\label{eq:160}
    M_{{\bf r}{\bf r}'}(t)\stackrel{t\gg t_{F(0)}}{\longrightarrow}\sum_{{\bf m}\in {\mathscr{F}_S}}|C_{{\bf m}}|^2\sum_\nu n_\nu C_\nu({\bf r},{\bf r}').
\end{eqnarray}
However, in the present case Eq.~(\ref{eq:25}) no longer holds: $\sum_\nu n_\nu C_\nu({\bf r},{\bf r}')$ depends explicitly on atypical ${\bf m}$ instead. Thus the right-hand side of Eq.~(\ref{eq:160}) cannot be simplified and the relaxed value depends on $F(0)$, i.e., is athermal.

Moreover, by repeating the analysis of multi-particle correlation functions in Sec.~\ref{sec:j_particle_correlation}, we find that Eq.~(\ref{eq:124}) can be generalized to
\begin{eqnarray}\label{eq:161}
  M^{(j)}_{\{{\bf r}\}\{{\bf r}'\}}(t)\stackrel{t\gg t_{F(0)}}{\longrightarrow}\sum_{{\bf m}\in {\mathscr{F}_S}}|C_{{\bf m}}|^2 \quad\quad\quad\quad\quad\quad\quad\nonumber\\
  \times\sum_P\sigma(P)\prod_{k=1}^j
  \sum_\nu n_\nu C_\nu({\bf r}_{{\bf r}_k},{\bf r}'_{P(k)}),\quad\quad
\end{eqnarray}
with Eq.~(\ref{eq:160}) as its special case. Similar to the discussions on the latter, the second line of Eq.~(\ref{eq:161}) depends explicitly on atypical ${\bf m}$. Thus the multi-particle correlation functions relax, but their relaxed values are $F(0)$ dependent, namely, athermal.

Then, by repeating the analysis in Sec.~\ref{sec:entanglement_entropy_short_time}, we find that Eq.~(\ref{eq:161}) gives rise to the relaxation of the RDM at the time scale of $t_{F(0)}$. The relaxed value is given by Eqs.~(\ref{eq:60}) and (\ref{eq:63}). That is,
\begin{eqnarray}\label{eq:162}
    \hat\rho_A(t)\stackrel{t\gg t_{F(0)}}{\longrightarrow}
    \sum_{{\bf m}\in\mathscr{F}_S}|C_{\bf m}|^2\sum_{\{O_i\}}\langle {\bf m}|{\prod_{i}} O^\dagger_{i}|{\bf m}\rangle\prod_{i} O_{i}.\quad\,\,
\end{eqnarray}
Correspondingly,
\begin{equation}\label{eq:163}
    S_A(t)\stackrel{t\gg t_{F(0)}}{\longrightarrow}const..
\end{equation}
We observe that in Eq.~(\ref{eq:162}), $\sum_{\{O_i\}}\langle {\bf m}|{\prod_{i}} O^\dagger_{i}|{\bf m}\rangle\prod_{i} O_{i}$ is a Gaussian state at given ${\bf m}$. However, such Gaussian state is very sensitive to ${\bf m}$, because in the present case ${\bf m}$ is atypical and $\langle {\bf m}|{\prod_{i}} O^\dagger_{i}|{\bf m}\rangle$ is very sensitive to ${\bf m}$. As the relaxed RDM is the algebraic mean of Gaussian states at distinct ${\bf m}$, it must be non-Gaussian, unlike the case with $F(0)\in \mathscr{P}$. So the relaxed RDM and thereby EE are $F(0)$ dependent. Note that because the relaxed RDM is non-Gaussian the calculations of the relaxed EE in Secs.~\ref{sec:entanglement_entropy_long_time_special} and \ref{sec:entanglement_entropy_long_time_general} cannot be generalized to the present case.

The results above show that for $F(0)\in \mathscr{P}'$, the ideal Fermi gas equilibrates at the time scale of $t_{F(0)}$, but is not thermalized.

\section{Comparison with canonical paradigm of thermalization}
\label{sec:comparison_standard_statistical_physics}

Standard statistical physics is built upon the ensemble distribution, which follows the Liouville equation (for a classical ensemble) or the von Neumann equation (for a quantum ensemble, namely, a mixed state). Bogoliubov postulated that after irregular transient processes, many-body chaos leads the dynamics of ensemble distribution to exhibit certain ``regularity'' \cite{Bogoliubov46}. That is, all many-particle distributions given by the Liouville or von Neumann equation are determined completely by the single-particle distribution given by the (generalized) Boltzmann equation \cite{Balescu75}. Then, following the latter equation, the single-particle distribution relaxes to a thermal (e.g., Maxwell-Boltzmann, Fermi-Dirac or Bose-Einstein) distribution, that in turn gives rise to the relaxation of all many-particle distributions. {As a consequence of this relaxation, a macroscopic observable, which is the average of a microscopic quantity with respect to the distribution function, {approaches} its thermal value. As such, the heat is generated by particle collisions.}

The results obtained in Secs.~\ref{sec:evolution_correlation_function} and \ref{sec:entanglement_entropy_evolution} enforce a completely different scenario for thermalization in many-particle systems (cf.~Table~\ref{tab:1} for comparison). Its building block is the quantum expectation value of observable at an evolving pure state $F(t)$, which is given by the momentum average of certain microscopic quantities with respect to the phase-space function $\mathfrak{M}^{(j)}(\{{\bf q}\},\{{\bf p}\};t)$. {(Note that this is {\it not} a probability distribution, because it can be negative.)} The latter evolves following the Moyal equation, whose classical limit is the Liouville equation. However, the Moyal equation does {\it not} {give rise to} the relaxation of $\mathfrak{M}^{(j)}(\{{\bf q}\},\{{\bf p}\};t)${. (At least we cannot prove this.)} Rather, the quantum expectation of observable relaxes at the time scale of $t_E$, and the relaxed value is thermal. In the meanwhile, the system evolves from a semiclassical state with low-level entanglement to a quantum state with high-level entanglement. This reflects the quantum origin of the emergent thermal equilibrium, and may be regarded as an analog of the second law.

In the new scenario, the roles of the exchange interaction, one-body chaos and (appropriately chosen) observables are ``additive''.
The exchange interaction creates coherence between particles at different energies, and embeds a virtual ``heat bath'' into the initial state {--- in this sense similar to a physical picture \cite{Rigol08} for numerical experiments on the eigenstate thermalization ---} via superposition by typical Fock states{; in fact, it has been shown \cite{Tian18} that without this interaction this virtual heat bath cannot arise}. The one-body chaos leads to the relaxation of the Fermi gas, after which the observable detects the heat bath. In contrast, for the initial state in $\mathscr{P}'$ the gas still relaxes, but the observable does not detect a heat bath, i.e., does not {approach a} thermal value, because the initial state is superposed by atypical Fock states.

\section{Concluding remarks}
\label{sec:conclusions_discussions}

In this work, we {analytically} studied the dynamics of the indistinguishability-{induced} entanglement of a truly ideal Fermi gas confined in a chaotic cavity, where the constituting particles have no direct interaction, but are subjected to the exchange interaction only. We found that the quantum-classical correspondence breakdown {of particle motion, via dramatically changing the spatial structure of many-body wavefunction,} has far-reaching impacts on the entanglement structure. In particular, it brings a semiclassical state with low-level entanglement to a quantum state with high-level entanglement. Moreover, for the class of initial states $\mathscr{P}$, this evolution of the entanglement structure gives rise to quantum thermalization of the {entire} Fermi gas {in the cavity. Various particle correlation functions at different spatial scales, which probe the global entanglement structure and thermal properties, and the RDM and the EE, which probe the entanglement between a subsystem and its complement and local thermal properties, all level} off at the thermal equilibrium value at the quantum state; for the class of initial states $\mathscr{P}'$, the evolution gives rise to equilibration of the Fermi gas, but not thermalization. {We should emphasize that global thermal equilibrium or thermal equilibrium of the entire system is more or less in the sense of von Neumann \cite{von Neumann29}, since it is diagnosed via the quantum expectation values of all multi-particle correlation functions at different spatial scales; whereas thermal equilibrium of subsystems deep inside the bulk is in the sense of standard statistical physics, since the RDM describes a genuine thermal ensemble, although the temperature and the chemical potential are determined by properties of the pure state describing the entire gas. In this sense, our findings provide a firm support to the conjecture of Garrison and Grover \cite{Grover18}.} Our findings suggest that the particle indistinguishability can lead to rich dynamical behaviors of quantum entanglement. They also shed new light on the foundational issue of statistical physics, namely, the emergence of thermal phenomena in an isolated system. In particular, they provide a new scenario for the emergence of thermal equilibrium phenomena from the pure-state evolution, with the exchange interaction and one-body chaos as {the key} components.

Many problems are open. Among the prominent ones are the following. First, it is obvious that the present results do not apply to a regular cavity, where the single-particle {motion} is integrable. However, kinematic studies \cite{Lai15,Tian18} have shown that the Fermi-Dirac distribution can still emerge from a typical Fock state of {such systems. Thus the modifications of entanglement dynamics are of fundamental interests}. Second, if the {weight of atypical Fock states is significant} initially, quantum thermalization does not occur. In this case, we conjecture that a weak direct interaction might be sufficient for all entanglement probes to approach thermal values at long time, and the process includes two stages. At the first stage, the weak direct interaction drives the state to the set $\mathscr{P}$ ({this process is similar to what is discussed in Appendix \ref{sec:discussions_experiment_initial_state}, and is likely a manifestation of eigenstate thermalization}); at the second stage, the direct interaction is negligible and the present results apply. Third, in this work we have focused on fermionic systems, it is {a natural problem} to generalize the present results to bosonic systems. We leave these problems for the future studies.

\section*{Acknowledgements}

We are grateful to G. Casati, S. Fishman, J. C. Garreau, I. Guarneri, D. Huse, H.-H. Lai, A. Polkovnikov and J. Wang for inspiring discussions at various stages of this work{, to P. Fang for preparing Fig.~\ref{fig:4}, and to F.-L. Lin and H.-J. Wang for reading the manuscript.} This work is supported by the National Natural Science Foundation of China (Grants No. 11925507
and No. 12047503) and the National Science Foundation (Grants No. DMR-1932796 and No. DMR-1644779).

\appendix

\section{Estimation of the quantum recurrence time}
\label{sec:quantum_recurrence_time}

In this appendix we discuss the quantum recurrence phenomenon and estimate the recurrence time $t_{rec}$ by generalizing the method of Ref.~\cite{Peres82}. When this phenomenon occurs $\hat M(t)$ is arbitrarily close to its initial (matrix) value, i.e.,
\begin{equation}\label{eq:S53}
    \|\hat M(t)-\hat M(0)\|\leq \epsilon,
\end{equation}
where $\epsilon$ is an arbitrarily small positive number. With the substitution of the explicit expression of $M_{{\bf r}{\bf r}'}(t)$ [see Eq.~(\ref{eq:S17}) below], we reduce this inequality to
\begin{eqnarray}\label{eq:S54}
    \|\hat M(t)-\hat M(0)\|&=&4\sum_{\nu\neq\nu'}|C_{\nu\nu'}|^2\sin^2\frac{(\varepsilon_\nu-\varepsilon_{\nu'})t}{2\hbar}\nonumber\\
    &\equiv& 4I(t)\leq \epsilon.
\end{eqnarray}
In order for this inequality to be met, it is required that all phases, $\frac{(\varepsilon_\nu-\varepsilon_{\nu'})t}{2\hbar}$, are close to multiple $\pi$. That is, there exist a set: $\{k_{\nu\nu'}\in \mathbb{Z}\}$ such that
\begin{eqnarray}
\sum_{(\nu,\nu')\in D(F)}|C_{\nu\nu'}|^2\left(\frac{(\varepsilon_\nu-\varepsilon_{\nu'})t}{2\pi\hbar}-k_{\nu\nu'}\right)^2\leq \left(\frac{\sqrt{\epsilon}}{2\pi}\right)^2,\quad\quad
\label{eq:S56}
\end{eqnarray}
where the set $D(F)$ depends on the initial state $F(0)$ and is defined as
\begin{equation}\label{eq:S68}
    D(F)\equiv\{(\nu,\nu')|C_{\nu\nu'}\neq 0,\nu\neq\nu'\}.
\end{equation}
Then $t_{rec}$ is the smallest $t(>0)$ when the inequality (\ref{eq:S56}) has a solution $\{k_{\nu\nu'}\}$ distinct from that at $t=0$. We call such $\{k_{\nu\nu'}\}$ a nontrivial solution.

Thanks to
\begin{eqnarray}\label{eq:S63}
    I(t)\geq |C|_{min}^2\sum_{(\nu,\nu')\in D(F)}\sin^2\frac{(\varepsilon_\nu-\varepsilon_{\nu'})t}{2\hbar},
\end{eqnarray}
where $|C|_{min}\equiv \min_{(\nu,\nu')\in D(F)}|C_{\nu\nu'}|$, the smallest time $t_-$ for the inequality,
\begin{eqnarray}
|C|_{min}^2\sum_{(\nu,\nu')\in D(F)}\left(\frac{(\varepsilon_\nu-\varepsilon_{\nu'})t}{2\pi\hbar}-k_{\nu\nu'}\right)^2\leq \left(\frac{\sqrt{\epsilon}}{2\pi}\right)^2,\quad\quad
\label{eq:S65}
\end{eqnarray}
to have a nontrivial solution $\{k_{\nu\nu'}\}$ bounds $t_{rec}$ from below, i.e., $t_{rec} \geq t_-$.

To find an explicit expression of $t_-$ we note that the equal sign of the inequality (\ref{eq:S65}) defines a sphere in $d_F$-dimensional space, where $d_F$ is the number of the elements in $D(F)$, with a moving center whose coordinate is $\{\frac{(\varepsilon_\nu-\varepsilon_{\nu'})t}{2\pi\hbar}\}$. The radius of the sphere is $\frac{\sqrt{\epsilon}}{2\pi|C|_{min}}$ and the cross section area
\begin{equation}\label{eq:S58}
    \sigma_-\equiv \pi^{\frac{d_F-1}{2}}\left(\frac{\sqrt{\epsilon}}{2\pi|C|_{min}}\right)^{d_F-1}{\bigg /}\Gamma\left(\frac{d_F+1}{2}\right),
\end{equation}
where $\Gamma(x)$ is the gamma function. As the sphere moves the cross section transverse to the velocity sweeps a cylinder. At $t=t_-$ this cylinder for the first time includes a lattice point. This gives
\begin{equation}\label{eq:S59}
d_F^{1/2}\frac{\Delta\varepsilon}{2\pi\hbar}\sigma_- t_-=1,
\end{equation}
where $\Delta\varepsilon$ is the mean squared value of $\left(\varepsilon_\nu-\varepsilon_{\nu'}\right)$. As a result,
\begin{equation}\label{eq:S60}
    t_- = \frac{2\pi\hbar}{\Delta \varepsilon} \frac{1}{d_F^{1/2}}\left(\frac{4\pi|C|_{min}^2}{\epsilon}\right)^{\frac{d_F-1}{2}}\Gamma\left(\frac{d_F+1}{2}\right).
\end{equation}

Thanks to
\begin{eqnarray}\label{eq:S43}
    I(t)\leq |C|_{max}^2\sum_{(\nu,\nu')\in D(F)}\sin^2\frac{(\varepsilon_\nu-\varepsilon_{\nu'})t}{2\hbar},
\end{eqnarray}
where $|C|_{max}\equiv \max_{(\nu,\nu')\in D(F)}|C_{\nu\nu'}|$, the smallest time $t_+$ for the inequality,
\begin{eqnarray}
|C|_{max}^2\sum_{(\nu,\nu')\in D(F)}\left(\frac{(\varepsilon_\nu-\varepsilon_{\nu'})t}{2\pi\hbar}-k_{\nu\nu'}\right)^2\leq \left(\frac{\sqrt{\epsilon}}{2\pi}\right)^2,\quad\quad
\label{eq:S44}
\end{eqnarray}
to have a nontrivial solution $\{k_{\nu\nu'}\}$ bounds $t_{rec}$ from above, i.e., $t_{rec} \leq t_+$. The procedures of calculating $t_+$ are the same as those of $t_-$. The result is
\begin{equation}\label{eq:S55}
    t_+ = \frac{2\pi\hbar}{\Delta \varepsilon} \frac{1}{d_F^{1/2}}\left(\frac{4\pi|C|_{max}^2}{\epsilon}\right)^{\frac{d_F-1}{2}}\Gamma\left(\frac{d_F+1}{2}\right).
\end{equation}

Combining Eqs.~(\ref{eq:S60}) and (\ref{eq:S55}), we find that interestingly, in the limiting case of $d_F=1$, $t_-=t_+$ and therefore $t_{rec}$ coincides with the Heisenberg time,
\begin{equation}\label{eq:S69}
    t_{rec}=\frac{2\pi\hbar}{\Delta\varepsilon},\quad for\,\, d_F=1.
\end{equation}
Note that in the semiclassical regime (i.e., $\hbar/A\ll 1$) the Heisenberg time $\sim \frac{\hbar}{\Delta\varepsilon}$ is much larger than $t_E$. As $d_F$ increases $t_{rec}$ grows very fast. For $d_F\gg 1$, Eqs.~(\ref{eq:S60}) and (\ref{eq:S55}) give
\begin{eqnarray}\label{eq:S67}
    &&\frac{4\pi^{3/2}\hbar}{\Delta \varepsilon} \frac{1}{{d_F}}\left(\frac{2\pi|C|_{min}^2d_F}{e\epsilon}\right)^{\frac{d_F}{2}}\nonumber\\
    &&\leq t_{rec} \leq \frac{4\pi^{3/2}\hbar}{\Delta \varepsilon} \frac{1}{{d_F}}\left(\frac{2\pi|C|_{max}^2d_F}{e\epsilon}\right)^{\frac{d_F}{2}}.
\end{eqnarray}
According to this, for large $d_F$ a very large $t_{rec}$ results. So the quantum recurrence phenomenon can be ignored practically.

\begin{figure}[t]
\includegraphics[width=8.6cm]{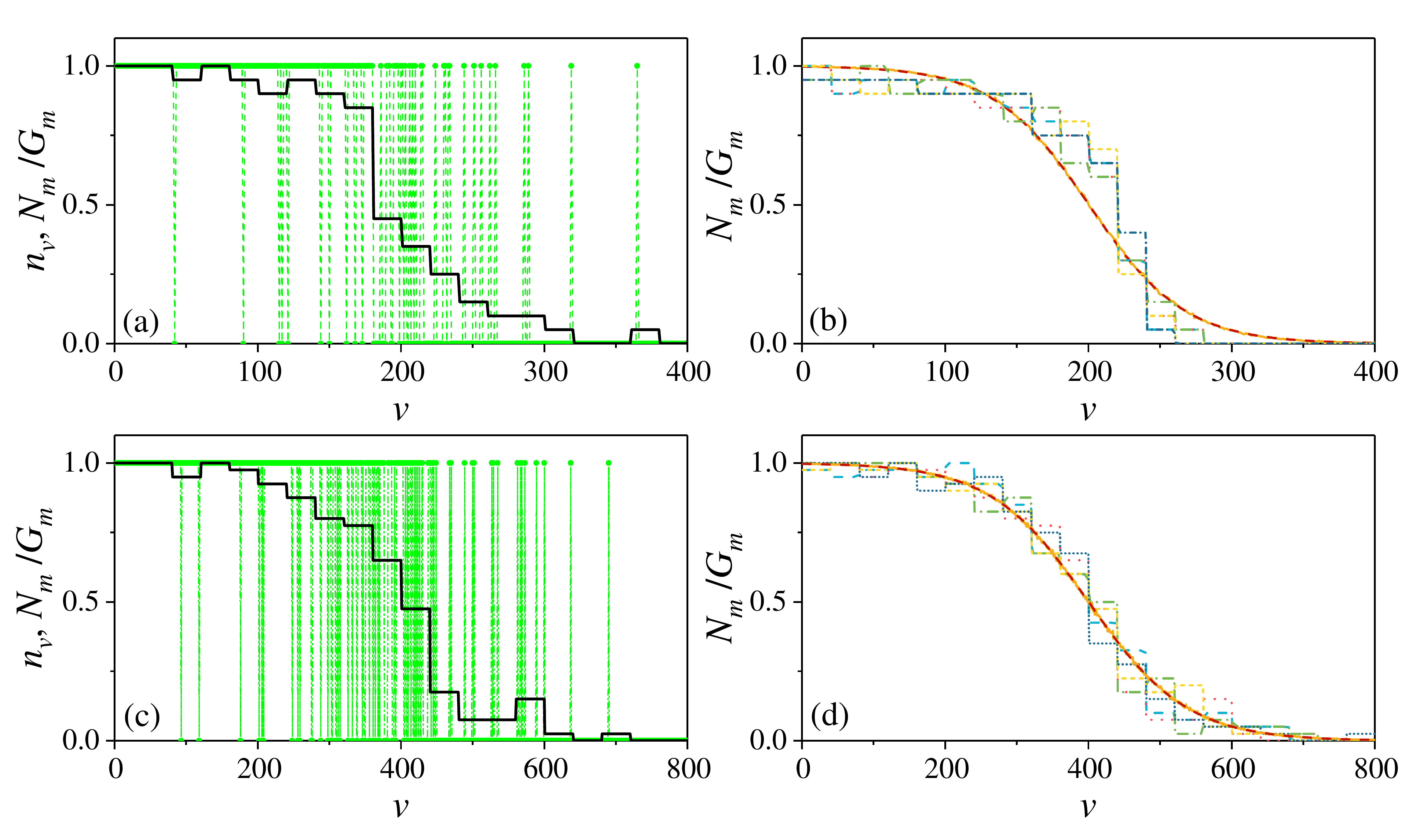}
\caption{{Results of numerical experiments on the integer partition. (a) A typical partition (green dashed line) and its observable-resolved structure namely the pattern $\{N_m/G_m\}$ (black solid line) are shown ($E=21900,N=200,G_m=20$). (b) $5$ patterns of $\{N_m/G_m\}$, which correspond to different partitions randomly drawn from the uniform probability measure and presented by different colors, all concentrate on a smooth curve (red dashed line), which is the average of $10000$ random partitions and is well fitted by the Fermi-Dirac distribution (yellow solid line): $n_{FD}(\nu)=\frac{1}{e^{\frac{\nu-\mu}{T}}+1}$ with fitting parameters $T=33.42,\mu=200.4$. (c,d) The same as (a,b), with $E=87800,N=400,G_m=40$ and fitting parameters $T=68.6,\mu=400.1$.}}
\label{fig:4}
\end{figure}

{\section{The observable-resolved structure in a simple system: simulation and mathematical results}
\label{sec:observable_resolved_Structure_fermionic_oscillator}}

{In this appendix, we review some simulation and rigorous mathematical results about the eigenstate typicality in a simple integrable many-particle system, where a large number of indistinguishable fermions are put in a harmonic oscillator \cite{Tian18}. The purpose is to help the readers to develop more intuitions about how the observable-resolved structure $\Lambda({\bf m})$, which differs dramatically from its parent structure, namely, the occupation number pattern $\{n_\nu\}$, gives rise to the emergence of the Fermi-Dirac distribution at the level of an individual Fock state ${\bf m}$, namely, many-body eigenstate of this system.}

{For this system the single-particle eigenstate has only one good quantum number, namely, the energy. Since the zero energy of the harmonic oscillator does not play any roles for present discussions, we shall ignore it henceforth. Then, with appropriate rescaling the single-particle eigenenergies are $\nu=1,2,\cdots$. For a many-body eigenstate ${\bf m}=\{n_\nu\}$,
we have
\begin{equation}\label{eq:193}
  \sum_{\nu=1}^\infty \nu n_\nu=E,\quad \sum_{\nu=1}^\infty n_\nu=N,\quad n_\nu=0,1.
\end{equation}
Interestingly, this maps the present problem to a celebrated problem in number theory, which is a partition of integer $E$ into $N$ distinct summands \cite{Andrews76}. In particular, the structure of the occupation number pattern $\{n_\nu\}$, when translated into the number theory language, is central to the so-called {\it random integer partition}, the studies of which were pioneered by Erd\"os and Lehner \cite{Erdoes41}. We shall see below that (under some conditions), when the set of partitions is equipped with a uniform probability measure, a typical random integer partition \cite{Vershik94,Vershik96,Vershik04} can have a limit shape, which is the very Fermi-Dirac distribution.}

{In Ref.~\cite{Tian18}, numerical experiments on the integer partition described by Eq.~(\ref{eq:193}) were performed for different $E,N$. The Monte Carlo method was used to draw randomly a partition from the uniform probability measure. The results are presented in Fig.~\ref{fig:4}. In panels (a) and (c), typical partitions $\{n_\nu\}$ (green dashed lines) are shown, which appear to be random. By dividing the natural set into subsets (labelled by $m$), each of which includes $G_m$ contiguous natural numbers, and counting the number $N_m=\sum_{\nu\in m}n_\nu$ in each subset, we obtain the pattern of $\{N_m/G_m\}$ (black solid lines). The latter gives the structure $\Lambda[{\bf m}]$. Panels (b) and (d) show that, for typical --- with respect to the uniform probability measure --- ${\bf m}$, their observable-resolved structure, namely, the pattern $\{N_m/G_m\}$, all concentrate on a smooth curve (red dashed lines), which is the average of $10000$ random partitions and is well fitted by the Fermi-Dirac distribution (yellow solid lines). Thus numerical experiment confirms the emergence of the Fermi-Dirac distribution from a typical ${\bf m}$.}

{In Ref.~\cite{Tian18}, the relations between the structure $\Lambda[{\bf m}]$ for this simple system and some rigorous mathematical results were uncovered. For simplicity we do not fix $N$, i.e., remove the second constraint in Eq.~(\ref{eq:193}). Consider the following one-body observable: $\hat{\mathfrak{N}}_u\equiv \sum_{\nu\geq u}a_\nu^\dagger a_\nu$, where $a_{\nu}$ ($a^\dagger_\nu$) being the annihilation (creation) operator at the single-particle eigenstate $\nu$. Its quantum expectation at ${\bf m}$ is
\begin{equation}\label{eq:194}
  \langle {\bf m}|\hat{\mathfrak{N}}_u|{\bf m}\rangle=\sum_{\nu\geq u} n_\nu\equiv \varphi_{\bf m}(u),
\end{equation}
which counts the number of summands $\geq u$ at given partition ${\bf m}$. This function, $\varphi_{\bf m}(u)$, is important in the studies of random integer partitions, and defines a random stepped curve \cite{Vershik94,Vershik96,Vershik04,Okounkov16}. According to Eq.~(\ref{eq:194}), the criterion Eq.~(\ref{eq:115}) is trivially satisfied by the observable $\hat{\mathfrak{N}}_u$. Therefore, we can use Eq.~(\ref{eq:120}) to obtain
\begin{equation}\label{eq:195}
  \varphi_{\bf m}(u)\stackrel{E\gg 1}{=}\int_{u}^{\infty}\frac{d\nu}{e^{\frac{\nu}{T}}+1}=T\ln(1+e^{-\frac{u}{T}}),\,T=\frac{\sqrt{12E}}{\pi}
\end{equation}
for an overwhelming number of ${\bf m}$, where $\mu=0$ because $N$ is not fixed. On the other hand, if the set of all partitions is equipped with a uniform probability measure $\mathbb{P}^E$, then Vershik's theorem \cite{Vershik96} follows,
\begin{equation}\label{eq:196}
  \forall \epsilon>0:\lim_{E\rightarrow\infty}\mathbb{P}^E\left\{{\bf m}:\left|\frac{1}{\sqrt{E}}\varphi_{\bf m}(\sqrt{E}u)+v(u)\right|<\epsilon\right\}=1.
\end{equation}
Here the function: $v(u)$ is defined through the Vershik curve,
\begin{equation}\label{eq:197}
  e^{-\frac{\pi v}{\sqrt{12}}}-e^{-\frac{\pi u}{\sqrt{12}}}=1.
\end{equation}
The theorem implies that, for a typical partition ${\bf m}$, the random stepped curve: $\frac{1}{\sqrt{E}}\varphi_{\bf m}(\sqrt{E}u)$ has a limit shape: $-v(u)$, i.e., the integrated Fermi-Dirac distribution. Equation (\ref{eq:195}) fully agrees with this theorem. Moreover, our numerical findings shown in Fig.~\ref{fig:4} (b) and (d) suggest that this theorem can be generalized to the case of large fixed $N$}.\\

\section{Further discussions on experimental preparation of initial states}
\label{sec:discussions_experiment_initial_state}

Let the state of the evolving interacting Fermi gas during the preparation of an initial state be $\tilde F(t)$. Note that throughout this appendix ${\tilde{t}}$ refers to the time in preparing the initial state, and should not be confused with $t$ used in other parts of this paper, which refers to the time in the evolution of ideal Fermi gas. $\tilde F({\tilde{t}})$ can be expanded in terms of Fock states as
\begin{equation}\label{eq:125}
  |\tilde F(\tilde{t})\rangle=\sum_{{\bf m}\in {\mathscr{F}_S}}\tilde C_{\bf m}(\tilde{t})|{\bf m}\rangle,
\end{equation}
where $\tilde C_{\bf m}(\tilde{t})$ are the evolving complex expansion coefficients. Note that, no matter whether the Fermi gas is interacting or noninteracting, the single-particle eigenstates used to construct the Fock states are the same, and so are the single-particle eigenenergies. It is important that because the particles have direct interaction, the Fock states ${\bf m}$ are no longer the many-body eigenstates of the gas. As such, $\tilde C_{\bf m}(\tilde{t})$ does not evolve in the way as that described by Eq.~(\ref{eq:S15}), but exhibits very complicated dynamical behaviors.

So for a specific $\tilde F(\tilde{t})$ we can introduce the average occupation number at given single-particle eigenstate $\nu$ or, equivalently, at single-particle eigenenergy $\varepsilon_\nu$, defined as
\begin{equation}\label{eq:127}
  \tilde N(\varepsilon_\nu,\tilde{t})\equiv \sum_{{\bf m}\in {\mathscr{F}_S}}|\tilde C_{\bf m}(\tilde{t})|^2 n_\nu,
\end{equation}
where the dependence of $n_\nu$ on ${\bf m}$ should be kept in mind. To study the dynamics of this quantity we note that the direct interaction between two particles creates a ``reaction'' in the (single-particle) spectral space, i.e.,
\begin{equation}\label{eq:128}
  \varepsilon_\nu + \varepsilon_{\nu'} \leftrightarrow (\varepsilon_\nu+\delta\varepsilon) + (\varepsilon_{\nu'}-\delta\varepsilon).
\end{equation}
That is, a two-particle state, with their eigenenergies being $\varepsilon_\nu$ and $\varepsilon_{\nu'}$, respectively, transits into another two-particle state, with their eigenenergies being $\varepsilon_\nu+\delta\varepsilon$ and $\varepsilon_{\nu'}-\delta\varepsilon$, respectively, and {\it vice versa}, where $\delta\varepsilon$ is the transferred energy. Because this system is chaotic, one may assume that transitions occurring at different times are independent. As a result, Eq.~(\ref{eq:128}) leads to the following equation satisfied by $\tilde N$,
\begin{eqnarray}
\label{eq:129}
\frac{\partial\tilde N(\varepsilon_\nu)}{\partial \tilde{t}}=\sum_{\delta\varepsilon}\sum_{\nu'}W(\delta\varepsilon)\quad\quad\quad\quad\quad\quad\quad\quad\quad\nonumber\\
\times\big(\tilde N(\varepsilon_\nu+\delta\varepsilon)\tilde N(\varepsilon_{\nu'}-\delta\varepsilon)(1-\tilde N(\varepsilon_\nu))(1-\tilde N(\varepsilon_{\nu'}))\quad\quad\nonumber\\
  -\tilde N(\varepsilon_\nu)\tilde N(\varepsilon_{\nu'})(1-\tilde N(\varepsilon_\nu+\delta\varepsilon))(1-\tilde N(\varepsilon_{\nu'}-\delta\varepsilon))\big),\,\,\,\,\quad
\end{eqnarray}
where $W(\delta\varepsilon)$ is the transition probability, and all the time arguments have been suppressed to make the formula compact.

By using Eq.~(\ref{eq:129}) it is easy to show that $\tilde N$ approaches the Fermi-Dirac distribution at long time, i.e.,
\begin{eqnarray}
\label{eq:130}
  \tilde N(\varepsilon_\nu,\tilde{t})\stackrel{\tilde{t}\rightarrow\infty}{\longrightarrow}n_{FD}(\varepsilon_\nu).
\end{eqnarray}
This implies that the majority of weight $|\tilde C_{\bf m}(\tilde{t}\rightarrow\infty)|^2$ goes to typical ${\bf m}$ and $\tilde F({\tilde t\rightarrow\infty})\in \mathscr{H}_{S1}$. In contrast, if the time $\tilde t$ is not sufficiently long, then the majority of weight $|\tilde C_{\bf m}(\tilde{t})|^2$ goes to atypical ${\bf m}$ and at this time $\tilde F({\tilde t})\in \mathscr{H}_{S2}$.

\section{Derivations of the von Neumann equation for correlation matrices}
\label{sec:derivations_von_Neumann}

Substituting Eq.~(\ref{eq:S15}) into Eq.~(\ref{eq:15}), we obtain
\begin{eqnarray}
\label{eq:S17}
  M_{{\bf r}{\bf r}'}(t) = \sum_{\nu'\nu}C_{\nu'\nu} \psi_{\nu'}({\bf r})\psi_\nu^*({\bf r}')e^{-i(\varepsilon_{\nu'}-\varepsilon_\nu)t/\hbar},
\end{eqnarray}
with the coefficient
\begin{eqnarray}
\label{eq:S19}
  C_{\nu'\nu}&=&\sum_{{{\bf m}}{{\bf m}}'}C_{{\bf m}}^*C_{{{\bf m}}'}\sqrt{n_\nu n'_{\nu'}}(-1)^{\sum_{\mu'<\nu'}n_{\mu'}-\sum_{\mu<\nu}n'_{\mu}}\nonumber\\
  &&\times \big((1-\delta_{\nu,{\nu}'})\delta_{n_\nu,n_{\nu}'+1}\delta_{n_{\nu'},n_{\nu'}'-1}\prod_{\mu\neq\nu,\nu'}\delta_{n_\mu,n_{\mu}'}\nonumber\\
  &&+ \delta_{\nu,{\nu}'}\prod_{\mu}\delta_{n_\mu,n_{\mu}'}\big).
\end{eqnarray}
Equation (\ref{eq:S17}) can be rewritten as
\begin{eqnarray}
\label{eq:6}
  M_{{\bf r}{\bf r}'}(t) = \langle {\bf r}|e^{-itH(\hat{{\bf q}},\hat{{\bf p}})/\hbar}\hat M(0)e^{itH(\hat{{\bf q}},\hat{{\bf p}})/\hbar}|{\bf r}'\rangle.
\end{eqnarray}
It is easy to check that this is the solution to Eq.~(\ref{eq:2}).

For a $j(\geq 2)$-particle correlation function we substitute Eq.~(\ref{eq:S15}) into Eq.~(\ref{eq:76}) to obtain
\begin{eqnarray}
\label{eq:77}
  &&M^{(j)}_{\{{\bf r}\}\{{\bf r}'\}}
  (t)=\sum_{\{\nu'_k\}\{\nu_k\}}C_{\{\nu'_k\}\{\nu_k\}}\nonumber\\
  &\times&\prod_{k=1}^j(\psi_{\nu'_k}({\bf r}_k)\psi_{\nu_k}^*({\bf r}'_k))e^{-i\sum_{k=1}^j(\varepsilon_{\nu'_k}-\varepsilon_{\nu_k})t/\hbar},
\end{eqnarray}
and the explicit expression of the coefficient $C_{\{\nu'_k\}\{\nu_k\}}$ is not important for present discussions. It can be rewritten as
\begin{eqnarray}
\label{eq:78}
  M^{(j)}_{\{{\bf r}\}\{{\bf r}'\}}
  (t)&=&\langle \{{\bf r}\}
  |e^{-\frac{it}{\hbar}\sum_{k=1}^j H(\hat{{\bf q}}_k,\hat{{\bf p}}_k)}\nonumber\\
  &\times&\hat M^{(j)}(0)e^{\frac{it}{\hbar}\sum_{k=1}^j H(\hat{{\bf q}}_k,\hat{{\bf p}}_k)}|\{{\bf r}'\}
  \rangle.\,\,\quad
\end{eqnarray}
It is easy to check that this is the solution to Eq.~(\ref{eq:79}).

\section{The autocorrelation of the single-particle eigenfunction}
\label{sec:S3}

In this appendix we derive the explicit form of $C_\nu({\bf r},{\bf r}')$. This was originally given in Ref.~\cite{Berry77} with some details of derivations missed. For the self-contained purpose here we give the detailed derivations for $d$-dimensional chaotic cavity. We denote the Wigner transformation of $\psi_\nu({\bf r})$ as $\Psi_{\nu}({\bf q},{\bf p})$. Since the classical single-particle motion is quantum chaotic, $\Psi_{\nu}({\bf q},{\bf p})$ is given by
\begin{equation}\label{eq:5}
    \Psi_{\nu}({\bf q},{\bf p})=\frac{\delta(\varepsilon-H({\bf q},{\bf p}))}{\int\!\!\int d{\bf q}d{\bf p}\delta(\varepsilon-H({\bf q},{\bf p}))}.
\end{equation}
Note that this microcanonical distribution is defined on the single-particle phase space. Using this result, we find
\begin{eqnarray}
C_\nu({\bf r},{\bf r}')=\frac{\int d{\bf \Omega}
e^{-\frac{i}{\lambda_{\varepsilon_\nu}}{\bf \Omega}\cdot({\bf r}-{\bf r}')}}{V\int d{\bf\Omega}}\equiv\frac{1}{V}f\left(\frac{|{\bf r}-{\bf r}'|}{\lambda_{\varepsilon_\nu}}\right).\quad\quad
\label{eq:112}
\end{eqnarray}
Here ${\bf\Omega}$ is the solid angle.

We proceed to calculate the ${\bf\Omega}$-integral. Without loss of generality we assume that ${\bf r}$ is in the $x_1$ direction, i.e., ${\bf r}=(r,0,\cdots,0)$. We parametrize the surface of a $d$-dimensional sphere, for which $\sum_{i=1}^d x_i^2=1$, by \cite{Prudnikoc}
\begin{eqnarray}
x_1&=&\cos\varphi_1,\nonumber\\
x_2&=&\sin\varphi_1\cos\varphi_2,\nonumber\\
x_3&=&\sin\varphi_1\sin\varphi_2\cos\varphi_3,\nonumber\\
&&\cdots\cdots\cdots\cdots\cdots\nonumber\\
x_{d-1}&=&\sin\varphi_1\sin\varphi_2\cdots\sin\varphi_{d-2}\cos\varphi_{d-1},\nonumber\\
x_{d}&=&\sin\varphi_1\sin\varphi_2\cdots\sin\varphi_{d-2}\sin\varphi_{d-1},
\label{eq:S7}
\end{eqnarray}
where
\begin{equation}\label{eq:S8}
    0\leq \varphi_1,\varphi_2,\cdots,\varphi_{d-2}\leq \pi,\quad 0\leq \varphi_{d-1}<2\pi.
\end{equation}
Correspondingly, the Jacobian
\begin{equation}\label{eq:S9}
    J=\sin^{d-2}\varphi_1\sin^{d-3}\varphi_2\cdots \sin^{2}\varphi_{d-3}\sin\varphi_{d-2}.
\end{equation}
Substituting Eqs.~(\ref{eq:S7})-(\ref{eq:S9}) into Eq.~(\ref{eq:112}) we obtain
\begin{eqnarray}
f\left(\frac{|{\bf r}-{\bf r}'|}{\lambda_\varepsilon}\right)
=\frac{\int_0^\pi d\varphi_1\sin^{d-2}\varphi_1e^{-i\frac{|{\bf r}-{\bf r}'|}{\lambda_\varepsilon}\cos \varphi_1}}{\int_0^\pi d\varphi_1\sin^{d-2}\varphi_1}.
\label{eq:S10}
\end{eqnarray}
To calculate the numerator of the right-hand side of Eq.~(\ref{eq:S10}) we use the Poisson integral expression for the Bessel function \cite{Prudnikoc},
\begin{equation}\label{eq:S11}
    J_{\bar{\nu}} (z)=\frac{\left(\frac{z}{2}\right)^{\bar{\nu}}}{\Gamma\left({\bar{\nu}}+\frac{1}{2}\right)\Gamma\left(\frac{1}{2}\right)}\int_0^\pi d\varphi \sin^{2{\bar{\nu}}}\varphi e^{iz\cos\varphi}
\end{equation}
for ${\rm Re}{\bar{\nu}} >-\frac{1}{2}$. To calculate the denominator we use the identity \cite{Gradshteyn},
\begin{equation}\label{eq:S12}
 \int_0^\pi d\varphi \sin^{{\bar{\nu}}-1}\varphi=\frac{\pi}{2^{{\bar{\nu}}-1}{\bar{\nu}} B\left(\frac{{\bar{\nu}}+1}{2},\frac{{\bar{\nu}}+1}{2}\right)}
\end{equation}
for ${\rm Re}{\bar{\nu}} >0$, where $B(x,y)$ is the beta function. By further using the identity,
\begin{equation}\label{eq:S13}
    B(x,x)=2^{1-2x}B(1/2,x)=2^{1-2x}\frac{\Gamma(x)\Gamma\left(\frac{1}{2}\right)}{\Gamma\left(x+\frac{1}{2}\right)},
\end{equation}
we obtain
\begin{equation}\label{eq:S1}
    f(x)=\Gamma\left(\frac{d}{2}\right)\left(\frac{x}{2}\right)^{-\frac{d-2}{2}}J_{\frac{d-2}{2}}(x).
\end{equation}
In the special case $d=2$, we have Eq.~(\ref{eq:26}).

\section{Some properties of the generating function $\boldsymbol{{\cal C}(\theta)}$}
\label{sec:generating_function}

In this appendix we derive several technical results for the generating function ${\cal C}(\theta)$. {Substituting the identity: $J_0(x)=\int_{-\pi}^\pi\frac{d\varphi}{2\pi}e^{-ix\sin\varphi}$ into Eq.~(\ref{eq:39})}, we obtain
\begin{eqnarray}\label{eq:40}
    {\cal C}(\theta)&=&\frac{a^2}{V}\int dm(\nu)n_{FD}(\varepsilon_\nu)\nonumber\\
    &\times&\left(\sum_{n\in \mathbb{Z}}\int_{-\pi}^\pi
    \frac{d\varphi}{2\pi}e^{in(\theta-\frac{a}{\lambda_{\varepsilon_\nu}}\sin\varphi)}\right).
\end{eqnarray}
From this
\begin{equation}\label{eq:46}
    {\cal C}(\theta)={\cal C}(-\theta)
\end{equation}
follows immediately.

With the help of the Poisson formula we rewrite Eq.~(\ref{eq:40}) as
\begin{eqnarray}\label{eq:41}
    {\cal C}(\theta)&=&\frac{a^2}{V}\int dm(\nu)n_{FD}(\varepsilon_\nu)\nonumber\\
    &\times&\left(\int_{-\pi}^\pi
    d\varphi\sum_{k\in \mathbb{Z}}\delta\left(\theta-\frac{a}{\lambda_{\varepsilon_\nu}}\sin\varphi-2\pi k\right)\right).
\end{eqnarray}
Performing the $\varphi$ integral we obtain Eq.~(\ref{eq:38}).

From Eq.~(\ref{eq:41}) it is obvious that ${\cal C}(\theta)\geq 0$. Now we wish to bound ${\cal C}(\theta)$ from above. We introduce the {\it essential upper bound}, defined as the least number $M$ for which the inequality: ${\cal C}(\theta)\leq M$ holds with the exception of a set of zero Lebesgue measure. Consider an arbitrary complex vector $\phi=\{\phi_j\}=\sum_{i=1}^{N_A}u_ie_i$, where $e_i$ are the mutually orthonormal eigenvectors of $\hat {\cal M}_{N_A}$ and $u_i\in \mathbb{C}$ are the expansion coefficients. Since the eigenvalues of $\hat {\cal M}_{N_A}$ are $\frac{1+v_i}{2}\in [0,1]$ and cannot all vanish, we have
\begin{eqnarray}
\label{eq:54}
  \frac{\phi^\dagger \hat {\cal M}_{N_A}\phi}{\phi^\dagger \phi} &=& \frac{\sum_{i=1}^{N_A}\frac{1+v_i}{2}|u_i|^2}{\sum_{i=1}^{N_A}|u_i|^2}
  \leq \lambda_{max}\frac{\sum_{i=1}^{N_A}|u_i|^2}{\sum_{i=1}^{N_A}|u_i|^2}\nonumber\\
  &=& \lambda_{max}(N_A),
\end{eqnarray}
where $\lambda_{max}(N_A)={\rm max}(\frac{1+v_1}{2},\cdots \frac{1+v_{N_A}}{2})$ and the equal sign is taken if and only if $\phi$ is in the direction of $e_i$ corresponding to $\lambda_{max}$. On the other hand, we have
\begin{eqnarray}
\label{eq:55}
  \frac{\phi^\dagger \hat {\cal M}_{N_A}\phi}{\phi^\dagger \phi} &=& \frac{\int_{-\pi}^\pi\frac{d\theta}{2\pi}|\sum_{j=1}^{N_A}\phi_je^{ij\theta}|^2{\cal C}(\theta)}{\phi^\dagger\phi}\nonumber\\
  &\leq& M\frac{\int_{-\pi}^\pi\frac{d\theta}{2\pi}|\sum_{j=1}^{N_A}\phi_je^{ij\theta}|^2}{\phi^\dagger\phi},\quad \forall \phi,
\end{eqnarray}
where in the first step we have used Eq.~(\ref{eq:37}) and in the second step we have used the definition of $M$. Since the maximal value of the left-hand side of the inequality (\ref{eq:55}) is $\lambda_{max}(N_A)$, we have $\lambda_{max}(N_A)\leq M$. In fact, a stronger result exists. By Weyl's theory of {\it equal distributions} \cite{Grenander53}, the meaning of which will be exposed in Appendix \ref{sec:generating_function_two_dimension} via a concrete example, we have
\begin{equation}\label{eq:102}
    \lim_{N_A\rightarrow\infty}\lambda_{max}(N_A)=M.
\end{equation}
From this it follows that
\begin{equation}\label{eq:57}
    {\cal C}(\theta)\leq 1,
\end{equation}
with exceptions at most constituting a set of zero Lebesgue measure.


\section{Some properties of the generating function $\boldsymbol{{\cal C}(\theta_1,\theta_2)}$}
\label{sec:generating_function_two_dimension}

In this appendix we derive several technical results for the generating function ${\cal C}(\theta_1,\theta_2)$.

For quantum particles moving in a $2$D chaotic cavity, $dm(\nu){\approx}\rho d\varepsilon$, with $\rho$ being a constant and $d\varepsilon$ being the Lebesgue measure of the energy axis. Combining this with Eq.~(\ref{eq:90}), we find that the Fourier component of ${\cal C}(\theta_1,\theta_2)$ is
\begin{eqnarray}
c_n=\frac{a^2\rho}{V}\int_0^\infty d\varepsilon J_0\left(\frac{a}{\lambda_{\varepsilon}}\sqrt{n_1^2+n_2^2}\right)n_{FD}(\varepsilon).
\label{eq:93}
\end{eqnarray}
Recall that $n\equiv (n_1,n_2)\in \mathbb{Z}^2$. Below we prove for this $c_n$ the following result:\\

{\it Lemma.} For finite temperature $T>0$ and $n_1^2+n_2^2\gg \frac{2\hbar^2}{ma^2T}$, we have
\begin{equation}\label{eq:94}
  c_n\approx\frac{a^2\rho T}{V}e^{\frac{\mu}{T}}e^{-\frac{ma^2T}{2\hbar^2}(n_1^2+n_2^2)}.
\end{equation}

{\it Proof.} Case I: $\mu>0$. This corresponds to a sufficiently low $T$. Upon rescaling and changing the integral variable, we can rewrite Eq.~(\ref{eq:93}) as
\begin{eqnarray}
c_n=\frac{a^2\rho T}{V}\int_0^1 dx J_0\left(\sqrt{-\tilde T\ln x}\right)\frac{1}{x+e^{-\frac{\mu}{T}}},
\label{eq:96}
\end{eqnarray}
where $\tilde T=2(n_1^2+n_2^2)ma^2T/\hbar^2$. Note that when $e^{-\frac{\mu}{T}}$ in the denominator vanishes, this integral diverges. Therefore, the integral in Eq.~(\ref{eq:96}) is dominated by $x$ near zero. So we further change the integral variable: $u\equiv \sqrt{-\ln x}$ to rewrite Eq.~(\ref{eq:96}) as
\begin{eqnarray}
c_n=-\frac{a^2\rho T}{V}\int_0^\infty d\ln (1+e^{-u^2+\frac{\mu}{T}}) J_0\left(\sqrt{\tilde T}u\right).
\label{eq:97}
\end{eqnarray}
Since the integral is dominated by large $u$ we can expand the logarithm and keep the leading term, which gives
\begin{eqnarray}
c_n\approx\frac{2a^2\rho T}{V}e^{\frac{\mu}{T}}\int_0^\infty du e^{-u^2}u J_0\left(\sqrt{\tilde T}u\right).
\label{eq:98}
\end{eqnarray}
By performing the integral \cite{Gradshteyn} we obtain Eq.~(\ref{eq:94}).

Case II: $\mu<0$. This corresponds to a sufficiently high $T$. The proof above can be generalized to this case straightforwardly. In fact, for very high $T$ we can derive Eq.~(\ref{eq:94}) in a simpler way. In this special high-$T$ case, the Fermi-Dirac distribution can be replaced by the Maxwell-Boltzmann distribution. With this replacement Eq.~(\ref{eq:93}) reduces to Eq.~(\ref{eq:98}) after rescaling and changing the integral variable. $\Box$

From this lemma, immediately, we have the following result for $\tilde c_n$, which is the Fourier component of $\tilde {\cal C}(\theta_1,\theta_2)$ defined by Eq.~(\ref{eq:90}):\\

{\it Theorem 1.} Let $|n|\equiv|n_1|+|n_2|$. Then
\begin{equation}\label{eq:95}
  \sum_{n_1,n_2\in \mathbb{Z}}(|\tilde c_n|+|n||\tilde c_n|^2)<\infty.
\end{equation}

Note that in the zero temperature case, Eq.~(\ref{eq:93}) gives $c_n=\frac{2a^2\rho}{V}\sqrt{\frac{1}{\tilde \mu}}J_1(\sqrt{\tilde \mu})$, where $\tilde \mu=2(n_1^2+n_2^2)ma^2\mu/\hbar^2$. With its substitution we find that the left-hand side of Eq.~(\ref{eq:95}) diverges. This implies that the relaxed value of the EE at $T=0$ behaves in a different way, as discussed in Sec.~\ref{sec:method_I}.

Now we return to finite temperature. Obviously, both the range of $\tilde {\cal C}(\theta_1,\theta_2)$ and the eigenvalues of the Toeplitz operator: $(\lambda+1)\mathbb{I}_{N_A}-2\hat {\cal M}_{N_A}$ reside in the real axis of the complex plane. Let us define their union as $\Lambda$. Then the following result is obvious:\\

{\it Theorem 2.} In the complex plane a path exists, which goes from $0$ to $\infty$ and does not intersect $\Lambda$.\\

Without loss of generality one may choose the path which does not pass $1$. Then, the logarithmic function involved in Doktorsky's theorem or Eq.~(\ref{eq:91}) is defined in the way so that it is analytic in the complex plane cut along this path and takes the value of zero at $1$.

Theorems 1 and 2 justify all conditions required by Doktorsky's theorem \cite{Doktorsky84}.

In the following we wish to bound ${\cal C}(\theta_1,\theta_2)$. By the identity \cite{Gradshteyn},
\begin{eqnarray}
\label{eq:99}
  J_0\left(\frac{a}{\lambda_{\varepsilon_\nu}}
  \sqrt{n_1^2+n_2^2}\right)&=&\frac{1}{2}\int_{-\pi}^\pi \frac{d\varphi}{2\pi} \Big(e^{i\frac{a}{\lambda_{\varepsilon_\nu}}(n_1\cos\varphi+n_2\sin\varphi)}\nonumber\\
  &+&e^{i\frac{a}{\lambda_{\varepsilon_\nu}}(n_1\cos\varphi-n_2\sin\varphi)}\Big),
\end{eqnarray}
we can rewrite ${\cal C}(\theta_1,\theta_2)$ given in Eq.~(\ref{eq:90}) as
\begin{eqnarray}\label{eq:100}
    {\cal C}(\theta_1,\theta_2)&=&\frac{a^2}{2V}\int dm(\nu)n_{FD}(\varepsilon_\nu)\int_{-\pi}^\pi\frac{d\varphi}{2\pi}\nonumber\\
    &\times&\sum_{n_{1,2}\in \mathbb{Z}}e^{i(n_1\theta_1+n_2\theta_2)}
    \Big(e^{i\frac{a}{\lambda_{\varepsilon_\nu}}(n_1\cos\varphi+n_2\sin\varphi)}\nonumber\\
  &+&e^{i\frac{a}{\lambda_{\varepsilon_\nu}}(n_1\cos\varphi-n_2\sin\varphi)}\Big).
\end{eqnarray}
Applying the Poisson formula to the summation, we obtain
\begin{eqnarray}\label{eq:101}
    {\cal C}(\theta_1,\theta_2)=\frac{\pi a^2}{V}\int dm(\nu)n_{FD}(\varepsilon_\nu)\int_{-\pi}^\pi d\varphi\sum_{s=\pm 1}\sum_{k_{1,2}\in \mathbb{Z}}\nonumber\\
    \delta(\theta_1+\frac{a}{\lambda_{\varepsilon_\nu}}\cos\varphi-2\pi k_1)\delta(\theta_2+s\frac{a}{\lambda_{\varepsilon_\nu}}\sin\varphi-2\pi k_2).\quad\quad
\end{eqnarray}
This shows that ${\cal C}(\theta_1,\theta_2)\geq 0$. In fact, the {\it essential lower bound}, which is defined in the way similar to the essential upper bound, of ${\cal C}(\theta_1,\theta_2)$ is zero.

To bound ${\cal C}(\theta_1,\theta_2)$ from above, we need the following theorem, which is a straightforward generalization of a classical result for the ordinary Toeplitz matrix \cite{Grenander53} to the Toeplitz matrix defined on a square of side length $\sqrt{N_A}$ and can be proven by replacing Szeg{\"o}'s theorem in the proof of that result by Doktorsky's theorem:\\

{\it Theorem 3.} Let $\{\lambda_i(N_A)\}$ be the eigenvalue spectrum of the Toeplitz matrix $\hat {\cal M}_{N_A}$ defined on a square of side length $\sqrt{N_A}$, and ${\cal C}(\theta_1,\theta_2)$ be integrable and its essential upper bound $M$ be finite. If ${\cal F}(\lambda)$ is any continuous function defined in the interval $[0, M]$, then
\begin{equation}\label{eq:103}
  \lim_{N_A\rightarrow \infty}\frac{\sum_{i=1}^{N_A}{\cal F}(\lambda_i(N_A))}{N_A}=\int\!\!\!\!\int_{-\pi}^\pi\frac{d\theta_1d\theta_2}{(2\pi)^2}{\cal F}[{\cal C}(\theta_1,\theta_2)].
\end{equation}

The right-hand side of Eq.~(\ref{eq:103}) can be written as
\begin{eqnarray}\label{eq:104}
  \int\!\!\!\!\int_{-\pi}^\pi\frac{d\theta_1d\theta_2}{(2\pi)^2}{\cal F}[{\cal C}(\theta_1,\theta_2)]=\lim_{N_A\rightarrow \infty}\frac{\sum_{\nu_1,\nu_2=1}^{N_A}{\cal F}[{\cal C}_{\nu_1\nu_2}]}{(\sqrt{N_A}+1)^2},\quad\nonumber\\
  {\cal C}_{\nu_1\nu_2}\equiv{\cal C}\left(-\pi+\frac{2\nu_1\pi}{\sqrt{N_A}+1},-\pi+\frac{2\nu_2\pi}{\sqrt{N_A}+1}\right).\,\,\quad\quad
\end{eqnarray}
Construct the following sequence,
\begin{equation}\label{eq:105}
  \{{\tilde \lambda}_i(N_A)\}_{i=1}^{N_A}\equiv \bigcup_{\nu_1=1}^{\sqrt{N_A}}\{{\cal C}_{\nu_1\nu_2}\}_{\nu_2=1}^{\sqrt{N_A}}.
\end{equation}
Upon substituting Eqs.~(\ref{eq:104}) and (\ref{eq:105}) into Eq.~(\ref{eq:103}), we find that
\begin{equation}\label{eq:106}
  \lim_{N_A\rightarrow \infty}\frac{\sum_{i=1}^{N_A}{\cal F}(\lambda_i(N_A))}{N_A}=\lim_{N_A\rightarrow \infty}\frac{\sum_{i=1}^{N_A}{\cal F}({\tilde \lambda}_i(N_A))}{N_A}.
\end{equation}
In the terminology of Weyl's theory \cite{Grenander53}, this means that the two sequences $\{\lambda_i(N_A)\}_{i=1}^{N_A}$ and $\{{\tilde \lambda}_i(N_A)\}_{i=1}^{N_A}$ are {\it equally distributed} in the interval $[0,M]$. To better understand this let us introduce the probability distribution $\rho_\lambda$ for the former sequence (which is, in the terminology of probability theory, in the {\it weak convergence} sense), read
\begin{eqnarray}\label{eq:107}
  &&\int_0^M {\cal F}(\lambda)\rho_\lambda d\lambda\nonumber\\
  &\equiv& \lim_{N_A\rightarrow \infty}\frac{1}{N_A}\int_0^M {\cal F}(\lambda)\sum_{i=1}^{N_A}\delta(\lambda-\lambda_i(N_A))d\lambda,
\end{eqnarray}
and $\rho_{{\tilde \lambda}}$ for the latter sequence, read
\begin{eqnarray}\label{eq:108}
  &&\int_0^M {\cal F}({\tilde \lambda})\rho_{\tilde \lambda} d{\tilde \lambda}\nonumber\\
  &\equiv& \lim_{N_A\rightarrow \infty}\frac{1}{N_A}\int_0^M {\cal F}({\tilde \lambda})\sum_{i=1}^{N_A}\delta({\tilde \lambda}-{\tilde \lambda}_i(N_A))d{\tilde \lambda}.
\end{eqnarray}
Letting ${\cal F}(\lambda)$ be $\lambda^s, s\in \mathbb{N}\cup \{0\}$, by Eq.~(\ref{eq:106}) we have
\begin{equation}\label{eq:109}
  \int_0^M \lambda^s\rho_\lambda d\lambda=\int_0^M {\tilde \lambda}^s\rho_{{\tilde \lambda}} d{\tilde \lambda},\, \forall s\in \mathbb{N}\cup \{0\}.
\end{equation}
So all the moments of $\rho_\lambda$ and $\rho_{\tilde \lambda}$ are identical, and thus the two distributions must be identical. Since the Toeplitz operator has the largest eigenvalue $1$, we have $M=1$, which gives
\begin{equation}\label{eq:56}
    {\cal C}(\theta_1,\theta_2)\leq 1
\end{equation}
up to a set of zero Lebesgue measure.

{\section{Discussions on subsystem geometry}
\label{sec:convexity_subsystem_geometry}}

{In this appendix, we use a simple example to appreciate the technical importance of the constraint on subsystem's geometry, namely, being either a polygon or convex. We assume that the temperature is high enough so that $\left({\cal M}_{N_A}\right)_{ij}$ defined by Eq.~(\ref{eq:148}) decays exponentially with the distance between two lattice points: $i,j$ as $\sim e^{-|i-j|/a}$.}

{Observing Eq.~(\ref{eq:151}), we find that in order to study the boundary effects it is necessary to consider the following general expression,
\begin{eqnarray}\label{eq:188}
  &&\sum_{i_1,\cdots,i_k}^{N_A}\left({\cal M}_{N_A}\right)_{ii_{1}}\left({\cal M}_{N_A}\right)_{i_{1}i_{2}}\cdots\left({\cal M}_{N_A}\right)_{i_{k}j}\nonumber\\
  &\sim&\sum_{i_1,\cdots,i_k}^{N_A}e^{-\frac{1}{a}(|i-i_1|+|i_1-i_2|+\cdots+|i_k-j|)},
\end{eqnarray}
where $i,j$ are on the subsystem boundary. Without loss of generality let $|i-j|\gg 1$. We call $i\rightarrow i_1\rightarrow i_2\rightarrow \cdots i_k\rightarrow j$ a path from $i$ to $j$, with the bracket in the exponent being its length. Thanks to the elementary triangle inequality, we have
\begin{eqnarray}\label{eq:189}
|i-i_1|+|i_1-i_2|+\cdots+|i_k-j|\geq |i-j|,
\end{eqnarray}
where the equal sign is taken only if the path coincides with the straight line connecting $i$ and $j$. So, were the sum in Eq.~(\ref{eq:188}) extended to $\mathbb{Z}^2$, it is dominated by those paths around that straight line. Thus if the straight line is in the interior of A or on its boundary, which requires A to be either a polygon or convex, the contributions to the sum due to those paths going out of and returning to A are negligibly small, and the sum can be extended to $\mathbb{Z}^2$.}\\

\section{Proof of Eq.~(\ref{eq:173})}
\label{sec:proof}

We write the left-hand side of Eq.~(\ref{eq:173}) as
\begin{eqnarray}\label{eq:178}
    \int_0^\infty dx x J_0\left(ax\right) J_0\left(bx\right)=\lim_{\gamma\rightarrow 0}J(\gamma),\nonumber\\
\end{eqnarray}
where
\begin{eqnarray}\label{eq:181}
J(\gamma)\equiv\int_0^\infty dx x J_0\left(ax\right) J_0\left(bx\right)e^{-\gamma^2x^2}.
\end{eqnarray}
For $\gamma>0$ this integral can be carried out \cite{Gradshteyn},
\begin{eqnarray}\label{eq:182}
J(\gamma)=\frac{1}{2\gamma^2}e^{-\frac{1}{4\gamma^2}(a^2+b^2)}I_0(\frac{ab}{2\gamma^2}),
\end{eqnarray}
with $I_0(x)=J_0(ix)$ is the zeroth-order modified Bessel function of imaginary argument. With the help of the asymptotic expression $I_0(x)$ \cite{Gradshteyn},
\begin{equation}\label{eq:179}
  I_0(x\gg 1)\approx \frac{e^x}{\sqrt{2\pi x}},
\end{equation}
we obtain
\begin{eqnarray}\label{eq:180}
    J(\gamma)&\stackrel{\gamma\rightarrow 0}{\longrightarrow}&\frac{1}{\sqrt{4\pi\gamma^2ab}}e^{-\frac{1}{4\gamma^2}(a-b)^2}\nonumber\\
    &\rightarrow& a^{-1}\delta(a-b).
\end{eqnarray}
Thus the identity Eq.~(\ref{eq:173}) is proved.


\begin{thebibliography}{}
\bibitem{note1} Throughout this work the Boltzmann constant is set to unity.
\bibitem{note6} In general, for finite particle number the temperature in the Fermi-Dirac distribution and the thermodynamic temperature can be different \cite{Borhonovi16}. But this is not the case here.
\end{thebibliography}
\end{document}